\documentclass[10pt,superscriptaddress,twocolumn,amsmath,amssymb,aps,prb,reprint]{revtex4-2}

\makeatletter
\renewcommand{\frontmatter@footnote@produce}[1]{\@empty}
\makeatother

\usepackage{times}
\usepackage{amsfonts}
\usepackage{mathrsfs}
\usepackage[version=3]{mhchem}
\usepackage{graphicx}
\usepackage{dcolumn}
\usepackage{bm}
\usepackage{color}

\usepackage[colorlinks,bookmarks=false,citecolor=blue,linkcolor=red,urlcolor=blue]{hyperref}
\allowdisplaybreaks

\usepackage{physics}
\usepackage{tikz-feynman}
\usepackage{tikz,tikz-feynhand} 
\usepackage{subfigure}
\usepackage{float} 

\def\be{\begin{equation}}       \def\ee{\end{equation}}
\def\bea{\begin{eqnarray}}      \def\eea{\end{eqnarray}}

\begin{document}

	\title{Impact of Nonlocal Coulomb Repulsion on Superconductivity and Density-Wave Orders in Bilayer Nickelates}
	
	\author{Jun Zhan}
	\affiliation{Beijing National Laboratory for Condensed Matter Physics and Institute of Physics, Chinese Academy of Sciences, Beijing 100190, China}
	\affiliation{School of Physical Sciences, University of Chinese Academy of Sciences, Beijing 100190, China}
	
	\author{Congcong Le}
	\affiliation{RIKEN Interdisciplinary Theoretical and Mathematical Sciences (iTHEMS), Wako, Saitama 351-0198, Japan}

	\author{Xianxin Wu}\email{xxwu@itp.ac.cn}
	\affiliation{CAS Key Laboratory of Theoretical Physics, Institute of Theoretical Physics,
		Chinese Academy of Sciences, Beijing 100190, China}

	\author{Jiangping Hu}\email{jphu@iphy.ac.cn}
	\affiliation{Beijing National Laboratory for Condensed Matter Physics and Institute of Physics, Chinese Academy of Sciences, Beijing 100190, China}
	\affiliation{Kavli Institute for Theoretical Sciences, University of Chinese Academy of Sciences, Beijing 100190, China}
	\affiliation{ New Cornerstone Science Laboratory, Beijing 100190, China}
	\begin{abstract}

The recent discovery of high-temperature superconductivity in pressurized bilayer nickelate La$_3$Ni$_2$O$_7$ and its thin films has generated significant interest in uncovering the underlying pairing mechanisms and correlated electronic states. While earlier theoretical studies have mainly focused on onsite Coulomb interactions, the role of nonlocal Coulomb repulsion has remained largely unexplored. In this work, we systematically investigate the effects of nonlocal Coulomb interactions, in the presence of onsite interactions, on both superconducting and density-wave instabilities using the functional renormalization group (FRG) approach. We find that the interlayer intraorbital repulsion suppresses the interlayer intraorbital $s_{\pm}$-wave pairing and spin-density-wave (SDW) order, while promoting a transition to an interlayer interorbital $d_{x^2-y^2}$-wave pairing state and a mirror-symmetry-breaking charge order. Remarkably, the critical scale of the interorbital $d_{x^2-y^2}$-wave superconductivity is significantly lower than that of the intraorbital $s_{\pm}$-wave superconductivity, indicating that the former is unlikely to account for the observed high-$T_c$ superconductivity. Moreover, the interlayer interorbital repulsion suppresses this $d_{x^2-y^2}$-wave pairing but enhances the $s_{\pm}$-wave pairing through strengthened interlayer charge fluctuations. In addition, the intralayer nearest-neighbor repulsion favors an in-plane charge-density-wave (CDW) order with wave vector $(\pi,\pi)$. Our findings reveal the profound impact of nonlocal Coulomb repulsion and underscore the robustness of interlayer pairing rooted in the bilayer structure and multi-orbital nature, thereby advancing the understanding of the intricate correlation effects in bilayer nickelates.
		
	\end{abstract}
	
	\maketitle
	
 	\section{Introduction}

Exploring high-temperature superconductivity and revealing its pairing mechanism are among the most important and challenging problems in condensed matter physics~\cite{Scalapinothread}. High-$T_c$ cuprates and iron-based superconductors (IBS) have long served as two paradigmatic systems with distinctive characteristics: cuprates feature a $d^9$ configuration in an octahedral complex with a single active $d_{x^2-y^2}$ orbital~\cite{RevModPhys.78.17}, while IBS exhibit a $d^6$ configuration in a tetrahedral complex with three active $t_{2g}$ orbitals~\cite{Hirschfeld_2011}. Surprisingly, a new type of bilayer nickelate La$_3$Ni$_2$O$_7$ (LNO) in the Ruddlesden-Popper phase has recently been discovered to exhibit superconductivity under pressure with an extraordinary $T_c$ of nearly 80 K~\cite{sun2023}, introducing a third high-$T_c$ family. Different from its predecessors, LNO hosts Ni$^{2.5+}$ with a $d^{7.5}$ configuration, where both $d_{x^2-y^2}$ and $d_{z^2}$ orbitals contribute to the low-energy physics, enabled by the presence of apical oxygens bridging the layers~\cite{YaoDX,YZhang2023,Lechermann2023,Hirofumi2023possible,XWu,XJZhou2023,HHWen2023,Geisler20241,Geisler20242}. Under ambient conditions, LNO crystallizes in the \textit{Amam} phase and exhibits density wave orders at low pressures~\cite{Liu2022,ShuLeiSDW,chen2024electronic,Kakoi2024,plokhikh2025,yashima2025,Khasanov2025,ZHAO2025,Ren2025}. With increasing pressure, superconductivity suddenly emerges following a structural transition to the \textit{Fmmm}/\textit{I4mmm} phases, accompanied by reduced interlayer spacing and a stretching of the apical Ni–O–Ni bond angle from 168$^\circ$ toward 180$^\circ$. Additionally, the trilayer nickelate La$_4$Ni$_3$O$_{10}$ also exhibits superconductivity with a lower $T_c$ of 20–30 K when its ambient density wave order is suppressed under pressure~\cite{Kuroki2023T,HHWen2024T,JZhao2023T,YQi2023T,MWang2023T}. Strikingly, superconductivity with a $T_c$ exceeding 40 K has been achieved in compressively strained LNO thin films grown on SrLaAlO$_4$ substrates at ambient pressure~\cite{Ko2025,Zhou2025}. So far, the essential ingredients for superconductivity remain unclear; however, recent experiments reveal an intimate relationship between superconductivity and spin density wave (SDW) order~\cite{shi2025}, implying an important role of magnetic fluctuations.

Pairing mechanism of LNO has been under intensive theoretical study since the discovery of superconductivity and a consensus is yet to be reached~\cite{Wang327prb, lu2024interlayer,HYZhangtype2,XWu,EPC,FangYang327prl,WeiLi327prl,Hirofumi2023possible,YifengYang327prb,YifengYang327prb2,YiZhuangYouSMG,tian2023correlation,Dagotto327prb,zhang2024structural,Jiang_2024,PhysRevB.108.L201121,liao2023electron,ryee2024quenched,luo2023hightc,fan2023superconductivity,KuWeiprl,KJiang:17402,Schloemer2024,Andres2024}. Theoretical calculations suggest that the pressure-driven Lifshitz transition, i.e., the $d_{z^2}$ interlayer bonding state crossing the Fermi level to form a hole pocket, is crucial for bulk superconductivity under pressure~\cite{sun2023}. From the perspective of weak to intermediate coupling, spin fluctuations are believed to promote an $s_{\pm}$-wave pairing following the Lifshitz transition~\cite{Hirofumi2023possible,XWu,Wang327prb,FangYang327prl}, where interlayer pairing in the $d_{z^2}$ orbital is dominant. In the strong-coupling regime, the itinerant $d_{x^2-y^2}$ orbital, acquiring an effective exchange coupling through Hund's coupling~\cite{lu2024interlayer,HYZhangtype2,WeiLi327prl} or via interorbital hybridization between $d_{x^2-y^2}$ and $d_{z^2}$ orbitals~\cite{YifengYang327prb}, plays an important role in driving interlayer pairing of the $d_{x^2-y^2}$ orbital.  
While these studies have focused primarily on on-site Coulomb repulsion, the nonlocal Coulomb repulsion has been largely neglected. However, this interaction can be particularly relevant in LNO for several reasons: (1) the interatomic distance is reduced under pressure, enhancing nonlocal repulsion; (2) the two active orbitals, $d_{x^2-y^2}$ and $d_{z^2}$, exhibit strong hybridization with in-plane and apical oxygens, leading to extended Wannier functions; (3) the screening effect is weak in the direction perpendicular to the NiO$_2$ planes, promoting interlayer repulsion. Indeed, constrained random phase approximation calculations reveal that nearest-neighbor Coulomb repulsion is sizable and cannot (cRPA) be neglected~\cite{PWcorre}. The presence of nonlocal Coulomb repulsion, particularly the interlayer interaction, can profoundly affect the interlayer pairing and the instability of charge/spin density waves within the bilayer structure. Consequently, it is essential to clarify the impact of nonlocal interaction on the correlated states in LNO.

In this work, we comprehensively investigate the effects of nonlocal Coulomb repulsion on both superconductivity and density-wave instabilities in the presence of onsite interactions by performing functional renormalization group (FRG) calculations for LNO. We consider both interlayer and intralayer nonlocal repulsions, for scenarios with and without the $d_{z^2}$-bonding $\gamma$ hole pocket.  
In both Fermi surface topologies, we find that interlayer intraorbital repulsion suppresses the interlayer intraorbital $s_{\pm}$-wave pairing and drives a transition to interlayer interorbital $d_{x^2 - y^2}$-wave pairing accompanied by a charge order that breaks the mirror symmetry of the bilayer structure. In contrast, the inclusion of interlayer interorbital repulsion suppresses this $d_{x^2 - y^2}$-wave pairing while enhancing charge fluctuations, thereby promoting and stabilizing the $s_{\pm}$-wave pairing.  
In the presence of the $\gamma$ pocket, increasing interlayer intraorbital repulsion induces a sequential transition from $s_{\pm}$-wave superconductivity to a metallic phase, and subsequently to $d_{x^2 - y^2}$-wave superconductivity under moderate Hund’s coupling. In the absence of the $\gamma$ pocket, however, the interlayer intra-orbital repulsion directly drives a transition from either $s_{\pm}$-wave superconductivity or spin-density-wave (SDW) order to $d_{x^2 - y^2}$-wave superconductivity. This transition occurs across a wide range of charge doping.  
Remarkably, the critical scale of the $d_{x^2 - y^2}$ pairing is typically much lower than that of the corresponding $s_{\pm}$ pairing, suggesting that the $d_{x^2 - y^2}$ state is unlikely to account for the experimentally observed high-$T_c$ superconductivity. With inclusion of both interlayer intra- and inter-orbital repulsion, the $s_{\pm}$ pairing remains robust under realistic interaction settings.  
Additionally, the intralayer nearest-neighbor repulsion favors an in-plane charge density wave (CDW) order with wave vector $(\pi,\pi)$ by suppressing both $s_{\pm}$-wave superconductivity and SDW tendencies. These results are relevant for LNO under both ambient and high-pressure conditions, as well as for thin films subjected to in-plane strain. Finally, we discuss potential experimental implications of our findings.

\begin{figure}
	\centering
	\includegraphics[width=0.45\textwidth]{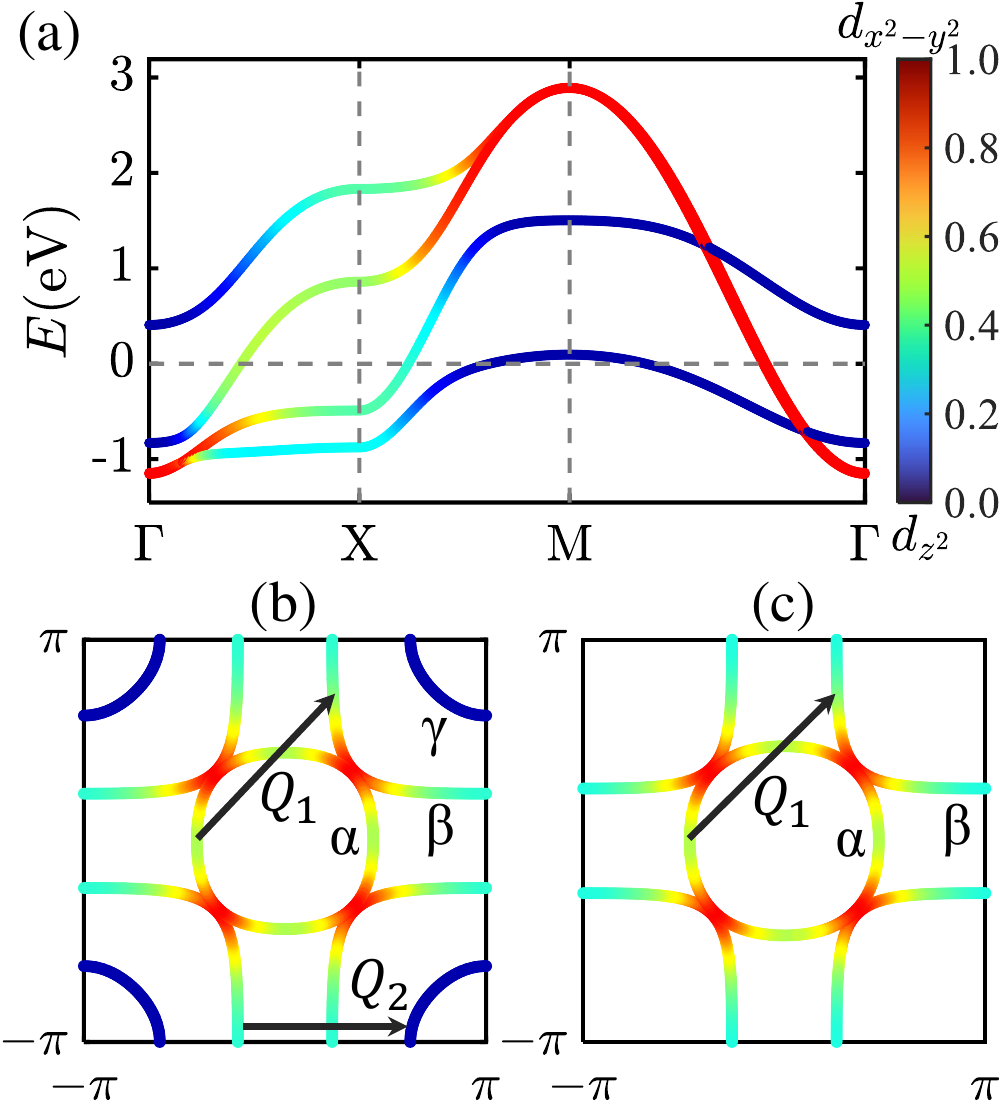}
	\caption{\textbf{Electronic structure and two representative Fermi surfaces of LNO.} 
		(a) Orbital-resolved band structure, with orbital contributions indicated by colors. 
		Fermi surfaces from the tight-binding model at fillings $n=3$ (b) and $n=3.33$ (c), 
		with the corresponding dominant nesting vectors labeled as $\mathbf{Q}_{1}$ and $\mathbf{Q}_{2}$.}
	\label{fig1}
\end{figure}

    \section{Model}
We start with the effective model for LNO. The low-energy electronic structure of bilayer nickelate LNO is dominated by Ni $d_{x^2-y^2}$ and $d_{z^2}$ orbitals according to theoretical calculations and experimental measurements~\cite{YaoDX,YZhang2023,Lechermann2023,Hirofumi2023possible,XWu,XJZhou2023,HHWen2023}. Therefore, the low-energy physics can be described by a two-orbital tight-binding (TB) Hamiltonian~\cite{YaoDX,YZhang2023,Lechermann2023,Hirofumi2023possible,XWu}, which reads
\begin{equation}
	\mathcal{H}_0 = \sum_{ij,\alpha\beta,\sigma} t_{\alpha\beta}^{ij} c_{i\alpha\sigma}^\dagger c_{j\beta\sigma} - \mu \sum_{i\alpha\sigma} c_{i\alpha\sigma}^\dagger c_{i\alpha\sigma}.
\end{equation}
Here, $i,j=(m,l)$ label the in-plane lattice site ($m$) and layer index ($l=t,b$), $\sigma$ labels spin, and $\alpha,\beta=x,z$ denote the Ni orbitals with $x$ representing $d_{x^2-y^2}$ and $z$ the $d_{z^2}$ orbital. $\mu$ is the chemical potential, and the hopping parameters are adopted from Ref.~\cite{XWu}. At high pressure, it is believed that the $d_{z^2}$ bonding state crosses the Fermi level and plays an important role in promoting superconductivity~\cite{sun2023}. The corresponding band structure is shown in Fig.~\ref{fig1}(a), where $d_{x^2-y^2}$ and $d_{z^2}$ orbitals are represented by red and blue colors, respectively. The Fermi surfaces (FS) at average filling $n=3$ per site (1.5 per Ni atom) are illustrated in Fig.~\ref{fig1}(b), which contain three pockets: the $\alpha$ electron pocket arising from the interlayer bonding state of the $d_{x^2-y^2}$ and $d_{z^2}$ orbitals, and the hole-like $\beta$ and $\gamma$ pockets originating mainly from the interlayer antibonding state of $d_{x^2-y^2}$ and the bonding state of $d_{z^2}$ orbitals, respectively. The antibonding $\beta$ pocket shows good nesting with the bonding $\alpha$ and $\gamma$ pockets at wave vectors $\mathbf{Q}_1$ and $\mathbf{Q}_2$, respectively. These nestings contribute to significant spin fluctuations at $\mathbf{Q}_{1,2}$ with interlayer antiferromagnetic coupling~\cite{CLe2025}. To study the correlated states with the same fermiology at ambient pressure, where only the $\alpha$ and $\beta$ pockets are present, we adjust the Fermi level to introduce \(1/3\) electron doping and the resulting FS are shown in Fig.~\ref{fig1}(c). The $\alpha$ and $\beta$ pockets still exhibit good nesting with wave vector $\mathbf{Q}_1$.

\begin{figure}[]
	\centering
	\includegraphics[width=0.48\textwidth]{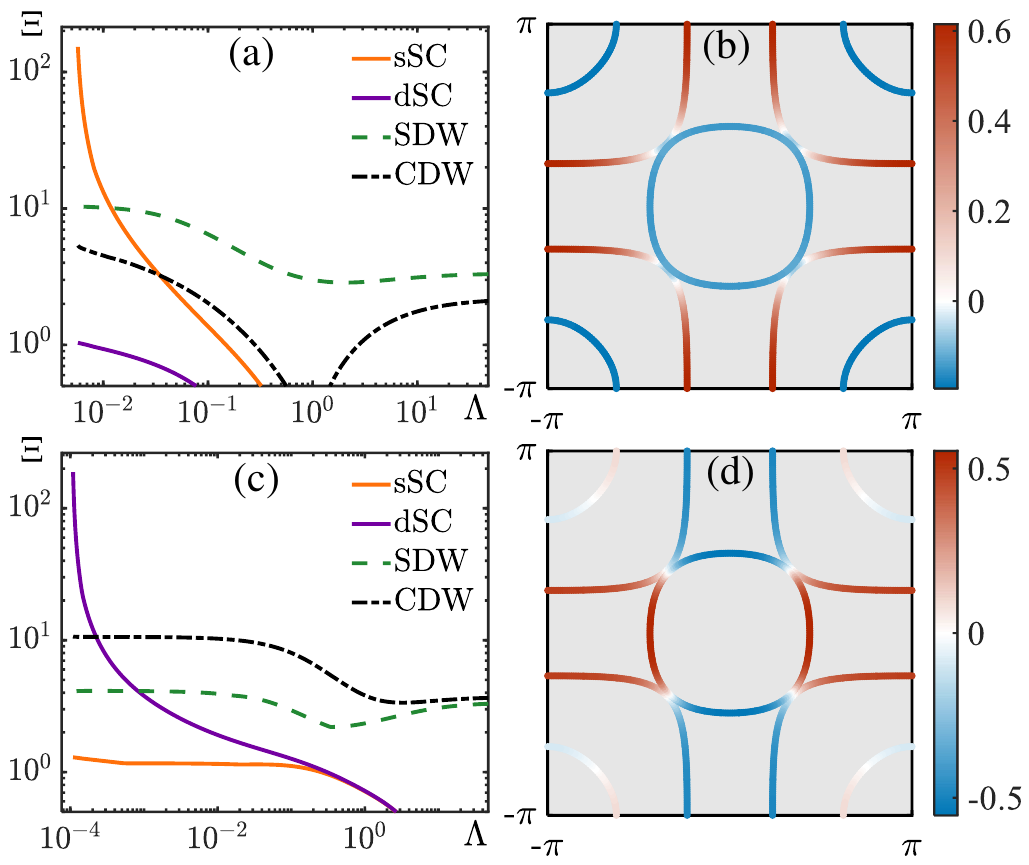}
	\caption{\textbf{FRG flow and leading SC gap with the $\gamma$ pocket.} 
		(a,b) FRG flow exhibiting the $s_{\pm}$ pairing instability and the leading superconducting gap functions on the Fermi surface for $J_{H}=0.3\,\mathrm{eV}$, $V_{\perp}=0$, and filling $n=3$. 
		(c,d) FRG flow exhibiting the $d_{x^2-y^2}$ pairing instability and the leading superconducting gap functions on the Fermi surface for $J_{H}=0.3\,\mathrm{eV}$, $V_{\perp}=1.4\,\mathrm{eV}$, and filling $n=3$. 
		Unless otherwise specified, $U=3\,\mathrm{eV}$ is used in the calculations.}
	\label{fig2}
\end{figure}

For the interacting part of the Hamiltonian, we consider the general multi-orbital on-site Hubbard interactions and nearest-neighbor Coulomb repulsions,
\begin{equation}
	\begin{aligned}
		\mathcal{H}_{\mathrm{EI}} &= \sum_{i\alpha} U n_{i\alpha\uparrow} n_{i\alpha\downarrow} 
		+ \sum_{i,\alpha\neq \beta} J_P c_{i\alpha\uparrow}^{\dagger} c_{i\alpha\downarrow}^{\dagger} c_{i\beta\downarrow} c_{i\beta\uparrow} \\
		&+ \sum_{i,\alpha< \beta,\sigma\sigma^{\prime}} \left( U^{\prime} n_{i\alpha\sigma} n_{i\beta\sigma^{\prime}} 
		+ J_H c_{i\alpha\sigma}^{\dagger} c_{i\beta\sigma} c_{i\beta\sigma^{\prime}}^{\dagger} c_{i\alpha\sigma^{\prime}} \right) \\
		&+ \sum_{\langle ij \rangle \alpha\beta\sigma\sigma^{\prime}} V_{ij}^{\alpha\beta} n_{i\alpha\sigma} n_{j\beta\sigma^{\prime}} .
	\end{aligned}
\end{equation}
Here, $U$ ($U^{\prime}$) denotes the on-site intra- (inter-)orbital Hubbard repulsion, $J_{H}$ is the Hund’s coupling, and $J_P$ is the pair-hopping interaction. The term $V_{ij}^{\alpha\beta}$ denotes the interlayer repulsion $V_{\perp}^{\alpha\beta}$ when $\langle ij \rangle$ is an out-of-plane nearest-neighbor bond, or the intralayer repulsion $V_{\parallel}^{\alpha\beta}$ when $\langle ij \rangle$ is an in-plane nearest-neighbor bond, between orbitals $\alpha$ and $\beta$. These nonlocal Coulomb repulsions, arising from the delocalized Wannier functions and weak screening effect, are the focus of this work. In the following, we adopt the standard Kanamori relations $U = U^{\prime} + 2J_{H}$ and $J_H = J_P$~\cite{Kanamori}. For the nonlocal intra-orbital repulsion, we take the approximate relations $V_{\perp}^{zz} = 2V_{\perp}^{xx} = V_{\perp}$ for interlayer terms and $V_{\parallel}^{xx} = 2V_{\parallel}^{zz} = V_{\parallel}$ for intralayer terms~\cite{PWcorre}, due to spatial orientation of these two orbitals. Our conclusions are robust against variations in these relations.

The correlated states have been studied from both weak- and strong-coupling perspectives, yet a consensus remains elusive. Recent experiments report that the magnetic order exhibits small local moments and itinerant characteristics~\cite{chen2024electronic,plokhikh2025,yashima2025,Ren2025}, reminiscent of IBS. Consequently, we employ the FRG approach, which treats all particle-hole and particle-particle channels on equal footing and provides a nuanced depiction of correlated states spanning from weak to intermediate coupling regimes~~\cite{10.1143/PTP.105.1,Metzner2012,Platt2013}. We implement FRG in the truncated-unity or singular-mode formalism~\cite{Lichtenstein2017,QHWangGraphene}, neglecting self-energy corrections, and use a hard Matsubara frequency cutoff as the regulator, with details given in Appendix~\ref{apA} and the Supplementary Material (SM). Compared with previous studies, our work systematically investigates the impact of interlayer and intralayer nonlocal repulsions on correlated states in the bilayer two-orbital model through the FRG approach.

\begin{figure}[t]
	\centering
	\includegraphics[width=0.5\textwidth]{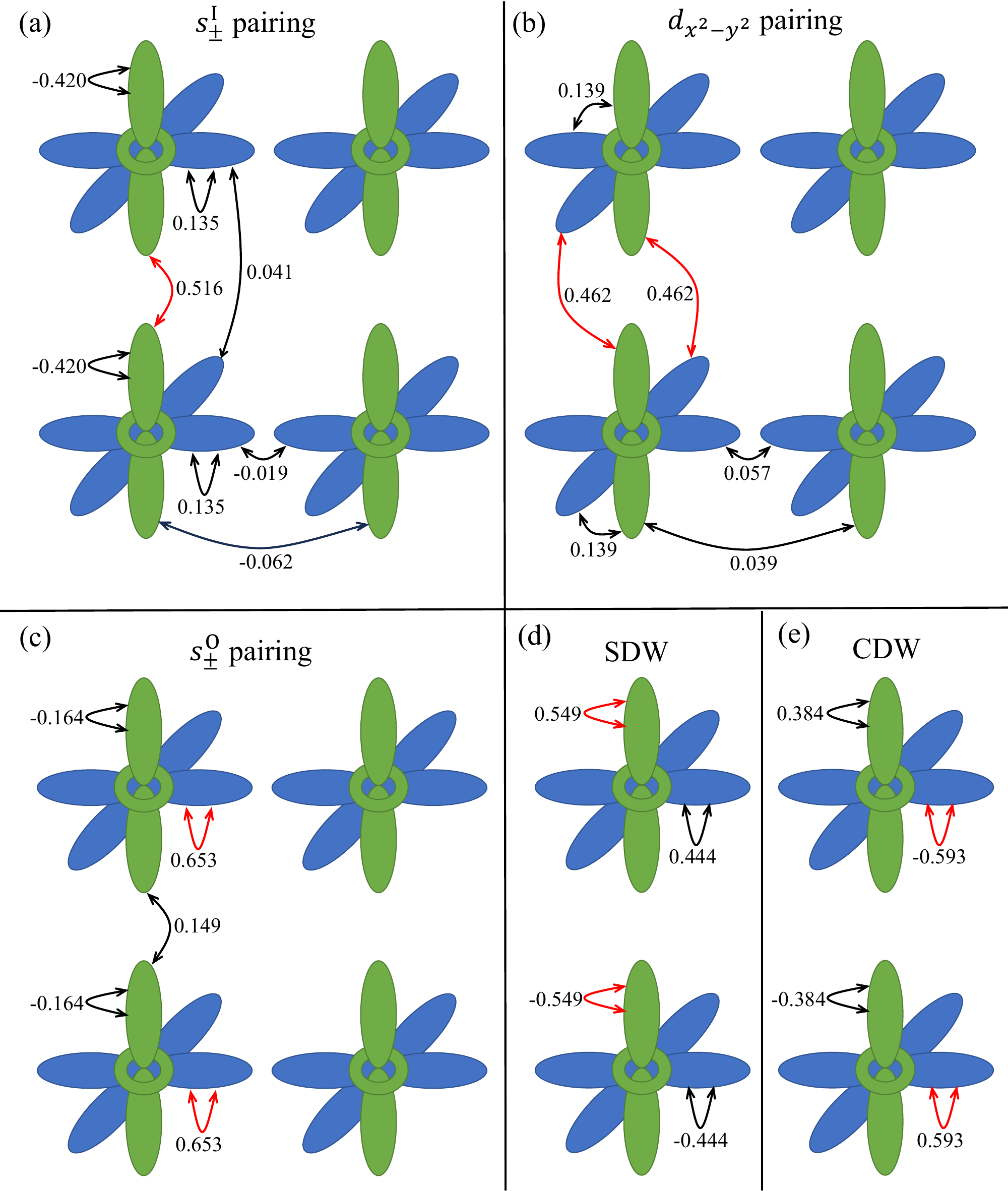}
	\caption{\textbf{Real-space order parameter components for particle-particle and particle-hole instabilities.} 
		Typical real-space order parameter patterns of 
		(a) interlayer-pairing-dominated $s_\pm^{\mathrm{I}}$ superconductivity at $J_{H}=0.3\,\mathrm{eV}$, $V_{\perp}=0$; 
		(b) $d_{x^2-y^2}$ pairing at $J_{H}=0.3\,\mathrm{eV}$, $V_{\perp}=1.4\,\mathrm{eV}$; 
		(c) intralayer-pairing-dominated $s_\pm^{\mathrm{O}}$ superconductivity at $J_{H}=0.3\,\mathrm{eV}$, $V_{\perp}=1.6\,\mathrm{eV}$; 
		(d) SDW at $J_{H}=0.6\,\mathrm{eV}$, $V_{\perp}=0$; 
		(e) charge order at $J_{H}=0.2\,\mathrm{eV}$, $V_{\perp}=1.4\,\mathrm{eV}$, $U=3\,\mathrm{eV}$, and $n=3$. 
		The superconducting order parameters are defined as 
		$\Delta^{\mathrm{SC}}_{i\alpha,j\beta}\left(c^{\dagger}_{i\alpha\uparrow}c^{\dagger}_{j\beta\downarrow} - c^{\dagger}_{i\alpha\downarrow}c^{\dagger}_{j\beta\uparrow}\right)$.
		The SDW order parameters are defined as 
		$\Delta^{\mathrm{SDW}}_{i\alpha} e^{i\bm{Q}\cdot \bm{R}_i} \left(c^{\dagger}_{i\alpha\uparrow}c_{i\alpha\uparrow} - c^{\dagger}_{i\alpha\downarrow}c_{i\alpha\downarrow}\right)$, 
		and the CDW order parameters as 
		$\Delta^{\mathrm{CDW}}_{i\alpha} e^{i\bm{Q}\cdot \bm{R}_i} \left(c^{\dagger}_{i\alpha\uparrow}c_{i\alpha\uparrow} + c^{\dagger}_{i\alpha\downarrow}c_{i\alpha\downarrow}\right)$. 
		Not all significant superconducting components are explicitly shown; the remaining components are obtained from lattice symmetry using the transformation properties of the corresponding irreducible representations. 
		Red lines indicate the dominant pairing components.}
	\label{figgap}
\end{figure}

\begin{figure*}[]
	\centering
	\includegraphics[width=0.85\textwidth]{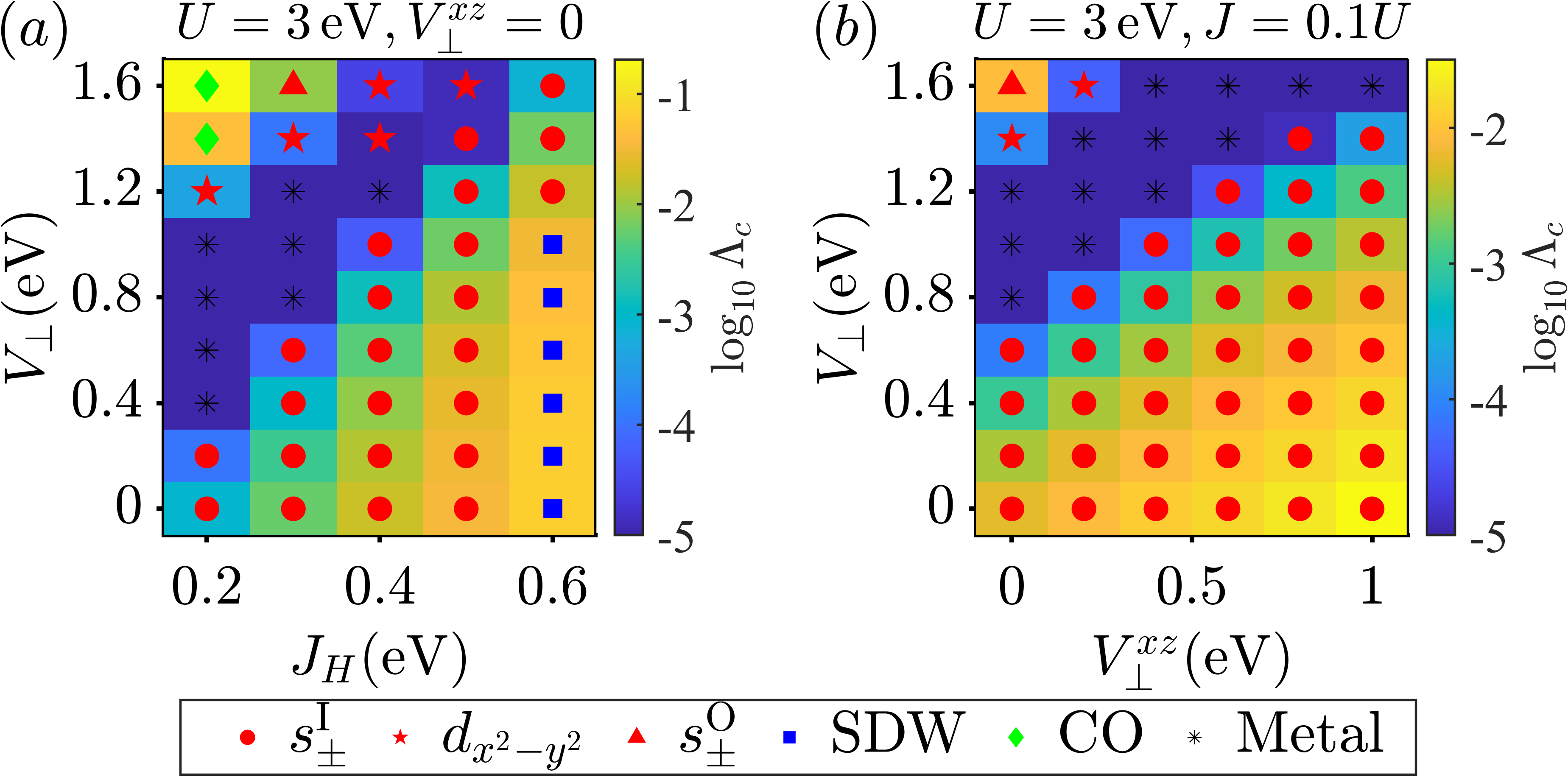}
	\caption{\textbf{FRG phase diagram with the $\gamma$ pocket.} 
		(a) Out-of-plane $V_{\perp}$--Hund's coupling $J_H$ phase diagram for $U=3\,\mathrm{eV}$ and $V_{\perp}^{zz}=2V_{\perp}^{xx}=V_{\perp}$ at filling $n=3$, with the color bar indicating the critical scale $\Lambda_c$ at which the corresponding FRG flow diverges. 
		(b) Interlayer interorbital repulsion $V_{\perp}^{xz}$ versus interlayer intraorbital repulsion $V_{\perp}$ phase diagram for $U=3\,\mathrm{eV}$, $J_{H}=0.1U$, and $n=3$. 
		Here, $s^\mathrm{I}_{\pm}$ and $s^\mathrm{O}_{\pm}$ denote interlayer-dominated and on-site-dominated $s_{\pm}$-wave superconductivity, respectively. 
		In the region marked by a black star, the FRG flow exhibits no divergence above the scale $10^{-5}$, and we label it as Metal.}
	\label{fig3}
\end{figure*}
	
	\section{Results}
	\subsection{Effect of interlayer repulsion with $\gamma$ pocket}
 
We first study the effect of interlayer repulsion for the case with the $\gamma$ pocket. With only on-site interactions, the typical FRG flow of leading eigenvalues $\Xi$ in different channels with representative parameters \( U = 3 \) eV and \( J_{H} = 0.1U \) is illustrated in Fig.~\ref{fig2}(a). 
The dominant instability is \( s_{\pm} \)-wave superconductivity, while SDW, CDW, and \( d_{x^2 - y^2} \)-wave superconductivity are subdominant. 
The \( s_{\pm} \) superconducting gap exhibits a sign change between the bonding \( \alpha, \gamma \) and antibonding \( \beta \) pockets, as depicted in Fig.~\ref{fig2}(b). This pairing state aligns with the scenario of spin fluctuation mediated pairing, where pockets connected by wave vectors \( \mathbf{Q}_1 \) and \( \mathbf{Q}_2 \) display sign-reversed gaps. In addition, the pairing dominantly involves interlayer pairing in the $d_{z^2}$ orbital, labeled as $s_{\pm}^{\mathrm{I}}$, according to the real-space pairing components shown in Fig.~\ref{figgap}(a). As the pairing gap on the bonding (antibonding) band is the sum (difference) of onsite and interlayer pairings, this interlayer-pairing dominated $s_{\pm}^{\mathrm{I}}$ pairing naturally leads to opposite-sign superconducting gaps on bonding and antibonding pockets.

The further inclusion of interlayer intraorbital repulsion suppresses the interlayer pairing and promotes interlayer intraorbital charge fluctuations. According to our calculations, $V_{\perp}$ gradually suppresses the $s^{\mathrm{I}}_{\pm}$-wave pairing, leading to a metallic phase in which the FRG flow exhibits no divergence above the scale of $10^{-5}$. With further increasing $V_{\perp}$, another \( d_{x^2 - y^2} \)-wave pairing emerges. Fig.~\ref{fig2}(c) shows a typical flow with an intermediate \( V_{\perp} \), where \( d_{x^2 - y^2} \) superconductivity diverges at a very low critical scale. It is apparent that SDW fluctuations are suppressed and CDW fluctuations are promoted, compared with the case of onsite interactions.  
As shown in Fig.~\ref{fig2}(d), the \( d_{x^2 - y^2} \)-wave gap function exhibits nodes along the $\Gamma-M$ direction and a nearly vanishing gap on the $\gamma$ pocket as well. Moreover, the gaps on the $\alpha$ and $\beta$ pockets are sign-reversed. To identify the origin, we calculate the real-space pairing components and plot them in Fig.~\ref{figgap}(b).
Remarkably, we observe that the in-plane pairing components are rather weak, and the dominant pairing occurs within the interlayer interorbital channel. This contrasts with the \( d_{x^2 - y^2} \) pairing in the single-orbital Hubbard model, where pairing occurs mainly between in-plane nearest-neighbor sites. The interorbital pairing results in maximum gaps on the $\alpha$ and $\beta$ pockets with mixed orbital characters, and a nearly vanishing gap on the $\gamma$ pocket due to its dominant $d_{z^2}$ weight, as shown in Fig.~\ref{fig1}(b). The interlayer pairing also induces sign-reversed gaps on the bonding ($\alpha$) and antibonding ($\beta$) pockets, dictated by the FS nesting at $\mathbf{Q}_1$. These explain the gap features in momentum space. 
The interlayer interorbital pairing, uniquely associated with the bilayer structure, becomes leading as the interlayer intraorbital pairing is suppressed and charge fluctuations are enhanced by the interlayer Coulomb repulsion. The enhanced charge fluctuations induced by $V_{\perp}$ mediate an effective attractive interaction between electrons in different orbitals which, in conjunction with spin fluctuations, promotes interlayer interorbital pairing (see SM for further details).

We present a representative phase diagram as a function of Hund’s coupling strength \( J_H \) and out-of-plane intraorbital repulsion \( V_{\perp} \) in Fig.~\ref{fig3}(a). The \( s^{\mathrm{I}}_{\pm} \) superconductivity dominates in the weak \( V_{\perp} \) regime. With increasing \( V_{\perp} \), the \( s^{\mathrm{I}}_{\pm} \) pairing gets suppressed to a metallic phase and further gives way to \( d_{x^2 - y^2} \) pairing for intermediate \( V_{\perp} \). This transition occurs in a wide \( J_H \) regime. 
There is also a competing spin-triplet \( p \)-wave pairing state near the phase boundary between the \( s_{\pm} \)-wave pairing or metal and the \( d_{x^2 - y^2} \)-wave pairing. This triplet pairing stems from strong ferromagnetic spin fluctuations associated with the flat \( d_{z^2} \) band around the M point.
In the regime of weak Hund's coupling \( J_H \), a charge order and \( s^{\mathrm{O}}_{\pm} \)-wave superconductivity can emerge as the leading instabilities. This charge order does not break the translational symmetry but breaks the out-of-plane mirror reflection, further introducing unequal occupancy between two orbitals on each site, as shown in Fig.~\ref{figgap}(e). 
The \( s^{\mathrm{O}}_{\pm} \)-wave pairing is characterized by dominant onsite pairing within the \( d_{x^2 - y^2} \) orbital, as shown in Fig.~\ref{figgap}(c), and arises due to strong charge and orbital fluctuations. Repulsive pair-hopping interaction leads to sign-reversed gaps on the \( d_{z^2} \) and \( d_{x^2 - y^2} \) orbitals. This results in the \( s^{\mathrm{O}}_{\pm} \) pairing exhibiting sign-changed gaps between the \( d_{z^2} \)-dominated \( \gamma \) pocket and the \( d_{x^2 - y^2} \)-dominated \( \alpha \) and \( \beta \) pockets.
In the regime of strong Hund's coupling \( J_H \), there is a direct transition from an SDW state to \( s^{\mathrm{I}}_{\pm} \)-wave superconductivity with increasing interlayer repulsion, originating from the suppressed interlayer antiferromagnetic coupling.

Furthermore, we investigate the effect of the interlayer interorbital repulsion \( V_{\perp}^{xz} \), as suggested by cRPA calculations~\cite{PWcorre}, which indicate that this type of repulsion is not negligible.  
Fig.~\ref{fig3}(b) presents the correlated phase diagram including both interlayer intraorbital repulsion \( V_{\perp}^{zz} = 2V_{\perp}^{xx} = V_{\perp} \) and interlayer interorbital repulsion \( V_{\perp}^{xz} \) with \( U = 3 \) eV and \( J_H = 0.1 U \). It is evident that the inclusion of \( V_{\perp}^{xz} \) enhances the interlayer \( s^{\mathrm{I}}_{\pm} \)-wave pairing, causing the transition into the metallic phase to occur at higher values of \( V_{\perp} \). Simultaneously, it suppresses the interlayer \( d_{x^2 - y^2} \)-wave pairing and the charge order. This behavior originates from enhanced interlayer charge fluctuations that strengthen the interlayer intraorbital pairing channel, assisted by \( V_{\perp}^{xz} \) (see SM).

	\subsection{Effect of interlayer repulsion without $\gamma$ pocket}

\begin{figure}[t]
	\centering
	\includegraphics[width=0.45\textwidth]{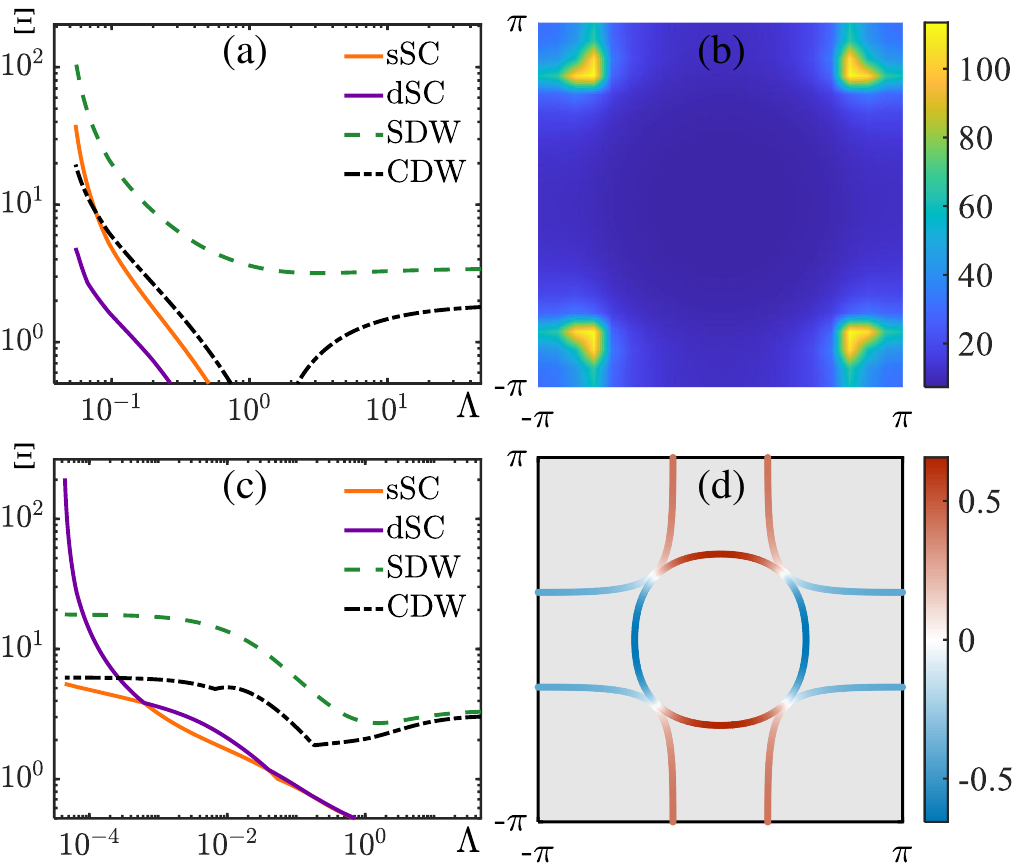}
	\caption{\textbf{FRG flow and leading SC gap without the $\gamma$ pocket.} 
		(a,b) FRG flow exhibiting the SDW instability and the $\mathbf{q}$ dependence of the leading effective interaction in the SDW channel for $U=3\,\mathrm{eV}$, $J_{H}=0.4\,\mathrm{eV}$, and $V_{\perp}=0$. 
		(c,d) FRG flow exhibiting the $d_{x^2-y^2}$ pairing instability and the leading superconducting gap functions on the Fermi surface for $U=3\,\mathrm{eV}$, $J_H=0.4\,\mathrm{eV}$, and $V_{\perp}=1.4\,\mathrm{eV}$.}
	\label{fig4}
\end{figure}

\begin{figure}[t]
	\centering
	\includegraphics[width=0.45\textwidth]{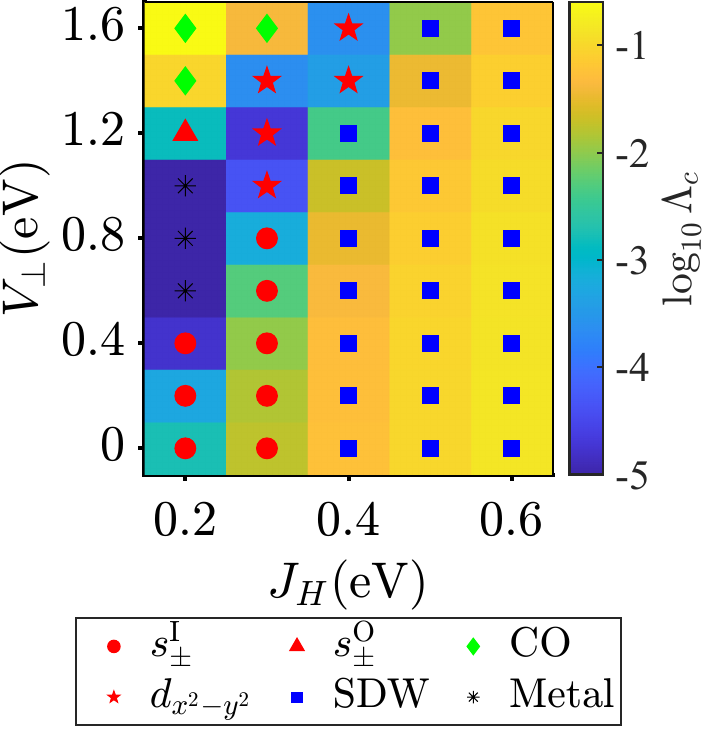}
	\caption{\textbf{FRG $J_{H}$--$V_{\perp}$ phase diagram without the $\gamma$ pocket.} 
		Out-of-plane $V_{\perp}$ versus Hund's coupling $J_{H}$ phase diagram for $U=3\,\mathrm{eV}$ and $V_{\perp}^{zz} = 2 V_{\perp}^{xx} = V_{\perp}$ at filling $n=3.33$.}
	\label{fig5}
\end{figure}

The $\gamma$ pocket is crucial to promoting the $s^\mathrm{I}_{\pm}$-wave pairing~\cite{CLe2025}. In this section, we study the impact of interlayer repulsion on the correlated state in the absence of the $\gamma$ pocket. To achieve this, we adjust the Fermi level to lie above the top of the $d_{z^2}$ bonding band, and the resulting Fermi surfaces are shown in Fig.~\ref{fig1}(c), where the $\alpha$ and $\beta$ pockets are almost identical to those in Fig.~\ref{fig1}(b).
In this case, the typical FRG flow with only onsite interactions is shown in Fig.~\ref{fig4}(a), where SDW fluctuations are significantly enhanced. The system develops an SDW instability with a wavevector of $\mathbf{Q}_1$ at a relatively high critical scale. The real-space pattern of this SDW is displayed in Fig.~\ref{figgap}(d), where the interlayer coupling is antiferromagnetic.
With the further inclusion of interlayer intraorbital repulsion, the SDW order is suppressed. Nonlocal repulsion indirectly suppresses SDW by enhancing charge fluctuations, which compete with spin fluctuations. When $V_{\perp}$ is above a critical value, the leading instability transitions directly from SDW to $d_{x^2 - y^2}$-wave superconductivity.
Fig.~\ref{fig4}(c) illustrates the FRG flow for a strong $V_{\perp}$, where charge fluctuations remain consistently weaker than magnetic fluctuations. The critical scale of $d_{x^2 - y^2}$-wave superconductivity is quite low, and the gap function is shown in Fig.~\ref{fig4}(d), both of which are similar to the case with the $\gamma$ pocket.
In the regime of weak Hund's coupling, as shown in Fig.~\ref{fig5}, $s^{\mathrm{I}}_{\pm}$-wave superconductivity emerges with only onsite interactions and transitions to $d_{x^2 - y^2}$-wave, $s^{\mathrm{O}}_{\pm}$-wave superconductivity, and out-of-plane charge order as $V_{\perp}$ increases. As the interorbital pairing vanishes on the $\gamma$ pocket, its absence has a negligible effect on $d_{x^2 - y^2}$-wave pairing.
With strong Hund's coupling, the SDW order dominates, and a large $V_{\perp}$ is needed to fully suppress it. Similar to the case with the $\gamma$ pocket, the further inclusion of interlayer interorbital repulsion $V_{\perp}^{xz}$ suppresses the $d_{x^2 - y^2}$-wave pairing and makes the $s_{\pm}^{\mathrm{I}}$-wave pairing more robust through enhanced interlayer charge fluctuations (see SM).

	\subsection{Effects of doping and intralayer repulsion}

\begin{figure}[t]
	\centering
	\includegraphics[width=0.45\textwidth]{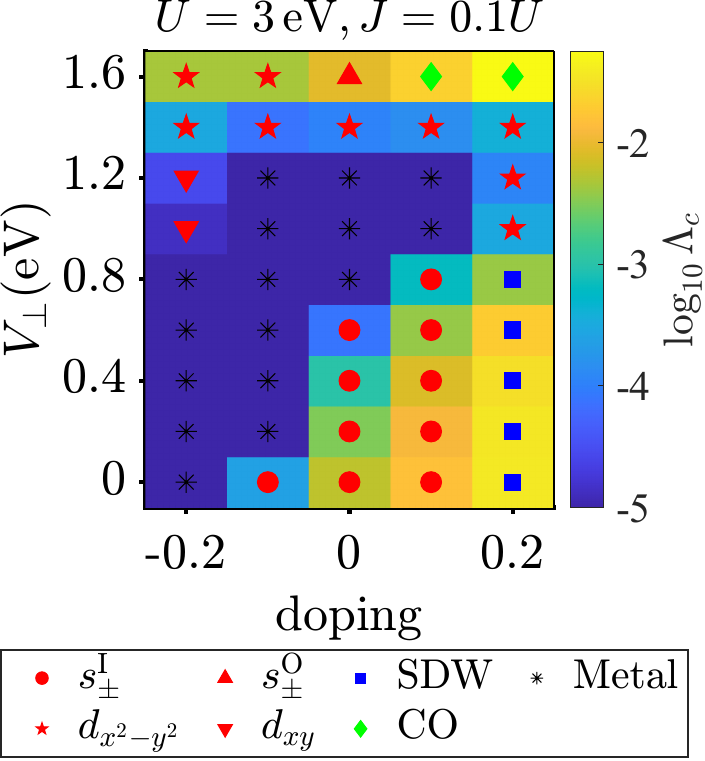}
	\caption{\textbf{FRG doping--$V_{\perp}$ phase diagram.} 
		Out-of-plane $V_{\perp}$ versus doping phase diagram for $U=3\,\mathrm{eV}$, $J_{H}=0.1U$, and $V_{\perp}^{zz} = 2 V_{\perp}^{xx} = V_{\perp}$.}
	\label{fig83}
\end{figure}

We further systematically study the effect of \( V_{\perp} \) as a function of doping. The resulting phase diagram of out-of-plane repulsion versus doping, with typical onsite interaction \( U = 3 \) eV and \( J = 0.1 U \), is shown in Fig.~\ref{fig83}. 
At low \( V_{\perp} \), electron doping enhances the transition temperature of the \( s^{\mathrm{I}}_{\pm} \)-wave pairing owing to the enhanced density of states and the suppression of pair-breaking fluctuations arising from nesting between the bonding $\alpha$ and $\gamma$ pockets~\cite{CLe2025}. However, excessive electron doping drives the system into the SDW state due to the elimination of the $\gamma$ pocket.
On the electron-doped side, the interlayer \( s_{\pm} \) pairing is relatively stable at low \( V_{\perp} \) regime. However, superconductivity can be destroyed with heavy hole doping. The transition from \( s^{\mathrm{I}}_{\pm} \)-wave pairing to \( d_{x^2 - y^2} \)-wave pairing with increasing \( V_{\perp} \) exists over a large range of charge doping.
As discussed above, the interlayer interorbital repulsion \( V_{\perp}^{xz} \) enhances the \( s^{\mathrm{I}}_{\pm} \)-wave pairing but suppresses the \( d_{x^2 - y^2} \)-wave pairing. Its inclusion will alter the phase boundaries between different phases.

\begin{figure}[t]
	\centering
	\includegraphics[width=0.42\textwidth]{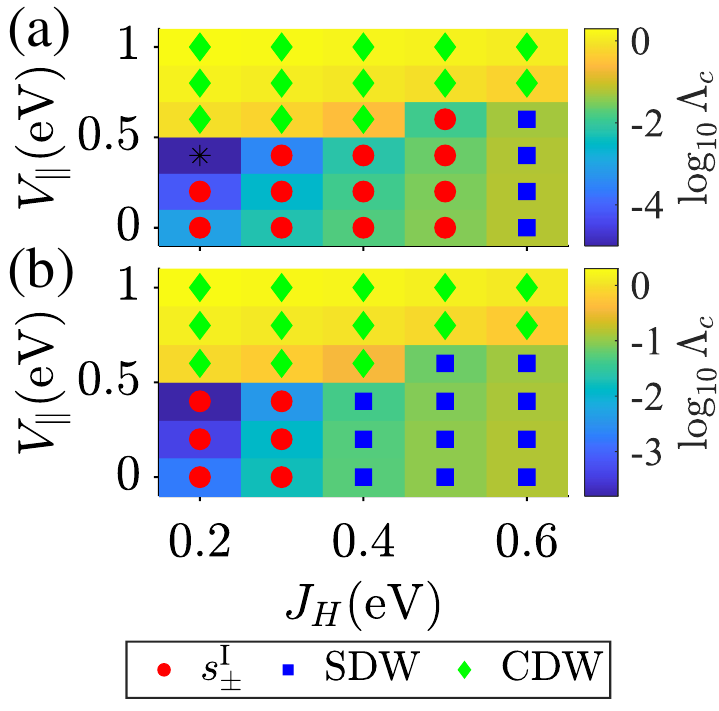}
	\caption{\textbf{FRG $J_{H}$--$V_{\parallel}$ phase diagram.} 
		In-plane repulsion $V_{\parallel}$ versus Hund's coupling $J_H$ phase diagram for $U=3\,\mathrm{eV}$, $V_{\parallel}^{xx} = 2 V_{\parallel}^{zz} = V_{\parallel}$ at fillings $n=3$ (a) and $n=3.33$ (b).}
	\label{fig6}
\end{figure}

As a comparison, we also investigate the effect of intralayer nearest-neighbor repulsion on the correlated state. Since all relevant pairings occur in the interlayer channel, the intralayer repulsion does not directly suppress pairing but rather weakens it indirectly by diminishing magnetic fluctuations. Our FRG calculations show that the intralayer repulsion suppresses both superconductivity and SDW, driving the system into a CDW order with wavevector $\mathbf{Q} = (\pi, \pi)$, as depicted in Fig.~\ref{fig6}. This occurs both with and without the $\gamma$ pocket.
The critical intralayer repulsion is approximately 0.5 eV for \( U = 3 \) eV and is independent of the presence of the $\gamma$ pocket. This value is significantly smaller than the critical interlayer repulsion required to drive a phase transition, owing to the larger coordination number associated with intralayer interactions.
The CDW exhibits an in-plane charge density modulation with $\mathbf{Q} = (\pi, \pi)$ to minimize the energy cost from the intralayer repulsion.

\section{Discussions and conclusions} 

Our findings reveal the profound influence of nonlocal Coulomb repulsion on the superconducting and density-wave instabilities as well as the intrinsic characteristics of the bilayer nickelate LNO. Conventionally, one would expect that interlayer repulsion suppresses interlayer pairing and pushes the pairing into the in-plane nearest-neighbor channel. However, owing to the multi-orbital nature, an interlayer interorbital pairing emerges to avoid strong interlayer intra-orbital repulsion. Additionally, even when both interlayer intraorbital and interorbital repulsions are included, the \( s^{\mathrm{I}}_{\pm} \)-wave pairing can be enhanced, and its transition to the \( d_{x^2 - y^2} \)-wave pairing persists, albeit at larger values of \( V_{\perp} \). 
The prominence of interlayer pairing under these interacting settings suggests its robustness within the bilayer structure, in contrast to cuprates and IBS. According to our calculations, the transition temperature of \( d_{x^2 - y^2} \)-wave superconductivity is usually much lower than that of \( s^{\mathrm{I}}_{\pm} \)-wave superconductivity, as the interlayer repulsion significantly suppresses interlayer magnetic couplings. This implies that \( d_{x^2 - y^2} \)-wave pairing may not be responsible for the observed high-\( T_c \) superconductivity in experiments. 
Moreover, according to cRPA calculations~\cite{PWcorre}, the relation between interlayer intraorbital and interorbital repulsion is estimated as \( V_{\perp}^{xz} \approx \frac{3}{4} V_{\perp}^{zz} \) under pressure. This suggests that the interlayer intraorbital \( s_{\pm} \)-wave pairing remains robust under realistic interlayer interaction settings.
Strong interlayer and intralayer repulsions can drive charge orders, breaking mirror reflection and translational symmetries, respectively. These orders can be entangled with SDW orders, generating the complex spin-charge stripe order observed in experiments~\cite{plokhikh2025,yashima2025,Ren2025}. The nonlocal Coulomb repulsion can further have an intricate interplay with interlayer electron-phonon coupling, influencing both superconductivity and density-wave orders~\cite{EPC}.

In conclusion, we have systematically explored the effect of nonlocal Coulomb repulsion on both superconductivity and density-wave orders in the presence of onsite interactions by performing FRG calculations. We find that interlayer repulsion suppresses the interlayer intra-orbital \( s_{\pm} \)-wave pairing and SDW order, driving a transition to interlayer inter-orbital \( d_{x^2 - y^2} \) pairing and a mirror-symmetry-breaking charge order. This transition occurs across a wide range of charge doping and is independent of the presence of the \(\gamma\) pocket.
However, the critical scale of the \( d_{x^2 - y^2} \)-wave superconductivity is significantly lower than that of the interlayer \( s_{\pm} \)-wave superconductivity, suggesting that this \( d_{x^2 - y^2} \)-wave pairing is unlikely to account for the experimentally observed high-\( T_c \) superconductivity. Intriguingly, the further inclusion of interlayer interorbital repulsion suppresses this \( d_{x^2 - y^2} \)-wave pairing but promotes the \( s_{\pm} \)-wave pairing due to enhanced interlayer charge fluctuations.
Additionally, the intralayer nearest-neighbor repulsion promotes an in-plane CDW order with wavevector \((\pi, \pi)\) by suppressing \( s_{\pm} \)-wave superconductivity and SDW order. Our results highlight the intriguing influence of nonlocal Coulomb repulsion and the robustness of interlayer pairing associated with the bilayer structure and multi-orbital nature, advancing the understanding of the complex electronic correlations in LNO.

{\it Acknowledgments}. We acknowledge the supports by National Natural Science Foundation of China (Grant No. 12494594, No. 11920101005, No. 11888101, No. 12047503, No. 12322405, No. 12104450), the Ministry of Science and Technology (Grant No. 2022YFA1403901), and the New Cornerstone Investigator Program. X.W. is supported by the National Key R\&D Program of China (Grant No. 2023YFA1407300) and the National Natural Science Foundation of China (Grant No. 12447103, No. 12447101). C.C.L. is supported by the RIKEN TRIP initiative (RIKEN Quantum).

{\it Note added}. During the preparation of this work, we became aware of independent studies of interlayer interaction effects in  La$_3$Ni$_2$O$_7$ using RPA~\cite{braz2025}, DMRG~\cite{borchia2025} FLEX~\cite{Xi2025}. The conclusion of \cite{Xi2025} that interlayer intraorbital Coulomb interactions can drive a transition from $s_{\pm}$ to $d_{x^2-y^2}$-wave pairing is consistent with our findings. In our work, we systematically explore the effects of nonlocal Coulomb repulsion, including both interlayer and intralayer interactions, on superconducting and density wave instabilities based on the unbiased FRG approach. Furthermore, we show that the additional inclusion of interlayer interorbital repulsion can enhance the stability of $s_{\pm}$ pairing.


\begin{thebibliography}{63}%
	\makeatletter
	\providecommand \@ifxundefined [1]{%
		\@ifx{#1\undefined}
	}%
	\providecommand \@ifnum [1]{%
		\ifnum #1\expandafter \@firstoftwo
		\else \expandafter \@secondoftwo
		\fi
	}%
	\providecommand \@ifx [1]{%
		\ifx #1\expandafter \@firstoftwo
		\else \expandafter \@secondoftwo
		\fi
	}%
	\providecommand \natexlab [1]{#1}%
	\providecommand \enquote  [1]{``#1''}%
	\providecommand \bibnamefont  [1]{#1}%
	\providecommand \bibfnamefont [1]{#1}%
	\providecommand \citenamefont [1]{#1}%
	\providecommand \href@noop [0]{\@secondoftwo}%
	\providecommand \href [0]{\begingroup \@sanitize@url \@href}%
	\providecommand \@href[1]{\@@startlink{#1}\@@href}%
	\providecommand \@@href[1]{\endgroup#1\@@endlink}%
	\providecommand \@sanitize@url [0]{\catcode `\\12\catcode `\$12\catcode
		`\&12\catcode `\#12\catcode `\^12\catcode `\_12\catcode `\%12\relax}%
	\providecommand \@@startlink[1]{}%
	\providecommand \@@endlink[0]{}%
	\providecommand \url  [0]{\begingroup\@sanitize@url \@url }%
	\providecommand \@url [1]{\endgroup\@href {#1}{\urlprefix }}%
	\providecommand \urlprefix  [0]{URL }%
	\providecommand \Eprint [0]{\href }%
	\providecommand \doibase [0]{http://dx.doi.org/}%
	\providecommand \selectlanguage [0]{\@gobble}%
	\providecommand \bibinfo  [0]{\@secondoftwo}%
	\providecommand \bibfield  [0]{\@secondoftwo}%
	\providecommand \translation [1]{[#1]}%
	\providecommand \BibitemOpen [0]{}%
	\providecommand \bibitemStop [0]{}%
	\providecommand \bibitemNoStop [0]{.\EOS\space}%
	\providecommand \EOS [0]{\spacefactor3000\relax}%
	\providecommand \BibitemShut  [1]{\csname bibitem#1\endcsname}%
	\let\auto@bib@innerbib\@empty
	\bibitem [{\citenamefont {Scalapino}(2012)}]{Scalapinothread}%
	\BibitemOpen
	\bibfield  {author} {\bibinfo {author} {\bibfnamefont {D.~J.}\ \bibnamefont
			{Scalapino}},\ }\href {\doibase 10.1103/RevModPhys.84.1383} {\bibfield
		{journal} {\bibinfo  {journal} {Rev. Mod. Phys.}\ }\textbf {\bibinfo {volume}
			{84}},\ \bibinfo {pages} {1383} (\bibinfo {year} {2012})}\BibitemShut
	{NoStop}%
	\bibitem [{\citenamefont {Lee}\ \emph {et~al.}(2006)\citenamefont {Lee},
		\citenamefont {Nagaosa},\ and\ \citenamefont {Wen}}]{RevModPhys.78.17}%
	\BibitemOpen
	\bibfield  {author} {\bibinfo {author} {\bibfnamefont {P.~A.}\ \bibnamefont
			{Lee}}, \bibinfo {author} {\bibfnamefont {N.}~\bibnamefont {Nagaosa}}, \ and\
		\bibinfo {author} {\bibfnamefont {X.-G.}\ \bibnamefont {Wen}},\ }\href
	{\doibase 10.1103/RevModPhys.78.17} {\bibfield  {journal} {\bibinfo
			{journal} {Rev. Mod. Phys.}\ }\textbf {\bibinfo {volume} {78}},\ \bibinfo
		{pages} {17} (\bibinfo {year} {2006})}\BibitemShut {NoStop}%
	\bibitem [{\citenamefont {Hirschfeld}\ \emph {et~al.}(2011)\citenamefont
		{Hirschfeld}, \citenamefont {Korshunov},\ and\ \citenamefont
		{Mazin}}]{Hirschfeld_2011}%
	\BibitemOpen
	\bibfield  {author} {\bibinfo {author} {\bibfnamefont {P.~J.}\ \bibnamefont
			{Hirschfeld}}, \bibinfo {author} {\bibfnamefont {M.~M.}\ \bibnamefont
			{Korshunov}}, \ and\ \bibinfo {author} {\bibfnamefont {I.~I.}\ \bibnamefont
			{Mazin}},\ }\href {\doibase 10.1088/0034-4885/74/12/124508} {\bibfield
		{journal} {\bibinfo  {journal} {Reports on Progress in Physics}\ }\textbf
		{\bibinfo {volume} {74}},\ \bibinfo {pages} {124508} (\bibinfo {year}
		{2011})}\BibitemShut {NoStop}%
	\bibitem [{\citenamefont {Sun}\ \emph {et~al.}(2023)\citenamefont {Sun},
		\citenamefont {Huo}, \citenamefont {Hu}, \citenamefont {Li}, \citenamefont
		{Liu}, \citenamefont {Han}, \citenamefont {Tang}, \citenamefont {Mao},
		\citenamefont {Yang}, \citenamefont {Wang}, \citenamefont {Cheng},
		\citenamefont {Yao}, \citenamefont {Zhang},\ and\ \citenamefont
		{Wang}}]{sun2023}%
	\BibitemOpen
	\bibfield  {author} {\bibinfo {author} {\bibfnamefont {H.}~\bibnamefont
			{Sun}}, \bibinfo {author} {\bibfnamefont {M.}~\bibnamefont {Huo}}, \bibinfo
		{author} {\bibfnamefont {X.}~\bibnamefont {Hu}}, \bibinfo {author}
		{\bibfnamefont {J.}~\bibnamefont {Li}}, \bibinfo {author} {\bibfnamefont
			{Z.}~\bibnamefont {Liu}}, \bibinfo {author} {\bibfnamefont {Y.}~\bibnamefont
			{Han}}, \bibinfo {author} {\bibfnamefont {L.}~\bibnamefont {Tang}}, \bibinfo
		{author} {\bibfnamefont {Z.}~\bibnamefont {Mao}}, \bibinfo {author}
		{\bibfnamefont {P.}~\bibnamefont {Yang}}, \bibinfo {author} {\bibfnamefont
			{B.}~\bibnamefont {Wang}}, \bibinfo {author} {\bibfnamefont {J.}~\bibnamefont
			{Cheng}}, \bibinfo {author} {\bibfnamefont {D.-X.}\ \bibnamefont {Yao}},
		\bibinfo {author} {\bibfnamefont {G.-M.}\ \bibnamefont {Zhang}}, \ and\
		\bibinfo {author} {\bibfnamefont {M.}~\bibnamefont {Wang}},\ }\href {\doibase
		10.1038/s41586-023-06408-7} {\bibfield  {journal} {\bibinfo  {journal}
			{Nature}\ }\textbf {\bibinfo {volume} {621}},\ \bibinfo {pages} {493}
		(\bibinfo {year} {2023})}\BibitemShut {NoStop}%
	\bibitem [{\citenamefont {Luo}\ \emph {et~al.}(2023)\citenamefont {Luo},
		\citenamefont {Hu}, \citenamefont {Wang}, \citenamefont {W\'u},\ and\
		\citenamefont {Yao}}]{YaoDX}%
	\BibitemOpen
	\bibfield  {author} {\bibinfo {author} {\bibfnamefont {Z.}~\bibnamefont
			{Luo}}, \bibinfo {author} {\bibfnamefont {X.}~\bibnamefont {Hu}}, \bibinfo
		{author} {\bibfnamefont {M.}~\bibnamefont {Wang}}, \bibinfo {author}
		{\bibfnamefont {W.}~\bibnamefont {W\'u}}, \ and\ \bibinfo {author}
		{\bibfnamefont {D.-X.}\ \bibnamefont {Yao}},\ }\href {\doibase
		10.1103/PhysRevLett.131.126001} {\bibfield  {journal} {\bibinfo  {journal}
			{Phys. Rev. Lett.}\ }\textbf {\bibinfo {volume} {131}},\ \bibinfo {pages}
		{126001} (\bibinfo {year} {2023})}\BibitemShut {NoStop}%
	\bibitem [{\citenamefont {Zhang}\ \emph
		{et~al.}(2023{\natexlab{a}})\citenamefont {Zhang}, \citenamefont {Lin},
		\citenamefont {Moreo},\ and\ \citenamefont {Dagotto}}]{YZhang2023}%
	\BibitemOpen
	\bibfield  {author} {\bibinfo {author} {\bibfnamefont {Y.}~\bibnamefont
			{Zhang}}, \bibinfo {author} {\bibfnamefont {L.-F.}\ \bibnamefont {Lin}},
		\bibinfo {author} {\bibfnamefont {A.}~\bibnamefont {Moreo}}, \ and\ \bibinfo
		{author} {\bibfnamefont {E.}~\bibnamefont {Dagotto}},\ }\href {\doibase
		10.1103/PhysRevB.108.L180510} {\bibfield  {journal} {\bibinfo  {journal}
			{Phys. Rev. B}\ }\textbf {\bibinfo {volume} {108}},\ \bibinfo {pages}
		{L180510} (\bibinfo {year} {2023}{\natexlab{a}})}\BibitemShut {NoStop}%
	\bibitem [{\citenamefont {Lechermann}\ \emph
		{et~al.}(2023{\natexlab{a}})\citenamefont {Lechermann}, \citenamefont
		{Gondolf}, \citenamefont {B\"otzel},\ and\ \citenamefont
		{Eremin}}]{Lechermann2023}%
	\BibitemOpen
	\bibfield  {author} {\bibinfo {author} {\bibfnamefont {F.}~\bibnamefont
			{Lechermann}}, \bibinfo {author} {\bibfnamefont {J.}~\bibnamefont {Gondolf}},
		\bibinfo {author} {\bibfnamefont {S.}~\bibnamefont {B\"otzel}}, \ and\
		\bibinfo {author} {\bibfnamefont {I.~M.}\ \bibnamefont {Eremin}},\ }\href
	{\doibase 10.1103/PhysRevB.108.L201121} {\bibfield  {journal} {\bibinfo
			{journal} {Phys. Rev. B}\ }\textbf {\bibinfo {volume} {108}},\ \bibinfo
		{pages} {L201121} (\bibinfo {year} {2023}{\natexlab{a}})}\BibitemShut
	{NoStop}%
	\bibitem [{\citenamefont {Sakakibara}\ \emph
		{et~al.}(2024{\natexlab{a}})\citenamefont {Sakakibara}, \citenamefont
		{Kitamine}, \citenamefont {Ochi},\ and\ \citenamefont
		{Kuroki}}]{Hirofumi2023possible}%
	\BibitemOpen
	\bibfield  {author} {\bibinfo {author} {\bibfnamefont {H.}~\bibnamefont
			{Sakakibara}}, \bibinfo {author} {\bibfnamefont {N.}~\bibnamefont
			{Kitamine}}, \bibinfo {author} {\bibfnamefont {M.}~\bibnamefont {Ochi}}, \
		and\ \bibinfo {author} {\bibfnamefont {K.}~\bibnamefont {Kuroki}},\ }\href
	{\doibase 10.1103/PhysRevLett.132.106002} {\bibfield  {journal} {\bibinfo
			{journal} {Phys. Rev. Lett.}\ }\textbf {\bibinfo {volume} {132}},\ \bibinfo
		{pages} {106002} (\bibinfo {year} {2024}{\natexlab{a}})}\BibitemShut
	{NoStop}%
	\bibitem [{\citenamefont {Gu}\ \emph {et~al.}(2025)\citenamefont {Gu},
		\citenamefont {Le}, \citenamefont {Yang}, \citenamefont {Wu},\ and\
		\citenamefont {Hu}}]{XWu}%
	\BibitemOpen
	\bibfield  {author} {\bibinfo {author} {\bibfnamefont {Y.}~\bibnamefont
			{Gu}}, \bibinfo {author} {\bibfnamefont {C.}~\bibnamefont {Le}}, \bibinfo
		{author} {\bibfnamefont {Z.}~\bibnamefont {Yang}}, \bibinfo {author}
		{\bibfnamefont {X.}~\bibnamefont {Wu}}, \ and\ \bibinfo {author}
		{\bibfnamefont {J.}~\bibnamefont {Hu}},\ }\href {\doibase
		10.1103/PhysRevB.111.174506} {\bibfield  {journal} {\bibinfo  {journal}
			{Phys. Rev. B}\ }\textbf {\bibinfo {volume} {111}},\ \bibinfo {pages}
		{174506} (\bibinfo {year} {2025})}\BibitemShut {NoStop}%
	\bibitem [{\citenamefont {Yang}\ \emph {et~al.}(2024)\citenamefont {Yang},
		\citenamefont {Sun}, \citenamefont {Hu}, \citenamefont {Xie}, \citenamefont
		{Miao}, \citenamefont {Luo}, \citenamefont {Chen}, \citenamefont {Liang},
		\citenamefont {Zhu}, \citenamefont {Qu}, \citenamefont {Chen}, \citenamefont
		{Huo}, \citenamefont {Huang}, \citenamefont {Zhang}, \citenamefont {Zhang},
		\citenamefont {Yang}, \citenamefont {Wang}, \citenamefont {Peng},
		\citenamefont {Mao}, \citenamefont {Liu}, \citenamefont {Xu}, \citenamefont
		{Qian}, \citenamefont {Yao}, \citenamefont {Wang}, \citenamefont {Zhao},\
		and\ \citenamefont {Zhou}}]{XJZhou2023}%
	\BibitemOpen
	\bibfield  {author} {\bibinfo {author} {\bibfnamefont {J.}~\bibnamefont
			{Yang}}, \bibinfo {author} {\bibfnamefont {H.}~\bibnamefont {Sun}}, \bibinfo
		{author} {\bibfnamefont {X.}~\bibnamefont {Hu}}, \bibinfo {author}
		{\bibfnamefont {Y.}~\bibnamefont {Xie}}, \bibinfo {author} {\bibfnamefont
			{T.}~\bibnamefont {Miao}}, \bibinfo {author} {\bibfnamefont {H.}~\bibnamefont
			{Luo}}, \bibinfo {author} {\bibfnamefont {H.}~\bibnamefont {Chen}}, \bibinfo
		{author} {\bibfnamefont {B.}~\bibnamefont {Liang}}, \bibinfo {author}
		{\bibfnamefont {W.}~\bibnamefont {Zhu}}, \bibinfo {author} {\bibfnamefont
			{G.}~\bibnamefont {Qu}}, \bibinfo {author} {\bibfnamefont {C.-Q.}\
			\bibnamefont {Chen}}, \bibinfo {author} {\bibfnamefont {M.}~\bibnamefont
			{Huo}}, \bibinfo {author} {\bibfnamefont {Y.}~\bibnamefont {Huang}}, \bibinfo
		{author} {\bibfnamefont {S.}~\bibnamefont {Zhang}}, \bibinfo {author}
		{\bibfnamefont {F.}~\bibnamefont {Zhang}}, \bibinfo {author} {\bibfnamefont
			{F.}~\bibnamefont {Yang}}, \bibinfo {author} {\bibfnamefont {Z.}~\bibnamefont
			{Wang}}, \bibinfo {author} {\bibfnamefont {Q.}~\bibnamefont {Peng}}, \bibinfo
		{author} {\bibfnamefont {H.}~\bibnamefont {Mao}}, \bibinfo {author}
		{\bibfnamefont {G.}~\bibnamefont {Liu}}, \bibinfo {author} {\bibfnamefont
			{Z.}~\bibnamefont {Xu}}, \bibinfo {author} {\bibfnamefont {T.}~\bibnamefont
			{Qian}}, \bibinfo {author} {\bibfnamefont {D.-X.}\ \bibnamefont {Yao}},
		\bibinfo {author} {\bibfnamefont {M.}~\bibnamefont {Wang}}, \bibinfo {author}
		{\bibfnamefont {L.}~\bibnamefont {Zhao}}, \ and\ \bibinfo {author}
		{\bibfnamefont {X.~J.}\ \bibnamefont {Zhou}},\ }\href {\doibase
		10.1038/s41467-024-48701-7} {\bibfield  {journal} {\bibinfo  {journal}
			{Nature Communications}\ }\textbf {\bibinfo {volume} {15}},\ \bibinfo {pages}
		{4373} (\bibinfo {year} {2024})}\BibitemShut {NoStop}%
	\bibitem [{\citenamefont {Liu}\ \emph {et~al.}(2024)\citenamefont {Liu},
		\citenamefont {Huo}, \citenamefont {Li}, \citenamefont {Li}, \citenamefont
		{Liu}, \citenamefont {Dai}, \citenamefont {Zhou}, \citenamefont {Hao},
		\citenamefont {Lu}, \citenamefont {Wang},\ and\ \citenamefont
		{Wen}}]{HHWen2023}%
	\BibitemOpen
	\bibfield  {author} {\bibinfo {author} {\bibfnamefont {Z.}~\bibnamefont
			{Liu}}, \bibinfo {author} {\bibfnamefont {M.}~\bibnamefont {Huo}}, \bibinfo
		{author} {\bibfnamefont {J.}~\bibnamefont {Li}}, \bibinfo {author}
		{\bibfnamefont {Q.}~\bibnamefont {Li}}, \bibinfo {author} {\bibfnamefont
			{Y.}~\bibnamefont {Liu}}, \bibinfo {author} {\bibfnamefont {Y.}~\bibnamefont
			{Dai}}, \bibinfo {author} {\bibfnamefont {X.}~\bibnamefont {Zhou}}, \bibinfo
		{author} {\bibfnamefont {J.}~\bibnamefont {Hao}}, \bibinfo {author}
		{\bibfnamefont {Y.}~\bibnamefont {Lu}}, \bibinfo {author} {\bibfnamefont
			{M.}~\bibnamefont {Wang}}, \ and\ \bibinfo {author} {\bibfnamefont {H.-H.}\
			\bibnamefont {Wen}},\ }\href {\doibase 10.1038/s41467-024-52001-5} {\bibfield
		{journal} {\bibinfo  {journal} {Nature Communications}\ }\textbf {\bibinfo
			{volume} {15}},\ \bibinfo {pages} {7570} (\bibinfo {year}
		{2024})}\BibitemShut {NoStop}%
	\bibitem [{\citenamefont {Geisler}\ \emph
		{et~al.}(2024{\natexlab{a}})\citenamefont {Geisler}, \citenamefont {Hamlin},
		\citenamefont {Stewart}, \citenamefont {Hennig},\ and\ \citenamefont
		{Hirschfeld}}]{Geisler20241}%
	\BibitemOpen
	\bibfield  {author} {\bibinfo {author} {\bibfnamefont {B.}~\bibnamefont
			{Geisler}}, \bibinfo {author} {\bibfnamefont {J.~J.}\ \bibnamefont {Hamlin}},
		\bibinfo {author} {\bibfnamefont {G.~R.}\ \bibnamefont {Stewart}}, \bibinfo
		{author} {\bibfnamefont {R.~G.}\ \bibnamefont {Hennig}}, \ and\ \bibinfo
		{author} {\bibfnamefont {P.~J.}\ \bibnamefont {Hirschfeld}},\ }\href
	{\doibase 10.1038/s41535-024-00648-0} {\bibfield  {journal} {\bibinfo
			{journal} {npj Quantum Materials}\ }\textbf {\bibinfo {volume} {9}},\
		\bibinfo {pages} {38} (\bibinfo {year} {2024}{\natexlab{a}})}\BibitemShut
	{NoStop}%
	\bibitem [{\citenamefont {Geisler}\ \emph
		{et~al.}(2024{\natexlab{b}})\citenamefont {Geisler}, \citenamefont
		{Fanfarillo}, \citenamefont {Hamlin}, \citenamefont {Stewart}, \citenamefont
		{Hennig},\ and\ \citenamefont {Hirschfeld}}]{Geisler20242}%
	\BibitemOpen
	\bibfield  {author} {\bibinfo {author} {\bibfnamefont {B.}~\bibnamefont
			{Geisler}}, \bibinfo {author} {\bibfnamefont {L.}~\bibnamefont {Fanfarillo}},
		\bibinfo {author} {\bibfnamefont {J.~J.}\ \bibnamefont {Hamlin}}, \bibinfo
		{author} {\bibfnamefont {G.~R.}\ \bibnamefont {Stewart}}, \bibinfo {author}
		{\bibfnamefont {R.~G.}\ \bibnamefont {Hennig}}, \ and\ \bibinfo {author}
		{\bibfnamefont {P.~J.}\ \bibnamefont {Hirschfeld}},\ }\href {\doibase
		10.1038/s41535-024-00690-y} {\bibfield  {journal} {\bibinfo  {journal} {npj
				Quantum Materials}\ }\textbf {\bibinfo {volume} {9}},\ \bibinfo {pages} {89}
		(\bibinfo {year} {2024}{\natexlab{b}})}\BibitemShut {NoStop}%
	\bibitem [{\citenamefont {Liu}\ \emph {et~al.}(2022)\citenamefont {Liu},
		\citenamefont {Sun}, \citenamefont {Huo}, \citenamefont {Ma}, \citenamefont
		{Ji}, \citenamefont {Yi}, \citenamefont {Li}, \citenamefont {Liu},
		\citenamefont {Yu}, \citenamefont {Zhang}, \citenamefont {Chen},
		\citenamefont {Liang}, \citenamefont {Dong}, \citenamefont {Guo},
		\citenamefont {Zhong}, \citenamefont {Shen}, \citenamefont {Li},\ and\
		\citenamefont {Wang}}]{Liu2022}%
	\BibitemOpen
	\bibfield  {author} {\bibinfo {author} {\bibfnamefont {Z.}~\bibnamefont
			{Liu}}, \bibinfo {author} {\bibfnamefont {H.}~\bibnamefont {Sun}}, \bibinfo
		{author} {\bibfnamefont {M.}~\bibnamefont {Huo}}, \bibinfo {author}
		{\bibfnamefont {X.}~\bibnamefont {Ma}}, \bibinfo {author} {\bibfnamefont
			{Y.}~\bibnamefont {Ji}}, \bibinfo {author} {\bibfnamefont {E.}~\bibnamefont
			{Yi}}, \bibinfo {author} {\bibfnamefont {L.}~\bibnamefont {Li}}, \bibinfo
		{author} {\bibfnamefont {H.}~\bibnamefont {Liu}}, \bibinfo {author}
		{\bibfnamefont {J.}~\bibnamefont {Yu}}, \bibinfo {author} {\bibfnamefont
			{Z.}~\bibnamefont {Zhang}}, \bibinfo {author} {\bibfnamefont
			{Z.}~\bibnamefont {Chen}}, \bibinfo {author} {\bibfnamefont {F.}~\bibnamefont
			{Liang}}, \bibinfo {author} {\bibfnamefont {H.}~\bibnamefont {Dong}},
		\bibinfo {author} {\bibfnamefont {H.}~\bibnamefont {Guo}}, \bibinfo {author}
		{\bibfnamefont {D.}~\bibnamefont {Zhong}}, \bibinfo {author} {\bibfnamefont
			{B.}~\bibnamefont {Shen}}, \bibinfo {author} {\bibfnamefont {S.}~\bibnamefont
			{Li}}, \ and\ \bibinfo {author} {\bibfnamefont {M.}~\bibnamefont {Wang}},\
	}\href {\doibase 10.1007/s11433-022-1962-4} {\bibfield  {journal} {\bibinfo
			{journal} {Science China Physics, Mechanics {\&} Astronomy}\ }\textbf
		{\bibinfo {volume} {66}},\ \bibinfo {pages} {217411} (\bibinfo {year}
		{2022})}\BibitemShut {NoStop}%
	\bibitem [{\citenamefont {Chen}\ \emph
		{et~al.}(2024{\natexlab{a}})\citenamefont {Chen}, \citenamefont {Liu},
		\citenamefont {Jiao}, \citenamefont {Zou}, \citenamefont {Jiang},
		\citenamefont {Li}, \citenamefont {Luo}, \citenamefont {Wu}, \citenamefont
		{Zhang}, \citenamefont {Guo},\ and\ \citenamefont {Shu}}]{ShuLeiSDW}%
	\BibitemOpen
	\bibfield  {author} {\bibinfo {author} {\bibfnamefont {K.}~\bibnamefont
			{Chen}}, \bibinfo {author} {\bibfnamefont {X.}~\bibnamefont {Liu}}, \bibinfo
		{author} {\bibfnamefont {J.}~\bibnamefont {Jiao}}, \bibinfo {author}
		{\bibfnamefont {M.}~\bibnamefont {Zou}}, \bibinfo {author} {\bibfnamefont
			{C.}~\bibnamefont {Jiang}}, \bibinfo {author} {\bibfnamefont
			{X.}~\bibnamefont {Li}}, \bibinfo {author} {\bibfnamefont {Y.}~\bibnamefont
			{Luo}}, \bibinfo {author} {\bibfnamefont {Q.}~\bibnamefont {Wu}}, \bibinfo
		{author} {\bibfnamefont {N.}~\bibnamefont {Zhang}}, \bibinfo {author}
		{\bibfnamefont {Y.}~\bibnamefont {Guo}}, \ and\ \bibinfo {author}
		{\bibfnamefont {L.}~\bibnamefont {Shu}},\ }\href {\doibase
		10.1103/PhysRevLett.132.256503} {\bibfield  {journal} {\bibinfo  {journal}
			{Phys. Rev. Lett.}\ }\textbf {\bibinfo {volume} {132}},\ \bibinfo {pages}
		{256503} (\bibinfo {year} {2024}{\natexlab{a}})}\BibitemShut {NoStop}%
	\bibitem [{\citenamefont {Chen}\ \emph
		{et~al.}(2024{\natexlab{b}})\citenamefont {Chen}, \citenamefont {Choi},
		\citenamefont {Jiang}, \citenamefont {Mei}, \citenamefont {Jiang},
		\citenamefont {Li}, \citenamefont {Agrestini}, \citenamefont
		{Garcia-Fernandez}, \citenamefont {Sun}, \citenamefont {Huang}, \citenamefont
		{Shen}, \citenamefont {Wang}, \citenamefont {Hu}, \citenamefont {Lu},
		\citenamefont {Zhou},\ and\ \citenamefont {Feng}}]{chen2024electronic}%
	\BibitemOpen
	\bibfield  {author} {\bibinfo {author} {\bibfnamefont {X.}~\bibnamefont
			{Chen}}, \bibinfo {author} {\bibfnamefont {J.}~\bibnamefont {Choi}}, \bibinfo
		{author} {\bibfnamefont {Z.}~\bibnamefont {Jiang}}, \bibinfo {author}
		{\bibfnamefont {J.}~\bibnamefont {Mei}}, \bibinfo {author} {\bibfnamefont
			{K.}~\bibnamefont {Jiang}}, \bibinfo {author} {\bibfnamefont
			{J.}~\bibnamefont {Li}}, \bibinfo {author} {\bibfnamefont {S.}~\bibnamefont
			{Agrestini}}, \bibinfo {author} {\bibfnamefont {M.}~\bibnamefont
			{Garcia-Fernandez}}, \bibinfo {author} {\bibfnamefont {H.}~\bibnamefont
			{Sun}}, \bibinfo {author} {\bibfnamefont {X.}~\bibnamefont {Huang}}, \bibinfo
		{author} {\bibfnamefont {D.}~\bibnamefont {Shen}}, \bibinfo {author}
		{\bibfnamefont {M.}~\bibnamefont {Wang}}, \bibinfo {author} {\bibfnamefont
			{J.}~\bibnamefont {Hu}}, \bibinfo {author} {\bibfnamefont {Y.}~\bibnamefont
			{Lu}}, \bibinfo {author} {\bibfnamefont {K.-J.}\ \bibnamefont {Zhou}}, \ and\
		\bibinfo {author} {\bibfnamefont {D.}~\bibnamefont {Feng}},\ }\href {\doibase
		10.1038/s41467-024-53863-5} {\bibfield  {journal} {\bibinfo  {journal}
			{Nature Communications}\ }\textbf {\bibinfo {volume} {15}},\ \bibinfo {pages}
		{9597} (\bibinfo {year} {2024}{\natexlab{b}})}\BibitemShut {NoStop}%
	\bibitem [{\citenamefont {Kakoi}\ \emph {et~al.}(2024)\citenamefont {Kakoi},
		\citenamefont {Oi}, \citenamefont {Ohshita}, \citenamefont {Yashima},
		\citenamefont {Kuroki}, \citenamefont {Kato}, \citenamefont {Takahashi},
		\citenamefont {Ishiwata}, \citenamefont {Adachi}, \citenamefont {Hatada},
		\citenamefont {Uda},\ and\ \citenamefont {Mukuda}}]{Kakoi2024}%
	\BibitemOpen
	\bibfield  {author} {\bibinfo {author} {\bibfnamefont {M.}~\bibnamefont
			{Kakoi}}, \bibinfo {author} {\bibfnamefont {T.}~\bibnamefont {Oi}}, \bibinfo
		{author} {\bibfnamefont {Y.}~\bibnamefont {Ohshita}}, \bibinfo {author}
		{\bibfnamefont {M.}~\bibnamefont {Yashima}}, \bibinfo {author} {\bibfnamefont
			{K.}~\bibnamefont {Kuroki}}, \bibinfo {author} {\bibfnamefont
			{T.}~\bibnamefont {Kato}}, \bibinfo {author} {\bibfnamefont {H.}~\bibnamefont
			{Takahashi}}, \bibinfo {author} {\bibfnamefont {S.}~\bibnamefont {Ishiwata}},
		\bibinfo {author} {\bibfnamefont {Y.}~\bibnamefont {Adachi}}, \bibinfo
		{author} {\bibfnamefont {N.}~\bibnamefont {Hatada}}, \bibinfo {author}
		{\bibfnamefont {T.}~\bibnamefont {Uda}}, \ and\ \bibinfo {author}
		{\bibfnamefont {H.}~\bibnamefont {Mukuda}},\ }\href {\doibase
		10.7566/JPSJ.93.053702} {\bibfield  {journal} {\bibinfo  {journal} {Journal
				of the Physical Society of Japan}\ }\textbf {\bibinfo {volume} {93}},\
		\bibinfo {pages} {053702} (\bibinfo {year} {2024})}\BibitemShut {NoStop}%
	\bibitem [{\citenamefont {Plokhikh}\ \emph {et~al.}(2025)\citenamefont
		{Plokhikh}, \citenamefont {Hicken}, \citenamefont {Keller}, \citenamefont
		{Pomjakushin}, \citenamefont {Moody}, \citenamefont {Foury-Leylekian},
		\citenamefont {Krieger}, \citenamefont {Luetkens}, \citenamefont {Guguchia},
		\citenamefont {Khasanov},\ and\ \citenamefont {Gawryluk}}]{plokhikh2025}%
	\BibitemOpen
	\bibfield  {author} {\bibinfo {author} {\bibfnamefont {I.}~\bibnamefont
			{Plokhikh}}, \bibinfo {author} {\bibfnamefont {T.~J.}\ \bibnamefont
			{Hicken}}, \bibinfo {author} {\bibfnamefont {L.}~\bibnamefont {Keller}},
		\bibinfo {author} {\bibfnamefont {V.}~\bibnamefont {Pomjakushin}}, \bibinfo
		{author} {\bibfnamefont {S.~H.}\ \bibnamefont {Moody}}, \bibinfo {author}
		{\bibfnamefont {P.}~\bibnamefont {Foury-Leylekian}}, \bibinfo {author}
		{\bibfnamefont {J.~J.}\ \bibnamefont {Krieger}}, \bibinfo {author}
		{\bibfnamefont {H.}~\bibnamefont {Luetkens}}, \bibinfo {author}
		{\bibfnamefont {Z.}~\bibnamefont {Guguchia}}, \bibinfo {author}
		{\bibfnamefont {R.}~\bibnamefont {Khasanov}}, \ and\ \bibinfo {author}
		{\bibfnamefont {D.~J.}\ \bibnamefont {Gawryluk}},\ }\href
	{https://arxiv.org/abs/2503.05287} {} (\bibinfo {year} {2025}),\ \Eprint
	{http://arxiv.org/abs/2503.05287} {arXiv:2503.05287} \BibitemShut {NoStop}%
	\bibitem [{\citenamefont {Yashima}\ \emph {et~al.}(2025)\citenamefont
		{Yashima}, \citenamefont {Seto}, \citenamefont {Oshita}, \citenamefont
		{Kakoi}, \citenamefont {Sakurai}, \citenamefont {Takano},\ and\ \citenamefont
		{Mukuda}}]{yashima2025}%
	\BibitemOpen
	\bibfield  {author} {\bibinfo {author} {\bibfnamefont {M.}~\bibnamefont
			{Yashima}}, \bibinfo {author} {\bibfnamefont {N.}~\bibnamefont {Seto}},
		\bibinfo {author} {\bibfnamefont {Y.}~\bibnamefont {Oshita}}, \bibinfo
		{author} {\bibfnamefont {M.}~\bibnamefont {Kakoi}}, \bibinfo {author}
		{\bibfnamefont {H.}~\bibnamefont {Sakurai}}, \bibinfo {author} {\bibfnamefont
			{Y.}~\bibnamefont {Takano}}, \ and\ \bibinfo {author} {\bibfnamefont
			{H.}~\bibnamefont {Mukuda}},\ }\href {\doibase 10.7566/JPSJ.94.054704}
	{\bibfield  {journal} {\bibinfo  {journal} {Journal of the Physical Society
				of Japan}\ }\textbf {\bibinfo {volume} {94}},\ \bibinfo {pages} {054704}
		(\bibinfo {year} {2025})}\BibitemShut {NoStop}%
	\bibitem [{\citenamefont {Khasanov}\ \emph {et~al.}(2025)\citenamefont
		{Khasanov}, \citenamefont {Hicken}, \citenamefont {Gawryluk}, \citenamefont
		{Sazgari}, \citenamefont {Plokhikh}, \citenamefont {Sorel}, \citenamefont
		{Bartkowiak}, \citenamefont {B{\"o}tzel}, \citenamefont {Lechermann},
		\citenamefont {Eremin}, \citenamefont {Luetkens},\ and\ \citenamefont
		{Guguchia}}]{Khasanov2025}%
	\BibitemOpen
	\bibfield  {author} {\bibinfo {author} {\bibfnamefont {R.}~\bibnamefont
			{Khasanov}}, \bibinfo {author} {\bibfnamefont {T.~J.}\ \bibnamefont
			{Hicken}}, \bibinfo {author} {\bibfnamefont {D.~J.}\ \bibnamefont
			{Gawryluk}}, \bibinfo {author} {\bibfnamefont {V.}~\bibnamefont {Sazgari}},
		\bibinfo {author} {\bibfnamefont {I.}~\bibnamefont {Plokhikh}}, \bibinfo
		{author} {\bibfnamefont {L.~P.}\ \bibnamefont {Sorel}}, \bibinfo {author}
		{\bibfnamefont {M.}~\bibnamefont {Bartkowiak}}, \bibinfo {author}
		{\bibfnamefont {S.}~\bibnamefont {B{\"o}tzel}}, \bibinfo {author}
		{\bibfnamefont {F.}~\bibnamefont {Lechermann}}, \bibinfo {author}
		{\bibfnamefont {I.~M.}\ \bibnamefont {Eremin}}, \bibinfo {author}
		{\bibfnamefont {H.}~\bibnamefont {Luetkens}}, \ and\ \bibinfo {author}
		{\bibfnamefont {Z.}~\bibnamefont {Guguchia}},\ }\href {\doibase
		10.1038/s41567-024-02754-z} {\bibfield  {journal} {\bibinfo  {journal}
			{Nature Physics}\ }\textbf {\bibinfo {volume} {21}},\ \bibinfo {pages} {430}
		(\bibinfo {year} {2025})}\BibitemShut {NoStop}%
	\bibitem [{\citenamefont {Zhao}\ \emph {et~al.}(2025)\citenamefont {Zhao},
		\citenamefont {Zhou}, \citenamefont {Huo}, \citenamefont {Wang},
		\citenamefont {Nie}, \citenamefont {Yang}, \citenamefont {Ying},
		\citenamefont {Wang}, \citenamefont {Wu},\ and\ \citenamefont
		{Chen}}]{ZHAO2025}%
	\BibitemOpen
	\bibfield  {author} {\bibinfo {author} {\bibfnamefont {D.}~\bibnamefont
			{Zhao}}, \bibinfo {author} {\bibfnamefont {Y.}~\bibnamefont {Zhou}}, \bibinfo
		{author} {\bibfnamefont {M.}~\bibnamefont {Huo}}, \bibinfo {author}
		{\bibfnamefont {Y.}~\bibnamefont {Wang}}, \bibinfo {author} {\bibfnamefont
			{L.}~\bibnamefont {Nie}}, \bibinfo {author} {\bibfnamefont {Y.}~\bibnamefont
			{Yang}}, \bibinfo {author} {\bibfnamefont {J.}~\bibnamefont {Ying}}, \bibinfo
		{author} {\bibfnamefont {M.}~\bibnamefont {Wang}}, \bibinfo {author}
		{\bibfnamefont {T.}~\bibnamefont {Wu}}, \ and\ \bibinfo {author}
		{\bibfnamefont {X.}~\bibnamefont {Chen}},\ }\href {\doibase
		https://doi.org/10.1016/j.scib.2025.02.019} {\bibfield  {journal} {\bibinfo
			{journal} {Science Bulletin}\ } (\bibinfo {year} {2025}),\
		https://doi.org/10.1016/j.scib.2025.02.019}\BibitemShut {NoStop}%
	\bibitem [{\citenamefont {Ren}\ \emph {et~al.}(2025)\citenamefont {Ren},
		\citenamefont {Sutarto}, \citenamefont {Wu}, \citenamefont {Zhang},
		\citenamefont {Huang}, \citenamefont {Xiang}, \citenamefont {Hu},
		\citenamefont {Comin}, \citenamefont {Zhou},\ and\ \citenamefont
		{Zhu}}]{Ren2025}%
	\BibitemOpen
	\bibfield  {author} {\bibinfo {author} {\bibfnamefont {X.}~\bibnamefont
			{Ren}}, \bibinfo {author} {\bibfnamefont {R.}~\bibnamefont {Sutarto}},
		\bibinfo {author} {\bibfnamefont {X.}~\bibnamefont {Wu}}, \bibinfo {author}
		{\bibfnamefont {J.}~\bibnamefont {Zhang}}, \bibinfo {author} {\bibfnamefont
			{H.}~\bibnamefont {Huang}}, \bibinfo {author} {\bibfnamefont
			{T.}~\bibnamefont {Xiang}}, \bibinfo {author} {\bibfnamefont
			{J.}~\bibnamefont {Hu}}, \bibinfo {author} {\bibfnamefont {R.}~\bibnamefont
			{Comin}}, \bibinfo {author} {\bibfnamefont {X.}~\bibnamefont {Zhou}}, \ and\
		\bibinfo {author} {\bibfnamefont {Z.}~\bibnamefont {Zhu}},\ }\href {\doibase
		10.1038/s42005-025-01971-z} {\bibfield  {journal} {\bibinfo  {journal}
			{Communications Physics}\ }\textbf {\bibinfo {volume} {8}},\ \bibinfo {pages}
		{52} (\bibinfo {year} {2025})}\BibitemShut {NoStop}%
	\bibitem [{\citenamefont {Sakakibara}\ \emph
		{et~al.}(2024{\natexlab{b}})\citenamefont {Sakakibara}, \citenamefont {Ochi},
		\citenamefont {Nagata}, \citenamefont {Ueki}, \citenamefont {Sakurai},
		\citenamefont {Matsumoto}, \citenamefont {Terashima}, \citenamefont {Hirose},
		\citenamefont {Ohta}, \citenamefont {Kato}, \citenamefont {Takano},\ and\
		\citenamefont {Kuroki}}]{Kuroki2023T}%
	\BibitemOpen
	\bibfield  {author} {\bibinfo {author} {\bibfnamefont {H.}~\bibnamefont
			{Sakakibara}}, \bibinfo {author} {\bibfnamefont {M.}~\bibnamefont {Ochi}},
		\bibinfo {author} {\bibfnamefont {H.}~\bibnamefont {Nagata}}, \bibinfo
		{author} {\bibfnamefont {Y.}~\bibnamefont {Ueki}}, \bibinfo {author}
		{\bibfnamefont {H.}~\bibnamefont {Sakurai}}, \bibinfo {author} {\bibfnamefont
			{R.}~\bibnamefont {Matsumoto}}, \bibinfo {author} {\bibfnamefont
			{K.}~\bibnamefont {Terashima}}, \bibinfo {author} {\bibfnamefont
			{K.}~\bibnamefont {Hirose}}, \bibinfo {author} {\bibfnamefont
			{H.}~\bibnamefont {Ohta}}, \bibinfo {author} {\bibfnamefont {M.}~\bibnamefont
			{Kato}}, \bibinfo {author} {\bibfnamefont {Y.}~\bibnamefont {Takano}}, \ and\
		\bibinfo {author} {\bibfnamefont {K.}~\bibnamefont {Kuroki}},\ }\href
	{\doibase 10.1103/PhysRevB.109.144511} {\bibfield  {journal} {\bibinfo
			{journal} {Phys. Rev. B}\ }\textbf {\bibinfo {volume} {109}},\ \bibinfo
		{pages} {144511} (\bibinfo {year} {2024}{\natexlab{b}})}\BibitemShut
	{NoStop}%
	\bibitem [{\citenamefont {Li}\ \emph {et~al.}(2024{\natexlab{a}})\citenamefont
		{Li}, \citenamefont {Zhang}, \citenamefont {Xiang}, \citenamefont {Zhang},
		\citenamefont {Zhu},\ and\ \citenamefont {Wen}}]{HHWen2024T}%
	\BibitemOpen
	\bibfield  {author} {\bibinfo {author} {\bibfnamefont {Q.}~\bibnamefont
			{Li}}, \bibinfo {author} {\bibfnamefont {Y.-J.}\ \bibnamefont {Zhang}},
		\bibinfo {author} {\bibfnamefont {Z.-N.}\ \bibnamefont {Xiang}}, \bibinfo
		{author} {\bibfnamefont {Y.}~\bibnamefont {Zhang}}, \bibinfo {author}
		{\bibfnamefont {X.}~\bibnamefont {Zhu}}, \ and\ \bibinfo {author}
		{\bibfnamefont {H.-H.}\ \bibnamefont {Wen}},\ }\href {\doibase
		10.1088/0256-307X/41/1/017401} {\bibfield  {journal} {\bibinfo  {journal}
			{Chinese Physics Letters}\ }\textbf {\bibinfo {volume} {41}},\ \bibinfo
		{pages} {017401} (\bibinfo {year} {2024}{\natexlab{a}})}\BibitemShut
	{NoStop}%
	\bibitem [{\citenamefont {Zhu}\ \emph {et~al.}(2024)\citenamefont {Zhu},
		\citenamefont {Peng}, \citenamefont {Zhang}, \citenamefont {Pan},
		\citenamefont {Chen}, \citenamefont {Chen}, \citenamefont {Ren},
		\citenamefont {Liu}, \citenamefont {Hao}, \citenamefont {Li}, \citenamefont
		{Xing}, \citenamefont {Lan}, \citenamefont {Han}, \citenamefont {Wang},
		\citenamefont {Jia}, \citenamefont {Wo}, \citenamefont {Gu}, \citenamefont
		{Gu}, \citenamefont {Ji}, \citenamefont {Wang}, \citenamefont {Gou},
		\citenamefont {Shen}, \citenamefont {Ying}, \citenamefont {Chen},
		\citenamefont {Yang}, \citenamefont {Cao}, \citenamefont {Zheng},
		\citenamefont {Zeng}, \citenamefont {Guo},\ and\ \citenamefont
		{Zhao}}]{JZhao2023T}%
	\BibitemOpen
	\bibfield  {author} {\bibinfo {author} {\bibfnamefont {Y.}~\bibnamefont
			{Zhu}}, \bibinfo {author} {\bibfnamefont {D.}~\bibnamefont {Peng}}, \bibinfo
		{author} {\bibfnamefont {E.}~\bibnamefont {Zhang}}, \bibinfo {author}
		{\bibfnamefont {B.}~\bibnamefont {Pan}}, \bibinfo {author} {\bibfnamefont
			{X.}~\bibnamefont {Chen}}, \bibinfo {author} {\bibfnamefont {L.}~\bibnamefont
			{Chen}}, \bibinfo {author} {\bibfnamefont {H.}~\bibnamefont {Ren}}, \bibinfo
		{author} {\bibfnamefont {F.}~\bibnamefont {Liu}}, \bibinfo {author}
		{\bibfnamefont {Y.}~\bibnamefont {Hao}}, \bibinfo {author} {\bibfnamefont
			{N.}~\bibnamefont {Li}}, \bibinfo {author} {\bibfnamefont {Z.}~\bibnamefont
			{Xing}}, \bibinfo {author} {\bibfnamefont {F.}~\bibnamefont {Lan}}, \bibinfo
		{author} {\bibfnamefont {J.}~\bibnamefont {Han}}, \bibinfo {author}
		{\bibfnamefont {J.}~\bibnamefont {Wang}}, \bibinfo {author} {\bibfnamefont
			{D.}~\bibnamefont {Jia}}, \bibinfo {author} {\bibfnamefont {H.}~\bibnamefont
			{Wo}}, \bibinfo {author} {\bibfnamefont {Y.}~\bibnamefont {Gu}}, \bibinfo
		{author} {\bibfnamefont {Y.}~\bibnamefont {Gu}}, \bibinfo {author}
		{\bibfnamefont {L.}~\bibnamefont {Ji}}, \bibinfo {author} {\bibfnamefont
			{W.}~\bibnamefont {Wang}}, \bibinfo {author} {\bibfnamefont {H.}~\bibnamefont
			{Gou}}, \bibinfo {author} {\bibfnamefont {Y.}~\bibnamefont {Shen}}, \bibinfo
		{author} {\bibfnamefont {T.}~\bibnamefont {Ying}}, \bibinfo {author}
		{\bibfnamefont {X.}~\bibnamefont {Chen}}, \bibinfo {author} {\bibfnamefont
			{W.}~\bibnamefont {Yang}}, \bibinfo {author} {\bibfnamefont {H.}~\bibnamefont
			{Cao}}, \bibinfo {author} {\bibfnamefont {C.}~\bibnamefont {Zheng}}, \bibinfo
		{author} {\bibfnamefont {Q.}~\bibnamefont {Zeng}}, \bibinfo {author}
		{\bibfnamefont {J.-g.}\ \bibnamefont {Guo}}, \ and\ \bibinfo {author}
		{\bibfnamefont {J.}~\bibnamefont {Zhao}},\ }\href {\doibase
		10.1038/s41586-024-07553-3} {\bibfield  {journal} {\bibinfo  {journal}
			{Nature}\ }\textbf {\bibinfo {volume} {631}},\ \bibinfo {pages} {531}
		(\bibinfo {year} {2024})}\BibitemShut {NoStop}%
	\bibitem [{\citenamefont {Zhang}\ \emph {et~al.}(2025)\citenamefont {Zhang},
		\citenamefont {Pei}, \citenamefont {Peng}, \citenamefont {Du}, \citenamefont
		{Hu}, \citenamefont {Cao}, \citenamefont {Wang}, \citenamefont {Wu},
		\citenamefont {Li}, \citenamefont {Liu}, \citenamefont {Wen}, \citenamefont
		{Song}, \citenamefont {Zhao}, \citenamefont {Li}, \citenamefont {Cao},
		\citenamefont {Zhu}, \citenamefont {Zhang}, \citenamefont {Yu}, \citenamefont
		{Cheng}, \citenamefont {Zhang}, \citenamefont {Li}, \citenamefont {Zhao},
		\citenamefont {Chen}, \citenamefont {Jin}, \citenamefont {Guo}, \citenamefont
		{Wu}, \citenamefont {Yang}, \citenamefont {Zeng}, \citenamefont {Yan},
		\citenamefont {Yang},\ and\ \citenamefont {Qi}}]{YQi2023T}%
	\BibitemOpen
	\bibfield  {author} {\bibinfo {author} {\bibfnamefont {M.}~\bibnamefont
			{Zhang}}, \bibinfo {author} {\bibfnamefont {C.}~\bibnamefont {Pei}}, \bibinfo
		{author} {\bibfnamefont {D.}~\bibnamefont {Peng}}, \bibinfo {author}
		{\bibfnamefont {X.}~\bibnamefont {Du}}, \bibinfo {author} {\bibfnamefont
			{W.}~\bibnamefont {Hu}}, \bibinfo {author} {\bibfnamefont {Y.}~\bibnamefont
			{Cao}}, \bibinfo {author} {\bibfnamefont {Q.}~\bibnamefont {Wang}}, \bibinfo
		{author} {\bibfnamefont {J.}~\bibnamefont {Wu}}, \bibinfo {author}
		{\bibfnamefont {Y.}~\bibnamefont {Li}}, \bibinfo {author} {\bibfnamefont
			{H.}~\bibnamefont {Liu}}, \bibinfo {author} {\bibfnamefont {C.}~\bibnamefont
			{Wen}}, \bibinfo {author} {\bibfnamefont {J.}~\bibnamefont {Song}}, \bibinfo
		{author} {\bibfnamefont {Y.}~\bibnamefont {Zhao}}, \bibinfo {author}
		{\bibfnamefont {C.}~\bibnamefont {Li}}, \bibinfo {author} {\bibfnamefont
			{W.}~\bibnamefont {Cao}}, \bibinfo {author} {\bibfnamefont {S.}~\bibnamefont
			{Zhu}}, \bibinfo {author} {\bibfnamefont {Q.}~\bibnamefont {Zhang}}, \bibinfo
		{author} {\bibfnamefont {N.}~\bibnamefont {Yu}}, \bibinfo {author}
		{\bibfnamefont {P.}~\bibnamefont {Cheng}}, \bibinfo {author} {\bibfnamefont
			{L.}~\bibnamefont {Zhang}}, \bibinfo {author} {\bibfnamefont
			{Z.}~\bibnamefont {Li}}, \bibinfo {author} {\bibfnamefont {J.}~\bibnamefont
			{Zhao}}, \bibinfo {author} {\bibfnamefont {Y.}~\bibnamefont {Chen}}, \bibinfo
		{author} {\bibfnamefont {C.}~\bibnamefont {Jin}}, \bibinfo {author}
		{\bibfnamefont {H.}~\bibnamefont {Guo}}, \bibinfo {author} {\bibfnamefont
			{C.}~\bibnamefont {Wu}}, \bibinfo {author} {\bibfnamefont {F.}~\bibnamefont
			{Yang}}, \bibinfo {author} {\bibfnamefont {Q.}~\bibnamefont {Zeng}}, \bibinfo
		{author} {\bibfnamefont {S.}~\bibnamefont {Yan}}, \bibinfo {author}
		{\bibfnamefont {L.}~\bibnamefont {Yang}}, \ and\ \bibinfo {author}
		{\bibfnamefont {Y.}~\bibnamefont {Qi}},\ }\href {\doibase
		10.1103/PhysRevX.15.021005} {\bibfield  {journal} {\bibinfo  {journal} {Phys.
				Rev. X}\ }\textbf {\bibinfo {volume} {15}},\ \bibinfo {pages} {021005}
		(\bibinfo {year} {2025})}\BibitemShut {NoStop}%
	\bibitem [{\citenamefont {Li}\ \emph {et~al.}(2024{\natexlab{b}})\citenamefont
		{Li}, \citenamefont {Chen}, \citenamefont {Huang}, \citenamefont {Han},
		\citenamefont {Huo}, \citenamefont {Huang}, \citenamefont {Ma}, \citenamefont
		{Qiu}, \citenamefont {Chen}, \citenamefont {Hu} \emph {et~al.}}]{MWang2023T}%
	\BibitemOpen
	\bibfield  {author} {\bibinfo {author} {\bibfnamefont {J.}~\bibnamefont
			{Li}}, \bibinfo {author} {\bibfnamefont {C.}~\bibnamefont {Chen}}, \bibinfo
		{author} {\bibfnamefont {C.}~\bibnamefont {Huang}}, \bibinfo {author}
		{\bibfnamefont {Y.}~\bibnamefont {Han}}, \bibinfo {author} {\bibfnamefont
			{M.}~\bibnamefont {Huo}}, \bibinfo {author} {\bibfnamefont {X.}~\bibnamefont
			{Huang}}, \bibinfo {author} {\bibfnamefont {P.}~\bibnamefont {Ma}}, \bibinfo
		{author} {\bibfnamefont {Z.}~\bibnamefont {Qiu}}, \bibinfo {author}
		{\bibfnamefont {J.}~\bibnamefont {Chen}}, \bibinfo {author} {\bibfnamefont
			{X.}~\bibnamefont {Hu}},  \emph {et~al.},\ }\href@noop {} {\bibfield
		{journal} {\bibinfo  {journal} {Sci. China Phys. Mech. Astron}\ }\textbf
		{\bibinfo {volume} {10}} (\bibinfo {year} {2024}{\natexlab{b}})}\BibitemShut
	{NoStop}%
	\bibitem [{\citenamefont {Ko}\ \emph {et~al.}(2025)\citenamefont {Ko},
		\citenamefont {Yu}, \citenamefont {Liu}, \citenamefont {Bhatt}, \citenamefont
		{Li}, \citenamefont {Thampy}, \citenamefont {Kuo}, \citenamefont {Wang},
		\citenamefont {Lee}, \citenamefont {Lee}, \citenamefont {Lee}, \citenamefont
		{Goodge}, \citenamefont {Muller},\ and\ \citenamefont {Hwang}}]{Ko2025}%
	\BibitemOpen
	\bibfield  {author} {\bibinfo {author} {\bibfnamefont {E.~K.}\ \bibnamefont
			{Ko}}, \bibinfo {author} {\bibfnamefont {Y.}~\bibnamefont {Yu}}, \bibinfo
		{author} {\bibfnamefont {Y.}~\bibnamefont {Liu}}, \bibinfo {author}
		{\bibfnamefont {L.}~\bibnamefont {Bhatt}}, \bibinfo {author} {\bibfnamefont
			{J.}~\bibnamefont {Li}}, \bibinfo {author} {\bibfnamefont {V.}~\bibnamefont
			{Thampy}}, \bibinfo {author} {\bibfnamefont {C.-T.}\ \bibnamefont {Kuo}},
		\bibinfo {author} {\bibfnamefont {B.~Y.}\ \bibnamefont {Wang}}, \bibinfo
		{author} {\bibfnamefont {Y.}~\bibnamefont {Lee}}, \bibinfo {author}
		{\bibfnamefont {K.}~\bibnamefont {Lee}}, \bibinfo {author} {\bibfnamefont
			{J.-S.}\ \bibnamefont {Lee}}, \bibinfo {author} {\bibfnamefont {B.~H.}\
			\bibnamefont {Goodge}}, \bibinfo {author} {\bibfnamefont {D.~A.}\
			\bibnamefont {Muller}}, \ and\ \bibinfo {author} {\bibfnamefont {H.~Y.}\
			\bibnamefont {Hwang}},\ }\href {\doibase 10.1038/s41586-024-08525-3}
	{\bibfield  {journal} {\bibinfo  {journal} {Nature}\ }\textbf {\bibinfo
			{volume} {638}},\ \bibinfo {pages} {935} (\bibinfo {year}
		{2025})}\BibitemShut {NoStop}%
	\bibitem [{\citenamefont {Zhou}\ \emph {et~al.}(2025)\citenamefont {Zhou},
		\citenamefont {Lv}, \citenamefont {Wang}, \citenamefont {Nie}, \citenamefont
		{Chen}, \citenamefont {Li}, \citenamefont {Huang}, \citenamefont {Chen},
		\citenamefont {Sun}, \citenamefont {Xue},\ and\ \citenamefont
		{Chen}}]{Zhou2025}%
	\BibitemOpen
	\bibfield  {author} {\bibinfo {author} {\bibfnamefont {G.}~\bibnamefont
			{Zhou}}, \bibinfo {author} {\bibfnamefont {W.}~\bibnamefont {Lv}}, \bibinfo
		{author} {\bibfnamefont {H.}~\bibnamefont {Wang}}, \bibinfo {author}
		{\bibfnamefont {Z.}~\bibnamefont {Nie}}, \bibinfo {author} {\bibfnamefont
			{Y.}~\bibnamefont {Chen}}, \bibinfo {author} {\bibfnamefont {Y.}~\bibnamefont
			{Li}}, \bibinfo {author} {\bibfnamefont {H.}~\bibnamefont {Huang}}, \bibinfo
		{author} {\bibfnamefont {W.}~\bibnamefont {Chen}}, \bibinfo {author}
		{\bibfnamefont {Y.}~\bibnamefont {Sun}}, \bibinfo {author} {\bibfnamefont
			{Q.-K.}\ \bibnamefont {Xue}}, \ and\ \bibinfo {author} {\bibfnamefont
			{Z.}~\bibnamefont {Chen}},\ }\href {\doibase 10.1038/s41586-025-08755-z}
	{\bibfield  {journal} {\bibinfo  {journal} {Nature}\ } (\bibinfo {year}
		{2025}),\ 10.1038/s41586-025-08755-z}\BibitemShut {NoStop}%
	\bibitem [{\citenamefont {Shi}\ \emph {et~al.}(2025)\citenamefont {Shi},
		\citenamefont {Peng}, \citenamefont {Li}, \citenamefont {Xing}, \citenamefont
		{Wang}, \citenamefont {Fan}, \citenamefont {Li}, \citenamefont {Wu},
		\citenamefont {Zeng}, \citenamefont {Zeng}, \citenamefont {Ying},
		\citenamefont {Wu},\ and\ \citenamefont {Chen}}]{shi2025}%
	\BibitemOpen
	\bibfield  {author} {\bibinfo {author} {\bibfnamefont {M.}~\bibnamefont
			{Shi}}, \bibinfo {author} {\bibfnamefont {D.}~\bibnamefont {Peng}}, \bibinfo
		{author} {\bibfnamefont {Y.}~\bibnamefont {Li}}, \bibinfo {author}
		{\bibfnamefont {Z.}~\bibnamefont {Xing}}, \bibinfo {author} {\bibfnamefont
			{Y.}~\bibnamefont {Wang}}, \bibinfo {author} {\bibfnamefont {K.}~\bibnamefont
			{Fan}}, \bibinfo {author} {\bibfnamefont {H.}~\bibnamefont {Li}}, \bibinfo
		{author} {\bibfnamefont {R.}~\bibnamefont {Wu}}, \bibinfo {author}
		{\bibfnamefont {Z.}~\bibnamefont {Zeng}}, \bibinfo {author} {\bibfnamefont
			{Q.}~\bibnamefont {Zeng}}, \bibinfo {author} {\bibfnamefont {J.}~\bibnamefont
			{Ying}}, \bibinfo {author} {\bibfnamefont {T.}~\bibnamefont {Wu}}, \ and\
		\bibinfo {author} {\bibfnamefont {X.}~\bibnamefont {Chen}},\ }\href
	{https://arxiv.org/abs/2501.14202} {} (\bibinfo {year} {2025}),\ \Eprint
	{http://arxiv.org/abs/2501.14202} {arXiv:2501.14202} \BibitemShut {NoStop}%
	\bibitem [{\citenamefont {Yang}\ \emph
		{et~al.}(2023{\natexlab{a}})\citenamefont {Yang}, \citenamefont {Wang},\ and\
		\citenamefont {Wang}}]{Wang327prb}%
	\BibitemOpen
	\bibfield  {author} {\bibinfo {author} {\bibfnamefont {Q.-G.}\ \bibnamefont
			{Yang}}, \bibinfo {author} {\bibfnamefont {D.}~\bibnamefont {Wang}}, \ and\
		\bibinfo {author} {\bibfnamefont {Q.-H.}\ \bibnamefont {Wang}},\ }\href
	{\doibase 10.1103/PhysRevB.108.L140505} {\bibfield  {journal} {\bibinfo
			{journal} {Phys. Rev. B}\ }\textbf {\bibinfo {volume} {108}},\ \bibinfo
		{pages} {L140505} (\bibinfo {year} {2023}{\natexlab{a}})}\BibitemShut
	{NoStop}%
	\bibitem [{\citenamefont {Lu}\ \emph {et~al.}(2024)\citenamefont {Lu},
		\citenamefont {Pan}, \citenamefont {Yang},\ and\ \citenamefont
		{Wu}}]{lu2024interlayer}%
	\BibitemOpen
	\bibfield  {author} {\bibinfo {author} {\bibfnamefont {C.}~\bibnamefont
			{Lu}}, \bibinfo {author} {\bibfnamefont {Z.}~\bibnamefont {Pan}}, \bibinfo
		{author} {\bibfnamefont {F.}~\bibnamefont {Yang}}, \ and\ \bibinfo {author}
		{\bibfnamefont {C.}~\bibnamefont {Wu}},\ }\href {\doibase
		10.1103/PhysRevLett.132.146002} {\bibfield  {journal} {\bibinfo  {journal}
			{Phys. Rev. Lett.}\ }\textbf {\bibinfo {volume} {132}},\ \bibinfo {pages}
		{146002} (\bibinfo {year} {2024})}\BibitemShut {NoStop}%
	\bibitem [{\citenamefont {Oh}\ and\ \citenamefont
		{Zhang}(2023)}]{HYZhangtype2}%
	\BibitemOpen
	\bibfield  {author} {\bibinfo {author} {\bibfnamefont {H.}~\bibnamefont
			{Oh}}\ and\ \bibinfo {author} {\bibfnamefont {Y.-H.}\ \bibnamefont {Zhang}},\
	}\href {\doibase 10.1103/PhysRevB.108.174511} {\bibfield  {journal} {\bibinfo
			{journal} {Phys. Rev. B}\ }\textbf {\bibinfo {volume} {108}},\ \bibinfo
		{pages} {174511} (\bibinfo {year} {2023})}\BibitemShut {NoStop}%
	\bibitem [{\citenamefont {Zhan}\ \emph {et~al.}(2025)\citenamefont {Zhan},
		\citenamefont {Gu}, \citenamefont {Wu},\ and\ \citenamefont {Hu}}]{EPC}%
	\BibitemOpen
	\bibfield  {author} {\bibinfo {author} {\bibfnamefont {J.}~\bibnamefont
			{Zhan}}, \bibinfo {author} {\bibfnamefont {Y.}~\bibnamefont {Gu}}, \bibinfo
		{author} {\bibfnamefont {X.}~\bibnamefont {Wu}}, \ and\ \bibinfo {author}
		{\bibfnamefont {J.}~\bibnamefont {Hu}},\ }\href {\doibase
		10.1103/PhysRevLett.134.136002} {\bibfield  {journal} {\bibinfo  {journal}
			{Phys. Rev. Lett.}\ }\textbf {\bibinfo {volume} {134}},\ \bibinfo {pages}
		{136002} (\bibinfo {year} {2025})}\BibitemShut {NoStop}%
	\bibitem [{\citenamefont {Liu}\ \emph {et~al.}(2023)\citenamefont {Liu},
		\citenamefont {Mei}, \citenamefont {Ye}, \citenamefont {Chen},\ and\
		\citenamefont {Yang}}]{FangYang327prl}%
	\BibitemOpen
	\bibfield  {author} {\bibinfo {author} {\bibfnamefont {Y.-B.}\ \bibnamefont
			{Liu}}, \bibinfo {author} {\bibfnamefont {J.-W.}\ \bibnamefont {Mei}},
		\bibinfo {author} {\bibfnamefont {F.}~\bibnamefont {Ye}}, \bibinfo {author}
		{\bibfnamefont {W.-Q.}\ \bibnamefont {Chen}}, \ and\ \bibinfo {author}
		{\bibfnamefont {F.}~\bibnamefont {Yang}},\ }\href {\doibase
		10.1103/PhysRevLett.131.236002} {\bibfield  {journal} {\bibinfo  {journal}
			{Phys. Rev. Lett.}\ }\textbf {\bibinfo {volume} {131}},\ \bibinfo {pages}
		{236002} (\bibinfo {year} {2023})}\BibitemShut {NoStop}%
	\bibitem [{\citenamefont {Qu}\ \emph {et~al.}(2024)\citenamefont {Qu},
		\citenamefont {Qu}, \citenamefont {Chen}, \citenamefont {Wu}, \citenamefont
		{Yang}, \citenamefont {Li},\ and\ \citenamefont {Su}}]{WeiLi327prl}%
	\BibitemOpen
	\bibfield  {author} {\bibinfo {author} {\bibfnamefont {X.-Z.}\ \bibnamefont
			{Qu}}, \bibinfo {author} {\bibfnamefont {D.-W.}\ \bibnamefont {Qu}}, \bibinfo
		{author} {\bibfnamefont {J.}~\bibnamefont {Chen}}, \bibinfo {author}
		{\bibfnamefont {C.}~\bibnamefont {Wu}}, \bibinfo {author} {\bibfnamefont
			{F.}~\bibnamefont {Yang}}, \bibinfo {author} {\bibfnamefont {W.}~\bibnamefont
			{Li}}, \ and\ \bibinfo {author} {\bibfnamefont {G.}~\bibnamefont {Su}},\
	}\href {\doibase 10.1103/PhysRevLett.132.036502} {\bibfield  {journal}
		{\bibinfo  {journal} {Phys. Rev. Lett.}\ }\textbf {\bibinfo {volume} {132}},\
		\bibinfo {pages} {036502} (\bibinfo {year} {2024})}\BibitemShut {NoStop}%
	\bibitem [{\citenamefont {Yang}\ \emph
		{et~al.}(2023{\natexlab{b}})\citenamefont {Yang}, \citenamefont {Zhang},\
		and\ \citenamefont {Zhang}}]{YifengYang327prb}%
	\BibitemOpen
	\bibfield  {author} {\bibinfo {author} {\bibfnamefont {Y.-f.}\ \bibnamefont
			{Yang}}, \bibinfo {author} {\bibfnamefont {G.-M.}\ \bibnamefont {Zhang}}, \
		and\ \bibinfo {author} {\bibfnamefont {F.-C.}\ \bibnamefont {Zhang}},\ }\href
	{\doibase 10.1103/PhysRevB.108.L201108} {\bibfield  {journal} {\bibinfo
			{journal} {Phys. Rev. B}\ }\textbf {\bibinfo {volume} {108}},\ \bibinfo
		{pages} {L201108} (\bibinfo {year} {2023}{\natexlab{b}})}\BibitemShut
	{NoStop}%
	\bibitem [{\citenamefont {Qin}\ and\ \citenamefont
		{Yang}(2023)}]{YifengYang327prb2}%
	\BibitemOpen
	\bibfield  {author} {\bibinfo {author} {\bibfnamefont {Q.}~\bibnamefont
			{Qin}}\ and\ \bibinfo {author} {\bibfnamefont {Y.-f.}\ \bibnamefont {Yang}},\
	}\href {\doibase 10.1103/PhysRevB.108.L140504} {\bibfield  {journal}
		{\bibinfo  {journal} {Phys. Rev. B}\ }\textbf {\bibinfo {volume} {108}},\
		\bibinfo {pages} {L140504} (\bibinfo {year} {2023})}\BibitemShut {NoStop}%
	\bibitem [{\citenamefont {Lu}\ \emph {et~al.}(2023)\citenamefont {Lu},
		\citenamefont {Li}, \citenamefont {Zeng}, \citenamefont {Hou}, \citenamefont
		{Wang}, \citenamefont {Yang},\ and\ \citenamefont {You}}]{YiZhuangYouSMG}%
	\BibitemOpen
	\bibfield  {author} {\bibinfo {author} {\bibfnamefont {D.-C.}\ \bibnamefont
			{Lu}}, \bibinfo {author} {\bibfnamefont {M.}~\bibnamefont {Li}}, \bibinfo
		{author} {\bibfnamefont {Z.-Y.}\ \bibnamefont {Zeng}}, \bibinfo {author}
		{\bibfnamefont {W.}~\bibnamefont {Hou}}, \bibinfo {author} {\bibfnamefont
			{J.}~\bibnamefont {Wang}}, \bibinfo {author} {\bibfnamefont {F.}~\bibnamefont
			{Yang}}, \ and\ \bibinfo {author} {\bibfnamefont {Y.-Z.}\ \bibnamefont
			{You}},\ }\href@noop {} {} (\bibinfo {year} {2023}),\ \Eprint
	{http://arxiv.org/abs/2308.11195} {arXiv:2308.11195} \BibitemShut {NoStop}%
	\bibitem [{\citenamefont {Tian}\ \emph {et~al.}(2024)\citenamefont {Tian},
		\citenamefont {Chen}, \citenamefont {Wang}, \citenamefont {He},\ and\
		\citenamefont {Lu}}]{tian2023correlation}%
	\BibitemOpen
	\bibfield  {author} {\bibinfo {author} {\bibfnamefont {Y.-H.}\ \bibnamefont
			{Tian}}, \bibinfo {author} {\bibfnamefont {Y.}~\bibnamefont {Chen}}, \bibinfo
		{author} {\bibfnamefont {J.-M.}\ \bibnamefont {Wang}}, \bibinfo {author}
		{\bibfnamefont {R.-Q.}\ \bibnamefont {He}}, \ and\ \bibinfo {author}
		{\bibfnamefont {Z.-Y.}\ \bibnamefont {Lu}},\ }\href {\doibase
		10.1103/PhysRevB.109.165154} {\bibfield  {journal} {\bibinfo  {journal}
			{Phys. Rev. B}\ }\textbf {\bibinfo {volume} {109}},\ \bibinfo {pages}
		{165154} (\bibinfo {year} {2024})}\BibitemShut {NoStop}%
	\bibitem [{\citenamefont {Zhang}\ \emph
		{et~al.}(2023{\natexlab{b}})\citenamefont {Zhang}, \citenamefont {Lin},
		\citenamefont {Moreo}, \citenamefont {Maier},\ and\ \citenamefont
		{Dagotto}}]{Dagotto327prb}%
	\BibitemOpen
	\bibfield  {author} {\bibinfo {author} {\bibfnamefont {Y.}~\bibnamefont
			{Zhang}}, \bibinfo {author} {\bibfnamefont {L.-F.}\ \bibnamefont {Lin}},
		\bibinfo {author} {\bibfnamefont {A.}~\bibnamefont {Moreo}}, \bibinfo
		{author} {\bibfnamefont {T.~A.}\ \bibnamefont {Maier}}, \ and\ \bibinfo
		{author} {\bibfnamefont {E.}~\bibnamefont {Dagotto}},\ }\href {\doibase
		10.1103/PhysRevB.108.165141} {\bibfield  {journal} {\bibinfo  {journal}
			{Phys. Rev. B}\ }\textbf {\bibinfo {volume} {108}},\ \bibinfo {pages}
		{165141} (\bibinfo {year} {2023}{\natexlab{b}})}\BibitemShut {NoStop}%
	\bibitem [{\citenamefont {Zhang}\ \emph {et~al.}(2024)\citenamefont {Zhang},
		\citenamefont {Lin}, \citenamefont {Moreo}, \citenamefont {Maier},\ and\
		\citenamefont {Dagotto}}]{zhang2024structural}%
	\BibitemOpen
	\bibfield  {author} {\bibinfo {author} {\bibfnamefont {Y.}~\bibnamefont
			{Zhang}}, \bibinfo {author} {\bibfnamefont {L.-F.}\ \bibnamefont {Lin}},
		\bibinfo {author} {\bibfnamefont {A.}~\bibnamefont {Moreo}}, \bibinfo
		{author} {\bibfnamefont {T.~A.}\ \bibnamefont {Maier}}, \ and\ \bibinfo
		{author} {\bibfnamefont {E.}~\bibnamefont {Dagotto}},\ }\href@noop {}
	{\bibfield  {journal} {\bibinfo  {journal} {Nature Communications}\ }\textbf
		{\bibinfo {volume} {15}},\ \bibinfo {pages} {2470} (\bibinfo {year}
		{2024})}\BibitemShut {NoStop}%
	\bibitem [{\citenamefont {Jiang}\ \emph
		{et~al.}(2024{\natexlab{a}})\citenamefont {Jiang}, \citenamefont {Wang},\
		and\ \citenamefont {Zhang}}]{Jiang_2024}%
	\BibitemOpen
	\bibfield  {author} {\bibinfo {author} {\bibfnamefont {K.}~\bibnamefont
			{Jiang}}, \bibinfo {author} {\bibfnamefont {Z.}~\bibnamefont {Wang}}, \ and\
		\bibinfo {author} {\bibfnamefont {F.-C.}\ \bibnamefont {Zhang}},\ }\href
	{\doibase 10.1088/0256-307X/41/1/017402} {\bibfield  {journal} {\bibinfo
			{journal} {Chinese Physics Letters}\ }\textbf {\bibinfo {volume} {41}},\
		\bibinfo {pages} {017402} (\bibinfo {year} {2024}{\natexlab{a}})}\BibitemShut
	{NoStop}%
	\bibitem [{\citenamefont {Lechermann}\ \emph
		{et~al.}(2023{\natexlab{b}})\citenamefont {Lechermann}, \citenamefont
		{Gondolf}, \citenamefont {B\"otzel},\ and\ \citenamefont
		{Eremin}}]{PhysRevB.108.L201121}%
	\BibitemOpen
	\bibfield  {author} {\bibinfo {author} {\bibfnamefont {F.}~\bibnamefont
			{Lechermann}}, \bibinfo {author} {\bibfnamefont {J.}~\bibnamefont {Gondolf}},
		\bibinfo {author} {\bibfnamefont {S.}~\bibnamefont {B\"otzel}}, \ and\
		\bibinfo {author} {\bibfnamefont {I.~M.}\ \bibnamefont {Eremin}},\ }\href
	{\doibase 10.1103/PhysRevB.108.L201121} {\bibfield  {journal} {\bibinfo
			{journal} {Phys. Rev. B}\ }\textbf {\bibinfo {volume} {108}},\ \bibinfo
		{pages} {L201121} (\bibinfo {year} {2023}{\natexlab{b}})}\BibitemShut
	{NoStop}%
	\bibitem [{\citenamefont {Liao}\ \emph {et~al.}(2023)\citenamefont {Liao},
		\citenamefont {Chen}, \citenamefont {Duan}, \citenamefont {Wang},
		\citenamefont {Liu}, \citenamefont {Yu},\ and\ \citenamefont
		{Si}}]{liao2023electron}%
	\BibitemOpen
	\bibfield  {author} {\bibinfo {author} {\bibfnamefont {Z.}~\bibnamefont
			{Liao}}, \bibinfo {author} {\bibfnamefont {L.}~\bibnamefont {Chen}}, \bibinfo
		{author} {\bibfnamefont {G.}~\bibnamefont {Duan}}, \bibinfo {author}
		{\bibfnamefont {Y.}~\bibnamefont {Wang}}, \bibinfo {author} {\bibfnamefont
			{C.}~\bibnamefont {Liu}}, \bibinfo {author} {\bibfnamefont {R.}~\bibnamefont
			{Yu}}, \ and\ \bibinfo {author} {\bibfnamefont {Q.}~\bibnamefont {Si}},\
	}\href {\doibase 10.1103/PhysRevB.108.214522} {\bibfield  {journal} {\bibinfo
			{journal} {Phys. Rev. B}\ }\textbf {\bibinfo {volume} {108}},\ \bibinfo
		{pages} {214522} (\bibinfo {year} {2023})}\BibitemShut {NoStop}%
	\bibitem [{\citenamefont {Ryee}\ \emph {et~al.}(2024)\citenamefont {Ryee},
		\citenamefont {Witt},\ and\ \citenamefont {Wehling}}]{ryee2024quenched}%
	\BibitemOpen
	\bibfield  {author} {\bibinfo {author} {\bibfnamefont {S.}~\bibnamefont
			{Ryee}}, \bibinfo {author} {\bibfnamefont {N.}~\bibnamefont {Witt}}, \ and\
		\bibinfo {author} {\bibfnamefont {T.~O.}\ \bibnamefont {Wehling}},\ }\href
	{\doibase 10.1103/PhysRevLett.133.096002} {\bibfield  {journal} {\bibinfo
			{journal} {Phys. Rev. Lett.}\ }\textbf {\bibinfo {volume} {133}},\ \bibinfo
		{pages} {096002} (\bibinfo {year} {2024})}\BibitemShut {NoStop}%
	\bibitem [{\citenamefont {Luo}\ \emph {et~al.}(2024)\citenamefont {Luo},
		\citenamefont {Lv}, \citenamefont {Wang}, \citenamefont {W{\'u}},\ and\
		\citenamefont {Yao}}]{luo2023hightc}%
	\BibitemOpen
	\bibfield  {author} {\bibinfo {author} {\bibfnamefont {Z.}~\bibnamefont
			{Luo}}, \bibinfo {author} {\bibfnamefont {B.}~\bibnamefont {Lv}}, \bibinfo
		{author} {\bibfnamefont {M.}~\bibnamefont {Wang}}, \bibinfo {author}
		{\bibfnamefont {W.}~\bibnamefont {W{\'u}}}, \ and\ \bibinfo {author}
		{\bibfnamefont {D.-X.}\ \bibnamefont {Yao}},\ }\href {\doibase
		10.1038/s41535-024-00668-w} {\bibfield  {journal} {\bibinfo  {journal} {npj
				Quantum Materials}\ }\textbf {\bibinfo {volume} {9}},\ \bibinfo {pages} {61}
		(\bibinfo {year} {2024})}\BibitemShut {NoStop}%
	\bibitem [{\citenamefont {Fan}\ \emph {et~al.}(2024)\citenamefont {Fan},
		\citenamefont {Zhang}, \citenamefont {Zhan}, \citenamefont {Lv},
		\citenamefont {Jiang}, \citenamefont {Normand},\ and\ \citenamefont
		{Xiang}}]{fan2023superconductivity}%
	\BibitemOpen
	\bibfield  {author} {\bibinfo {author} {\bibfnamefont {Z.}~\bibnamefont
			{Fan}}, \bibinfo {author} {\bibfnamefont {J.-F.}\ \bibnamefont {Zhang}},
		\bibinfo {author} {\bibfnamefont {B.}~\bibnamefont {Zhan}}, \bibinfo {author}
		{\bibfnamefont {D.}~\bibnamefont {Lv}}, \bibinfo {author} {\bibfnamefont
			{X.-Y.}\ \bibnamefont {Jiang}}, \bibinfo {author} {\bibfnamefont
			{B.}~\bibnamefont {Normand}}, \ and\ \bibinfo {author} {\bibfnamefont
			{T.}~\bibnamefont {Xiang}},\ }\href {\doibase 10.1103/PhysRevB.110.024514}
	{\bibfield  {journal} {\bibinfo  {journal} {Phys. Rev. B}\ }\textbf {\bibinfo
			{volume} {110}},\ \bibinfo {pages} {024514} (\bibinfo {year}
		{2024})}\BibitemShut {NoStop}%
	\bibitem [{\citenamefont {Jiang}\ \emph
		{et~al.}(2024{\natexlab{b}})\citenamefont {Jiang}, \citenamefont {Hou},
		\citenamefont {Fan}, \citenamefont {Lang},\ and\ \citenamefont
		{Ku}}]{KuWeiprl}%
	\BibitemOpen
	\bibfield  {author} {\bibinfo {author} {\bibfnamefont {R.}~\bibnamefont
			{Jiang}}, \bibinfo {author} {\bibfnamefont {J.}~\bibnamefont {Hou}}, \bibinfo
		{author} {\bibfnamefont {Z.}~\bibnamefont {Fan}}, \bibinfo {author}
		{\bibfnamefont {Z.-J.}\ \bibnamefont {Lang}}, \ and\ \bibinfo {author}
		{\bibfnamefont {W.}~\bibnamefont {Ku}},\ }\href {\doibase
		10.1103/PhysRevLett.132.126503} {\bibfield  {journal} {\bibinfo  {journal}
			{Phys. Rev. Lett.}\ }\textbf {\bibinfo {volume} {132}},\ \bibinfo {pages}
		{126503} (\bibinfo {year} {2024}{\natexlab{b}})}\BibitemShut {NoStop}%
	\bibitem [{\citenamefont {Jiang}\ \emph
		{et~al.}(2024{\natexlab{c}})\citenamefont {Jiang}, \citenamefont {Wang},\
		and\ \citenamefont {Zhang}}]{KJiang:17402}%
	\BibitemOpen
	\bibfield  {author} {\bibinfo {author} {\bibfnamefont {K.}~\bibnamefont
			{Jiang}}, \bibinfo {author} {\bibfnamefont {Z.}~\bibnamefont {Wang}}, \ and\
		\bibinfo {author} {\bibfnamefont {F.-C.}\ \bibnamefont {Zhang}},\ }\href
	{\doibase 10.1088/0256-307X/41/1/017402} {\bibfield  {journal} {\bibinfo
			{journal} {Chinese Physics Letters}\ }\textbf {\bibinfo {volume} {41}},\
		\bibinfo {eid} {017402} (\bibinfo {year} {2024}{\natexlab{c}})}\BibitemShut
	{NoStop}%
	\bibitem [{\citenamefont {Schl{\"o}mer}\ \emph {et~al.}(2024)\citenamefont
		{Schl{\"o}mer}, \citenamefont {Schollw{\"o}ck}, \citenamefont {Grusdt},\ and\
		\citenamefont {Bohrdt}}]{Schloemer2024}%
	\BibitemOpen
	\bibfield  {author} {\bibinfo {author} {\bibfnamefont {H.}~\bibnamefont
			{Schl{\"o}mer}}, \bibinfo {author} {\bibfnamefont {U.}~\bibnamefont
			{Schollw{\"o}ck}}, \bibinfo {author} {\bibfnamefont {F.}~\bibnamefont
			{Grusdt}}, \ and\ \bibinfo {author} {\bibfnamefont {A.}~\bibnamefont
			{Bohrdt}},\ }\href {\doibase 10.1038/s42005-024-01854-9} {\bibfield
		{journal} {\bibinfo  {journal} {Communications Physics}\ }\textbf {\bibinfo
			{volume} {7}},\ \bibinfo {pages} {366} (\bibinfo {year} {2024})}\BibitemShut
	{NoStop}%
	\bibitem [{\citenamefont {Bejas}\ \emph {et~al.}(2025)\citenamefont {Bejas},
		\citenamefont {Wu}, \citenamefont {Chakraborty}, \citenamefont {Schnyder},\
		and\ \citenamefont {Greco}}]{Andres2024}%
	\BibitemOpen
	\bibfield  {author} {\bibinfo {author} {\bibfnamefont {M.}~\bibnamefont
			{Bejas}}, \bibinfo {author} {\bibfnamefont {X.}~\bibnamefont {Wu}}, \bibinfo
		{author} {\bibfnamefont {D.}~\bibnamefont {Chakraborty}}, \bibinfo {author}
		{\bibfnamefont {A.~P.}\ \bibnamefont {Schnyder}}, \ and\ \bibinfo {author}
		{\bibfnamefont {A.}~\bibnamefont {Greco}},\ }\href {\doibase
		10.1103/PhysRevB.111.144514} {\bibfield  {journal} {\bibinfo  {journal}
			{Phys. Rev. B}\ }\textbf {\bibinfo {volume} {111}},\ \bibinfo {pages}
		{144514} (\bibinfo {year} {2025})}\BibitemShut {NoStop}%
	\bibitem [{\citenamefont {Christiansson}\ \emph {et~al.}(2023)\citenamefont
		{Christiansson}, \citenamefont {Petocchi},\ and\ \citenamefont
		{Werner}}]{PWcorre}%
	\BibitemOpen
	\bibfield  {author} {\bibinfo {author} {\bibfnamefont {V.}~\bibnamefont
			{Christiansson}}, \bibinfo {author} {\bibfnamefont {F.}~\bibnamefont
			{Petocchi}}, \ and\ \bibinfo {author} {\bibfnamefont {P.}~\bibnamefont
			{Werner}},\ }\href {\doibase 10.1103/PhysRevLett.131.206501} {\bibfield
		{journal} {\bibinfo  {journal} {Phys. Rev. Lett.}\ }\textbf {\bibinfo
			{volume} {131}},\ \bibinfo {pages} {206501} (\bibinfo {year}
		{2023})}\BibitemShut {NoStop}%
	\bibitem [{\citenamefont {Le}\ \emph {et~al.}(2025)\citenamefont {Le},
		\citenamefont {Zhan}, \citenamefont {Wu},\ and\ \citenamefont
		{Hu}}]{CLe2025}%
	\BibitemOpen
	\bibfield  {author} {\bibinfo {author} {\bibfnamefont {C.}~\bibnamefont
			{Le}}, \bibinfo {author} {\bibfnamefont {J.}~\bibnamefont {Zhan}}, \bibinfo
		{author} {\bibfnamefont {X.}~\bibnamefont {Wu}}, \ and\ \bibinfo {author}
		{\bibfnamefont {J.}~\bibnamefont {Hu}},\ }\href
	{https://arxiv.org/abs/2501.14665} {} (\bibinfo {year} {2025}),\ \Eprint
	{http://arxiv.org/abs/2501.14665} {arXiv:2501.14665} \BibitemShut {NoStop}%
	\bibitem [{\citenamefont {Castellani}\ \emph {et~al.}(1978)\citenamefont
		{Castellani}, \citenamefont {Natoli},\ and\ \citenamefont
		{Ranninger}}]{Kanamori}%
	\BibitemOpen
	\bibfield  {author} {\bibinfo {author} {\bibfnamefont {C.}~\bibnamefont
			{Castellani}}, \bibinfo {author} {\bibfnamefont {C.~R.}\ \bibnamefont
			{Natoli}}, \ and\ \bibinfo {author} {\bibfnamefont {J.}~\bibnamefont
			{Ranninger}},\ }\href {\doibase 10.1103/PhysRevB.18.4945} {\bibfield
		{journal} {\bibinfo  {journal} {Phys. Rev. B}\ }\textbf {\bibinfo {volume}
			{18}},\ \bibinfo {pages} {4945} (\bibinfo {year} {1978})}\BibitemShut
	{NoStop}%
	\bibitem [{\citenamefont {Salmhofer}\ and\ \citenamefont
		{Honerkamp}(2001)}]{10.1143/PTP.105.1}%
	\BibitemOpen
	\bibfield  {author} {\bibinfo {author} {\bibfnamefont {M.}~\bibnamefont
			{Salmhofer}}\ and\ \bibinfo {author} {\bibfnamefont {C.}~\bibnamefont
			{Honerkamp}},\ }\href {\doibase 10.1143/PTP.105.1} {\bibfield  {journal}
		{\bibinfo  {journal} {Progress of Theoretical Physics}\ }\textbf {\bibinfo
			{volume} {105}},\ \bibinfo {pages} {1} (\bibinfo {year} {2001})}\BibitemShut
	{NoStop}%
	\bibitem [{\citenamefont {Metzner}\ \emph {et~al.}(2012)\citenamefont
		{Metzner}, \citenamefont {Salmhofer}, \citenamefont {Honerkamp},
		\citenamefont {Meden},\ and\ \citenamefont {Sch\"onhammer}}]{Metzner2012}%
	\BibitemOpen
	\bibfield  {author} {\bibinfo {author} {\bibfnamefont {W.}~\bibnamefont
			{Metzner}}, \bibinfo {author} {\bibfnamefont {M.}~\bibnamefont {Salmhofer}},
		\bibinfo {author} {\bibfnamefont {C.}~\bibnamefont {Honerkamp}}, \bibinfo
		{author} {\bibfnamefont {V.}~\bibnamefont {Meden}}, \ and\ \bibinfo {author}
		{\bibfnamefont {K.}~\bibnamefont {Sch\"onhammer}},\ }\href {\doibase
		10.1103/RevModPhys.84.299} {\bibfield  {journal} {\bibinfo  {journal} {Rev.
				Mod. Phys.}\ }\textbf {\bibinfo {volume} {84}},\ \bibinfo {pages} {299}
		(\bibinfo {year} {2012})}\BibitemShut {NoStop}%
	\bibitem [{\citenamefont {Platt}\ \emph {et~al.}(2013)\citenamefont {Platt},
		\citenamefont {Hanke},\ and\ \citenamefont {Thomale}}]{Platt2013}%
	\BibitemOpen
	\bibfield  {author} {\bibinfo {author} {\bibfnamefont {C.}~\bibnamefont
			{Platt}}, \bibinfo {author} {\bibfnamefont {W.}~\bibnamefont {Hanke}}, \ and\
		\bibinfo {author} {\bibfnamefont {R.}~\bibnamefont {Thomale}},\ }\href
	{\doibase 10.1080/00018732.2013.862020} {\bibfield  {journal} {\bibinfo
			{journal} {Advances in Physics}\ }\textbf {\bibinfo {volume} {62}},\ \bibinfo
		{pages} {453} (\bibinfo {year} {2013})}\BibitemShut {NoStop}%
	\bibitem [{\citenamefont {Lichtenstein}\ \emph {et~al.}(2017)\citenamefont
		{Lichtenstein}, \citenamefont {S{\'a}nchez de~la Pe{\~{n}}a}, \citenamefont
		{Rohe}, \citenamefont {Di~Napoli}, \citenamefont {Honerkamp},\ and\
		\citenamefont {Maier}}]{Lichtenstein2017}%
	\BibitemOpen
	\bibfield  {author} {\bibinfo {author} {\bibfnamefont {J.}~\bibnamefont
			{Lichtenstein}}, \bibinfo {author} {\bibfnamefont {D.}~\bibnamefont
			{S{\'a}nchez de~la Pe{\~{n}}a}}, \bibinfo {author} {\bibfnamefont
			{D.}~\bibnamefont {Rohe}}, \bibinfo {author} {\bibfnamefont {E.}~\bibnamefont
			{Di~Napoli}}, \bibinfo {author} {\bibfnamefont {C.}~\bibnamefont
			{Honerkamp}}, \ and\ \bibinfo {author} {\bibfnamefont {S.~A.}\ \bibnamefont
			{Maier}},\ }\href
	{https://www.sciencedirect.com/science/article/pii/S0010465516303927}
	{\bibfield  {journal} {\bibinfo  {journal} {Computer Physics Communications}\
		}\textbf {\bibinfo {volume} {213}},\ \bibinfo {pages} {100} (\bibinfo {year}
		{2017})}\BibitemShut {NoStop}%
	\bibitem [{\citenamefont {Wang}\ \emph {et~al.}(2012)\citenamefont {Wang},
		\citenamefont {Xiang}, \citenamefont {Wang}, \citenamefont {Wang},
		\citenamefont {Yang},\ and\ \citenamefont {Lee}}]{QHWangGraphene}%
	\BibitemOpen
	\bibfield  {author} {\bibinfo {author} {\bibfnamefont {W.-S.}\ \bibnamefont
			{Wang}}, \bibinfo {author} {\bibfnamefont {Y.-Y.}\ \bibnamefont {Xiang}},
		\bibinfo {author} {\bibfnamefont {Q.-H.}\ \bibnamefont {Wang}}, \bibinfo
		{author} {\bibfnamefont {F.}~\bibnamefont {Wang}}, \bibinfo {author}
		{\bibfnamefont {F.}~\bibnamefont {Yang}}, \ and\ \bibinfo {author}
		{\bibfnamefont {D.-H.}\ \bibnamefont {Lee}},\ }\href {\doibase
		10.1103/PhysRevB.85.035414} {\bibfield  {journal} {\bibinfo  {journal} {Phys.
				Rev. B}\ }\textbf {\bibinfo {volume} {85}},\ \bibinfo {pages} {035414}
		(\bibinfo {year} {2012})}\BibitemShut {NoStop}%
	\bibitem [{\citenamefont {Braz}\ \emph {et~al.}(2025)\citenamefont {Braz},
		\citenamefont {Martins},\ and\ \citenamefont {da~Silva}}]{braz2025}%
	\BibitemOpen
	\bibfield  {author} {\bibinfo {author} {\bibfnamefont {L.~B.}\ \bibnamefont
			{Braz}}, \bibinfo {author} {\bibfnamefont {G.~B.}\ \bibnamefont {Martins}}, \
		and\ \bibinfo {author} {\bibfnamefont {L.~G. G. V.~D.}\ \bibnamefont
			{da~Silva}},\ }\href {https://arxiv.org/abs/2502.08425} {} (\bibinfo {year}
	{2025}),\ \Eprint {http://arxiv.org/abs/2502.08425} {arXiv:2502.08425}
	\BibitemShut {NoStop}%
	\bibitem [{\citenamefont {Borchia}\ \emph {et~al.}(2025)\citenamefont
		{Borchia}, \citenamefont {Lange},\ and\ \citenamefont
		{Grusdt}}]{borchia2025}%
	\BibitemOpen
	\bibfield  {author} {\bibinfo {author} {\bibfnamefont {P.}~\bibnamefont
			{Borchia}}, \bibinfo {author} {\bibfnamefont {H.}~\bibnamefont {Lange}}, \
		and\ \bibinfo {author} {\bibfnamefont {F.}~\bibnamefont {Grusdt}},\ }\href
	{https://arxiv.org/abs/2502.13960} {} (\bibinfo {year} {2025}),\ \Eprint
	{http://arxiv.org/abs/2502.13960} {arXiv:2502.13960} \BibitemShut {NoStop}%
	\bibitem [{\citenamefont {Xi}\ \emph {et~al.}(2025)\citenamefont {Xi},
		\citenamefont {Yu},\ and\ \citenamefont {Li}}]{Xi2025}%
	\BibitemOpen
	\bibfield  {author} {\bibinfo {author} {\bibfnamefont {W.}~\bibnamefont
			{Xi}}, \bibinfo {author} {\bibfnamefont {S.-L.}\ \bibnamefont {Yu}}, \ and\
		\bibinfo {author} {\bibfnamefont {J.-X.}\ \bibnamefont {Li}},\ }\href
	{\doibase 10.1103/PhysRevB.111.104505} {\bibfield  {journal} {\bibinfo
			{journal} {Phys. Rev. B}\ }\textbf {\bibinfo {volume} {111}},\ \bibinfo
		{pages} {104505} (\bibinfo {year} {2025})}\BibitemShut {NoStop}%
\end{thebibliography}

\begin{thebibliography}{4}%
		\makeatletter
		\providecommand \@ifxundefined [1]{%
			\@ifx{#1\undefined}
		}%
		\providecommand \@ifnum [1]{%
			\ifnum #1\expandafter \@firstoftwo
			\else \expandafter \@secondoftwo
			\fi
		}%
		\providecommand \@ifx [1]{%
			\ifx #1\expandafter \@firstoftwo
			\else \expandafter \@secondoftwo
			\fi
		}%
		\providecommand \natexlab [1]{#1}%
		\providecommand \enquote  [1]{``#1''}%
		\providecommand \bibnamefont  [1]{#1}%
		\providecommand \bibfnamefont [1]{#1}%
		\providecommand \citenamefont [1]{#1}%
		\providecommand \href@noop [0]{\@secondoftwo}%
		\providecommand \href [0]{\begingroup \@sanitize@url \@href}%
		\providecommand \@href[1]{\@@startlink{#1}\@@href}%
		\providecommand \@@href[1]{\endgroup#1\@@endlink}%
		\providecommand \@sanitize@url [0]{\catcode `\\12\catcode `\$12\catcode
			`\&12\catcode `\#12\catcode `\^12\catcode `\_12\catcode `\%12\relax}%
		\providecommand \@@startlink[1]{}%
		\providecommand \@@endlink[0]{}%
		\providecommand \url  [0]{\begingroup\@sanitize@url \@url }%
		\providecommand \@url [1]{\endgroup\@href {#1}{\urlprefix }}%
		\providecommand \urlprefix  [0]{URL }%
		\providecommand \Eprint [0]{\href }%
		\providecommand \doibase [0]{http://dx.doi.org/}%
		\providecommand \selectlanguage [0]{\@gobble}%
		\providecommand \bibinfo  [0]{\@secondoftwo}%
		\providecommand \bibfield  [0]{\@secondoftwo}%
		\providecommand \translation [1]{[#1]}%
		\providecommand \BibitemOpen [0]{}%
		\providecommand \bibitemStop [0]{}%
		\providecommand \bibitemNoStop [0]{.\EOS\space}%
		\providecommand \EOS [0]{\spacefactor3000\relax}%
		\providecommand \BibitemShut  [1]{\csname bibitem#1\endcsname}%
		\let\auto@bib@innerbib\@empty
		\bibitem [{\citenamefont {Salmhofer}\ and\ \citenamefont
			{Honerkamp}(2001)}]{10.1143/PTP.105.1}%
		\BibitemOpen
		\bibfield  {author} {\bibinfo {author} {\bibfnamefont {M.}~\bibnamefont
				{Salmhofer}}\ and\ \bibinfo {author} {\bibfnamefont {C.}~\bibnamefont
				{Honerkamp}},\ }\href {\doibase 10.1143/PTP.105.1} {\bibfield  {journal}
			{\bibinfo  {journal} {Progress of Theoretical Physics}\ }\textbf {\bibinfo
				{volume} {105}},\ \bibinfo {pages} {1} (\bibinfo {year} {2001})}\BibitemShut
		{NoStop}%
		\bibitem [{\citenamefont {Metzner}\ \emph {et~al.}(2012)\citenamefont
			{Metzner}, \citenamefont {Salmhofer}, \citenamefont {Honerkamp},
			\citenamefont {Meden},\ and\ \citenamefont {Sch\"onhammer}}]{Metzner2012}%
		\BibitemOpen
		\bibfield  {author} {\bibinfo {author} {\bibfnamefont {W.}~\bibnamefont
				{Metzner}}, \bibinfo {author} {\bibfnamefont {M.}~\bibnamefont {Salmhofer}},
			\bibinfo {author} {\bibfnamefont {C.}~\bibnamefont {Honerkamp}}, \bibinfo
			{author} {\bibfnamefont {V.}~\bibnamefont {Meden}}, \ and\ \bibinfo {author}
			{\bibfnamefont {K.}~\bibnamefont {Sch\"onhammer}},\ }\href {\doibase
			10.1103/RevModPhys.84.299} {\bibfield  {journal} {\bibinfo  {journal} {Rev.
					Mod. Phys.}\ }\textbf {\bibinfo {volume} {84}},\ \bibinfo {pages} {299}
			(\bibinfo {year} {2012})}\BibitemShut {NoStop}%
		\bibitem [{\citenamefont {Lichtenstein}\ \emph {et~al.}(2017)\citenamefont
			{Lichtenstein}, \citenamefont {S{\'a}nchez de~la Pe{\~{n}}a}, \citenamefont
			{Rohe}, \citenamefont {Di~Napoli}, \citenamefont {Honerkamp},\ and\
			\citenamefont {Maier}}]{Lichtenstein2017}%
		\BibitemOpen
		\bibfield  {author} {\bibinfo {author} {\bibfnamefont {J.}~\bibnamefont
				{Lichtenstein}}, \bibinfo {author} {\bibfnamefont {D.}~\bibnamefont
				{S{\'a}nchez de~la Pe{\~{n}}a}}, \bibinfo {author} {\bibfnamefont
				{D.}~\bibnamefont {Rohe}}, \bibinfo {author} {\bibfnamefont {E.}~\bibnamefont
				{Di~Napoli}}, \bibinfo {author} {\bibfnamefont {C.}~\bibnamefont
				{Honerkamp}}, \ and\ \bibinfo {author} {\bibfnamefont {S.~A.}\ \bibnamefont
				{Maier}},\ }\href
		{https://www.sciencedirect.com/science/article/pii/S0010465516303927}
		{\bibfield  {journal} {\bibinfo  {journal} {Computer Physics Communications}\
			}\textbf {\bibinfo {volume} {213}},\ \bibinfo {pages} {100} (\bibinfo {year}
			{2017})}\BibitemShut {NoStop}%
		\bibitem [{\citenamefont {Wang}\ \emph {et~al.}(2012)\citenamefont {Wang},
			\citenamefont {Xiang}, \citenamefont {Wang}, \citenamefont {Wang},
			\citenamefont {Yang},\ and\ \citenamefont {Lee}}]{QHWangGraphene}%
		\BibitemOpen
		\bibfield  {author} {\bibinfo {author} {\bibfnamefont {W.-S.}\ \bibnamefont
				{Wang}}, \bibinfo {author} {\bibfnamefont {Y.-Y.}\ \bibnamefont {Xiang}},
			\bibinfo {author} {\bibfnamefont {Q.-H.}\ \bibnamefont {Wang}}, \bibinfo
			{author} {\bibfnamefont {F.}~\bibnamefont {Wang}}, \bibinfo {author}
			{\bibfnamefont {F.}~\bibnamefont {Yang}}, \ and\ \bibinfo {author}
			{\bibfnamefont {D.-H.}\ \bibnamefont {Lee}},\ }\href {\doibase
			10.1103/PhysRevB.85.035414} {\bibfield  {journal} {\bibinfo  {journal} {Phys.
					Rev. B}\ }\textbf {\bibinfo {volume} {85}},\ \bibinfo {pages} {035414}
			(\bibinfo {year} {2012})}\BibitemShut {NoStop}%
	\end{thebibliography}

%

%
%

	\clearpage             
	\onecolumngrid         
	
	\setcounter{equation}{0}
	\setcounter{figure}{0}
	\setcounter{table}{0}
	\renewcommand{\theequation}{S\arabic{equation}}
	\renewcommand{\thefigure}{S\arabic{figure}}
	\renewcommand{\thetable}{S\arabic{table}}
	

%
%
%
%
%
%
	
	\begin{center}
		{\bfseries\large
			Supplementary materials for: “Impact of Nonlocal Coulomb Repulsion on Superconductivity and Density-Wave Orders in Bilayer Nickelates”}\\[0.75em]
		\normalsize
		Jun Zhan,$^{1,2}$ Congcong Le,$^{3}$ Xianxin Wu,$^{4,*}$ and Jiangping Hu$^{1,5,6,\dagger}$\\[0.4em]
		\textit{
			$^{1}$Beijing National Laboratory for Condensed Matter Physics and Institute of Physics, Chinese Academy of Sciences, Beijing 100190, China\\
			$^{2}$School of Physical Sciences, University of Chinese Academy of Sciences, Beijing 100190, China\\
			$^{3}$RIKEN Interdisciplinary Theoretical and Mathematical Sciences (iTHEMS), Wako, Saitama 351-0198, Japan\\
			$^{4}$CAS Key Laboratory of Theoretical Physics, Institute of Theoretical Physics, Chinese Academy of Sciences, Beijing 100190, China\\
			$^{5}$Kavli Institute for Theoretical Sciences, University of Chinese Academy of Sciences, Beijing 100190, China\\
			$^{6}$New Cornerstone Science Laboratory, Beijing 100190, China
		}\\[1em]
	\end{center}
	
	

	\setcounter{section}{0}
	\section{FRG details} \label{apA}
\begin{figure}[tpb]
		\centerline{\includegraphics[width=0.99\textwidth]{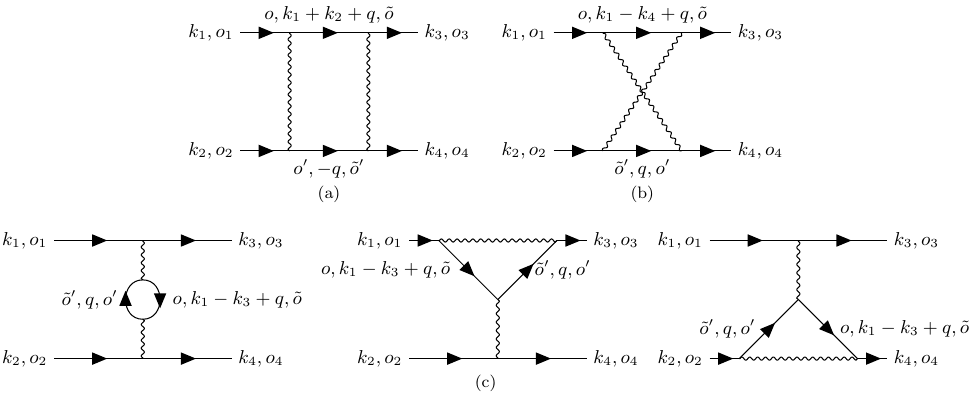}}
		\caption{One-loop contributions to the renormalization of effective interaction $V^{\Lambda}_{\left\lbrace  o_i \right\rbrace  }(k_1,k_2;k_3,k_4)$ in orbit basis of multi-orbit system: (a)The particle-particle term. (b)The crossed particle-hole term. (c)The direct particle-hole terms.}
		\label{fig:fRG}
	\end{figure}

	The FRG is an unbiased method to determine Fermi liquid instability of interacting fermionic systems from weak to moderate coupling regimes~\cite{10.1143/PTP.105.1, Metzner2012}. 
	The fundamental idea of the FRG is to introduce scale dependence into the effective action or generating functional of one-particle irreducible (1PI) vertex functions, denoted as $\Gamma \to \Gamma^\Lambda$, by incorporating an infrared cutoff into the bare propagator, $G_0 \to G_0^\Lambda$. By differentiating $\Gamma^\Lambda$ with respect to $\Lambda$, one derives exact functional flow equations. These equations enable an interpolation between a microscopic bare action at $\Lambda \to \infty$ and a low-energy effective action at small $\Lambda$. Expanding in powers of the fields yields an exact hierarchy of flow equations for the 1PI vertex functions. We adopt standard approximations that truncate the hierarchy after the two-particle vertex function and neglect self-energy feedback, which is suitable for weak to intermediate coupling regimes.
	For multi-orbit spin $SU(2)$ invariant system, the two particle part of effective action can be expressed by effective interaction $V^{\Lambda}$ and Grassmann filed $\psi$ and $\bar{\psi}$ as
	\begin{equation}
		\begin{aligned}
			\Gamma^{(4) \Lambda}&[\bar{\psi},\psi] =\frac{1}{2!} \int \prod_{i=1}^{4} \dd \xi_{i}
			V^{\Lambda}_{o_1o_2o_3o_4}(k_1,k_2;k_3,k_4)  
			\delta(k_1+k_2-k_3-k_4)\bar{\psi}_{\sigma}(\xi_1)\bar{\psi}_{\bar{\sigma}'}(\xi_2)\psi_{\bar{\sigma}'}(\xi_4)\psi_{\sigma}(\xi_3)
		\end{aligned}
	\end{equation}
	where $k_i=(\omega_{i},\mathbf{k}_{i})$ and $\xi_i = (\omega_i, \mathbf{k}_i, o_i )$ are multi-indices
	gathering a Matsubara frequency $\omega_i$, wave vector $\mathbf{k}_i$,
	and orbital index $o_i$ and
	$\dd \xi_i$ stands for $\int \frac{d \mathbf{k}_i}{S_{B Z}} \frac{1}{\beta} \sum \omega_i \sum o_i$ with the Brillouin zone area $S_{BZ}$ and inverse
	temperature $\beta$.
	The flow equation for effective two particle interaction reads
	
	\begin{equation}\label{spinless}
		\begin{aligned}
			\frac{\dd}{\dd \Lambda}&V^{\Lambda}_{\left\lbrace  o_i \right\rbrace  }(k_1,k_2;k_3,k_4) = \mathcal{T}^{\mathrm{pp}}_{\left\lbrace  o_i \right\rbrace  }\left( k_1,k_2;k_3,k_4\right)  
			+\mathcal{T}^{\mathrm{cph}}_{\left\lbrace  o_i \right\rbrace  }\left( k_1,k_2;k_3,k_4\right)  +\mathcal{T}^{\mathrm{dph}}_{\left\lbrace  o_i \right\rbrace  } \left(k_1,k_2;k_3,k_4\right)
		\end{aligned}
	\end{equation}
	The three terms of right hand side of flow equation are given by
		\begin{equation}
			\begin{aligned}
				\mathcal{T}_{\left\lbrace o_i \right\rbrace}^{\mathrm{pp}}\left( {k}_{1},  {k}_{2};  {k}_{3}, {k}_{4}\right) &= -\int_{q} \frac{\mathrm{d}}{\mathrm{d}\Lambda}\left[ G^{\Lambda}_{o\tilde{o}}({k}_1+ {k}_2- {q}) G^{\Lambda}_{o'\tilde{o}'}(- {q}) \right] \\
				&\quad \quad \times V^{\Lambda}_{o_1o_2oo'}\left( {k}_1, {k}_2; {k}_1+ {k}_2+ {q},- {q}\right)  V^{\Lambda}_{\tilde{o}\tilde{o}'o_3o_4}\left(  {k}_1+ {k}_2+ {q},- {q} ;  {k}_3, {k}_4 \right) ,\\
				\mathcal{T}_{\left\lbrace o_i \right\rbrace}^{\mathrm{cph}}\left( {k}_{1},  {k}_{2};  {k}_{3}, {k}_{4}\right) &= -\int_{q}
				\frac{\mathrm{d}}{\mathrm{d}\Lambda} \left[ G^{\Lambda}_{o\tilde{o}}({k}_1- {k}_4+ {q}) G^{\Lambda}_{\tilde{o}'o'}({q}) \right] \\
				&\quad \quad \times
				V^{\Lambda}_{o_1o'oo_4}\left( {k}_1, {q}; {k}_1- {k}_4+ {q}, {k}_4\right)  V^{\Lambda}_{\tilde{o}o_2o_3\tilde{o}'}\left(  {k}_1- {k}_4+ {q}, {k}_2;  {k}_3, {q} \right),\\
				\mathcal{T}_{\left\lbrace o_i \right\rbrace}^{\mathrm{dph}}\left( {k}_{1},  {k}_{2};  {k}_{3}, {k}_{4}\right) &= \int_{q}
				\frac{\mathrm{d}}{\mathrm{d}\Lambda} \left[ G^{\Lambda}_{o\tilde{o}}({k}_1- {k}_3+ {q}) G^{\Lambda}_{\tilde{o}'o'}({q}) \right] \\
				&\quad \quad \times
				\left[ 2V^{\Lambda}_{o_1o'o_3o}\left( {k}_1, {q}; {k}_3, {k}_1- {k}_3+ {q}\right)   V^{\Lambda}_{\tilde{o}o_2\tilde{o}'o_4}\left(  {k}_1- {k}_3+ {q}, {k}_2;  {q}, {k}_4 \right)\right. \\
				&\left. \quad \quad \quad -V^{\Lambda}_{o_1o'oo_3}\left( {k}_1, {q}, {k}_1- {k}_3+ {q}, {k}_3\right)   V^{\Lambda}_{\tilde{o}o_2\tilde{o}'o_4}\left(  {k}_1- {k}_3+ {q}, {k}_2;  {q}, {k}_4 \right)\right. \\
				&\left. \quad \quad \quad -V^{\Lambda}_{o_1o'o_3o}\left( {k}_1, {q}; {k}_3, {k}_1- {k}_3+ {q}\right)   V^{\Lambda}_{\tilde{o}o_2o_4\tilde{o}'}\left(  {k}_1- {k}_3+ {q}, {k}_2;  {k}_4, {q} \right)\right] ,\\
			\end{aligned}
		\end{equation}
	which are contributions to the flow of effective interactions in particle-particle, crossed particle-hole and direct particle-hole channels respectively 
	as shown in Fig. \ref{fig:fRG}. 
	Here $G^{\Lambda}_{oo'}$ is bare propagator in orbit basis at scale $\Lambda$ and $\int_q=\sum_{oo' \tilde{o}\tilde{o}'}\frac{1}{\beta}\sum_{\omega}\int \frac{\dd \mathbf{q}}{S_{BZ}}$. 
	
	By integrating the flow equations with reducing the RG scale $\Lambda$, one can obtain the renormalized effective interaction. The flow is terminated at a critical scale $\Lambda_c$ when a divergence in a vertex element is encountered, signaling that the normal metallic phase becomes unstable toward a symmetry-broken state. The critical scale $\Lambda_c$ provides an estimate for the critical temperature $T_c$, and the most divergent vertex component indicates the type of symmetry-broken phase the system is likely to enter. In our FRG calculations, we have neglected six-point and higher-order vertices self-energy effects. While these approximations may influence the instabilities, they are considered reasonable within the weak to moderate coupling regime and are not expected to significantly alter the dominant instability.
	We implement our FRG calculations in the truncated unity functional renormalizational group (TUFRG) formalism~\cite{Lichtenstein2017, QHWangGraphene}. In TUFRG, the static four point vertex function is decomposed into a scale independent initial bare interaction $V^{(0)}$ plus three coupling functions $\Phi^{X}, X\in \left\lbrace P,C,D \right\rbrace $,
	\begin{equation} \label{TUdecomposition}
		\begin{aligned}
			V^{\Lambda} & =V^{(0)}+\Phi^{\mathrm{P}}(\Lambda)+\Phi^{\mathrm{C}}(\Lambda)+\Phi^{\mathrm{D}}(\Lambda).
		\end{aligned}
	\end{equation}
	Here 
	$V^{(0)} \equiv V^{\Lambda_0}$ is the initial bare interaction,
	and $\Phi^{\mathrm{P}}$, $\Phi^{\mathrm{C}}$, and $\Phi^{\mathrm{D}}$ are the single-channel coupling function in the particle-particle, crossed
	particle-hole, and direct particle-hole channels respectively.
	The static single channel coupling functions flow according to
	\begin{equation}
		\begin{aligned}
			&\frac{\dd}{\dd \Lambda}\Phi^{\mathrm{P}}_{\left\lbrace  o_i \right\rbrace  }\left( \mathbf{k}_1,\mathbf{k}_2;\mathbf{k}_3,\mathbf{k}_4 \right) = \mathcal{T}_{\left\lbrace o_i \right\rbrace}^{\mathrm{pp}}\left(\mathbf{k}_1,\mathbf{k}_2;\mathbf{k}_3,\mathbf{k}_4\right) \\
			&\frac{\dd}{\dd \Lambda}\Phi^{\mathrm{C}}_{\left\lbrace  o_i \right\rbrace  }\left( \mathbf{k}_1,\mathbf{k}_2;\mathbf{k}_3,\mathbf{k}_4 \right) =\mathcal{T}_{\left\lbrace o_i \right\rbrace}^{\mathrm{cph}}\left(\mathbf{k}_1,\mathbf{k}_2;\mathbf{k}_3,\mathbf{k}_4\right) \\
			&\frac{\dd}{\dd \Lambda}\Phi^{\mathrm{D}}_{\left\lbrace  o_i \right\rbrace  }\left( \mathbf{k}_1,\mathbf{k}_2;\mathbf{k}_3,\mathbf{k}_4 \right) =\mathcal{T}_{\left\lbrace o_i \right\rbrace}^{\mathrm{dph}}\left(\mathbf{k}_1,\mathbf{k}_2;\mathbf{k}_3,\mathbf{k}_4\right)
		\end{aligned}
	\end{equation}
	where $\mathcal{T}_{\left\lbrace o_i \right\rbrace}^{\mathrm{pp}},, \mathcal{T}_{\left\lbrace o_i \right\rbrace}^{\mathrm{pp}} \mathcal{T}_{\left\lbrace o_i \right\rbrace}^{\mathrm{cph}}, \mathcal{T}_{\left\lbrace o_i \right\rbrace}^{\mathrm{dph}}$ are contributions to the flow of interaction in particle-particle, direct particle-hole, crossed particle-hole channels respectively with Feynman diagrams shown in Fig.~\ref{fig:fRG}. 
	For a general coupling function $\mathcal{F}_{\left\lbrace o_i \right\rbrace}\left(\mathbf{k}_{1}, \mathbf{k}_{2}; \mathbf{k}_{3},\mathbf{k}_{4}\right)$, projections onto three channels P, C and D are defined as 
	
	\begin{figure}[t]
		\centerline{\includegraphics[width=0.75\textwidth]{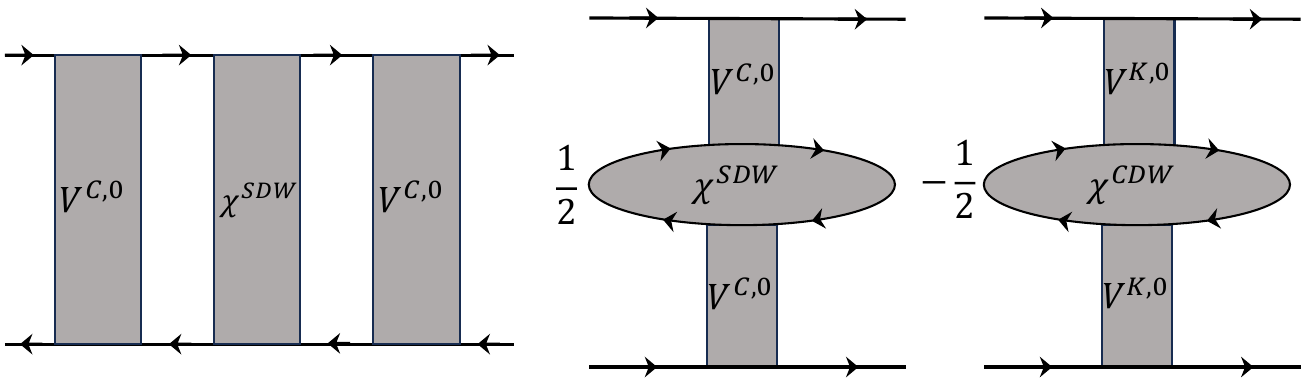}}
		\caption{Effective pairing interaction mediated by spin and charge fluctuations. \label{fig992}}
	\end{figure}

		\begin{gather}
			\mathcal{F}^{\text{P}}= \hat{\mathrm{P}}\left[\mathcal{F} \right] ,\quad \mathcal{F}^{\text{C}}= \hat{\mathrm{C}}\left[\mathcal{F} \right] ,\quad
			\mathcal{F}^{\text{D}}= \hat{\mathrm{D}}\left[\mathcal{F} \right] , \notag\\
			\mathcal{F}^{\text{P}}_{o_1o_2m,o_3o_4n}(\mathbf{q})= S_{\text{BZ}}^{-2} \int \dd \mathbf{k} \dd \mathbf{k}' f_m(\mathbf{k})\mathcal{F}_{o_1o_2o_3o_4  }(\mathbf{k}+\mathbf{q},-\mathbf{k};\mathbf{k}'+\mathbf{q},-\mathbf{k}') f_n^*(\mathbf{k}') ,\\
			\mathcal{F}^{\text{C}}_{o_1o_4m,o_3o_2n}(\mathbf{q})  =S_{\text{BZ}}^{-2} \int \dd \mathbf{k} \dd \mathbf{k}' f_m(\mathbf{k})\mathcal{F}_{o_1o_2o_3o_4  }(\mathbf{k}+\mathbf{q},\mathbf{k}';\mathbf{k}'+\mathbf{q},\mathbf{k}) f_n^*(\mathbf{k}'), \\
			\mathcal{F}^{\text{D}}_{o_1o_3m,o_4o_2n}(\mathbf{q}) =S_{\text{BZ}}^{-2} \int \dd \mathbf{k} \dd \mathbf{k}' f_m(\mathbf{k})\mathcal{F}_{o_1o_2o_3o_4  }(\mathbf{k}+\mathbf{q},\mathbf{k}';\mathbf{k},\mathbf{k}'+\mathbf{q})f_n^*(\mathbf{k}').
		\end{gather}
		Here $f_{m}(\mathbf{k})$ are some formfactor basis satisfying orthogonality and completeness relations
		\begin{equation}
			\begin{aligned}
				&S_{\text{BZ}}^{-1}\sum_{m}f_m(\mathbf{k}) f_m^*(\mathbf{k'})=\delta(\mathbf{k}-\mathbf{k}')
				,\quad S_{\text{BZ}}^{-1}\int \dd \mathbf{k} f_m(\mathbf{k}) f_n^*(\mathbf{k})=\delta_{mn} .
			\end{aligned}
		\end{equation}
		The inverse transformations of above projections read
		\begin{equation}\label{Projections}
			\begin{aligned}
				&\mathcal{F}_{o_1o_2o_3o_4  }(\mathbf{k}+\mathbf{q},-\mathbf{k};\mathbf{k}'+\mathbf{q},-\mathbf{k}') =\sum_{mn} f_m^*(\mathbf{k}) \mathcal{F}^{\text{P}}_{o_1o_2m,o_3o_4n}(\mathbf{q})  f_n(\mathbf{k}') ,\\
				& \mathcal{F}_{o_1o_2o_3o_4  }(\mathbf{k}+\mathbf{q},\mathbf{k}';\mathbf{k}'+\mathbf{q},\mathbf{k}) =\sum_{mn} f_m^*(\mathbf{k}) \mathcal{F}^{\text{C}}_{o_1o_4m,o_3o_2n}(\mathbf{q})  f_n(\mathbf{k}'), \\
				&\mathcal{F}_{o_1o_2o_3o_4  }(\mathbf{k}+\mathbf{q},\mathbf{k}';\mathbf{k},\mathbf{k}'+\mathbf{q})= \sum_{mn} f_m^*(\mathbf{k})\mathcal{F}^{\text{D}}_{o_1o_3m,o_4o_2n}(\mathbf{q})
				f_n(\mathbf{k}').
			\end{aligned}
		\end{equation}

		In the singular mode-FRG (SMFRG)~\cite{QHWangGraphene} or truncated unity-FRG (TUFRG) ~\cite{Lichtenstein2017}  formalism, three bosonic propagators are defined by projecting single channel coupling functions onto three channels, 
		\begin{equation}\label{ProjectionsEP}
			\begin{aligned}
				P= \hat{\mathrm{P}}\left[\Phi^{\mathrm{P}} \right] ,\quad C= \hat{\mathrm{C}}\left[\Phi^{\mathrm{C}} \right] ,\quad
				D= \hat{\mathrm{D}}\left[\Phi^{\mathrm{D}} \right] .
			\end{aligned}
		\end{equation}
		One also define $V^{\mathrm{P}, \mathrm{C},\mathrm{D}} $ as projections of the effective interactions $V^{\Lambda}$ onto three channels,
		\begin{equation}\label{ProjectionsV}
			\begin{aligned}
				V^{\mathrm{P}}= \hat{\mathrm{P}}\left[V \right] ,\quad V^{ \mathrm{C}}= \hat{\mathrm{C}}\left[V \right] ,\quad
				V^{\mathrm{D}}= \hat{\mathrm{D}}\left[V \right] .
			\end{aligned}
		\end{equation}
		The TUFRG flow equations for the bosonic propagators read
		\begin{equation} \label{TUfloweq1}
			\begin{aligned}
				& \frac{\dd}{\dd \Lambda}P(\boldsymbol{q})=V^{\mathrm{P}}(\boldsymbol{q}) \left[ \frac{\dd}{\dd \Lambda}\chi^{\mathrm{pp}}(\boldsymbol{q}) \right] V^{\mathrm{P}}(\boldsymbol{q}), \quad \frac{\dd}{\dd \Lambda}C(\boldsymbol{q})=V^{\mathrm{C}}(\boldsymbol{q}) \left[ \frac{\dd}{\dd \Lambda}\chi^{\mathrm{ph}}(\boldsymbol{q}) \right] V^{\mathrm{C}}(\boldsymbol{q}), \\
				& \frac{\dd}{\dd \Lambda}D(\boldsymbol{q})=\left[V^{\mathrm{C}}(\boldsymbol{q})-V^{\mathrm{D}}(\boldsymbol{q})\right] \left[ \frac{\dd}{\dd \Lambda}\chi^{\mathrm{ph}}(\boldsymbol{q}) \right] V^{\mathrm{D}}(\boldsymbol{q})+V^{\mathrm{D}}(\boldsymbol{q}) \left[ \frac{\dd}{\dd \Lambda}\chi^{\mathrm{ph}}(\boldsymbol{q}) \right] \left[V^{\mathrm{C}}(\boldsymbol{q})-V^{\mathrm{D}}(\boldsymbol{q})\right].
			\end{aligned}
		\end{equation}
		Here $\chi^{\mathrm{pp/ph, \Lambda}}$ is particle-particle/particle-hole susceptibility matrix in orbit and formfactor basis,
		\begin{equation}
			\begin{aligned}
				& \chi_{oo'\tilde{o}\tilde{o}',mn}^{\text{pp}}(\mathbf{q})=-\int_p  f_m(\mathbf{p}) \left[G^{\Lambda}_{o\tilde{o}}(i \omega, \mathbf{q}+\mathbf{p})\right. \left. G^{\Lambda}_{o'\tilde{o}'}( - i \omega, -\mathbf{p})\right]  f_{n}^*(\mathbf{p}),\\
				& \chi_{oo'\tilde{o}\tilde{o}',mn}^{\text{ph}}(\mathbf{q})=-\int_p  f_m(\mathbf{p}) \left[G^{\Lambda}_{o\tilde{o}}(i \omega, \mathbf{q}+\mathbf{p})\right. \left. G^{\Lambda}_{\tilde{o}'o'}(  i \omega,  \mathbf{p})\right]  f_{n}^*(\mathbf{p}).
			\end{aligned}
		\end{equation}
		
		\begin{figure}[]
			\centerline{\includegraphics[width=0.75\textwidth]{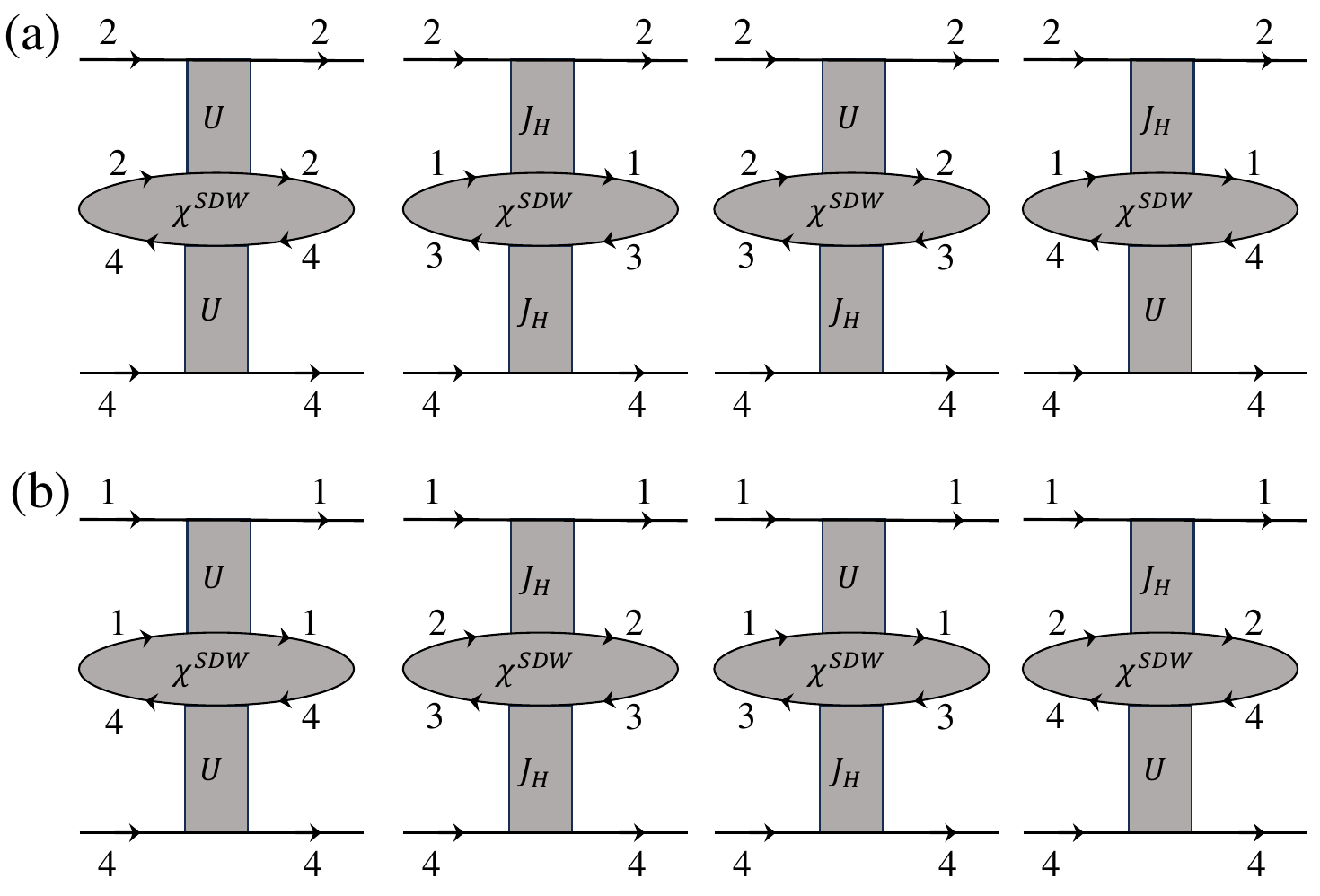}}
			\caption{Dominant contributions to effective pairing interaction $V^{\mathrm{P}}_{24,24}$ (a) and $V^{\mathrm{P}}_{14,14}$ (b) mediated by  spin fluctuations. \label{fig993}}
		\end{figure}
		
		Using Eqs. (\ref{TUdecomposition}), (\ref{ProjectionsEP}) and (\ref{ProjectionsV}) the projections can be expressed by $P, C $
		and $D$,
		\begin{equation}\label{Proj}
			\begin{aligned}
				V^{\mathrm{P}}&=\hat{\mathrm{P}}\left[ V \right] =\hat{\mathrm{P}}\left[ V^{(0)}+\hat{\mathrm{P}}^{-1}\left[P \right]+\hat{\mathrm{C}}^{-1}\left[C \right] +\hat{\mathrm{D}}^{-1}\left[D \right] \right]=V^{\mathrm{P},0}+P+V^{\mathrm{P} \leftarrow \mathrm{C}}+V^{\mathrm{P} \leftarrow \mathrm{D}}, \\
				V^{\mathrm{C}}&=\hat{\mathrm{C}}\left[ V \right] =\hat{\mathrm{C}}\left[ V^{(0)}+\hat{\mathrm{P}}^{-1}\left[P \right]+\hat{\mathrm{C}}^{-1}\left[C \right] +\hat{\mathrm{D}}^{-1}\left[D \right] \right]=V^{\mathrm{C},0}+V^{\mathrm{C} \leftarrow \mathrm{P}}+C+V^{\mathrm{C} \leftarrow \mathrm{D}}, \\
				V^{\mathrm{D}}&=\hat{\mathrm{D}}\left[ V \right] =\hat{\mathrm{D}}\left[ V^{(0)}+\hat{\mathrm{P}}^{-1}\left[P \right]+\hat{\mathrm{C}}^{-1}\left[C \right] +\hat{\mathrm{D}}^{-1}\left[D \right] \right]=V^{\mathrm{D},0}+V^{\mathrm{D} \leftarrow \mathrm{P}}+V^{\mathrm{D} \leftarrow \mathrm{C}}+D.
			\end{aligned}
		\end{equation}
	Thus (\ref{TUfloweq1}) become closed differential equations for exchange propagators. 
	We then solve above TUfRG flow equations with a hard Matsubara frequency cutoff in bare propagator, $G^{\Lambda}_{oo'}(k)=G_{oo'}\theta(|\omega|-\Lambda)$.
	The kernel of the particle-particle (-) and particle-hole (+) bubbles is given by
	\begin{equation}
		\begin{aligned}
			\frac{\dd}{\dd \Lambda} \chi_{\pm}^{\Lambda}(\vb{k_1},\vb{k_2})&=\frac{1}{2\pi}[( G(i\Lambda,\vb{k_1} )G(\pm i\Lambda,\vb{k_2} ) 
			+ G(-i\Lambda,\vb{k_1} )G(\mp i\Lambda,\vb{k_2} ) 
		\end{aligned}
	\end{equation}
	where orbit indexes are not written explicitly. The form factor basis is truncated to 9 terms, encompassing on-site, first-nearest-neighbor, and second-nearest-neighbor bonds.
	
	The effective interactions in pairing, magnetic and charge channels are $-V^{\mathrm{P}}, V^{\mathrm{C}}, V^{\mathrm{C}}-2V^{\mathrm{D}}$ respectively. 
	By diagonalizing the interaction matrices in different channels, expressed as $V^{\mathrm{ch}} = \sum_{i} \Xi^{\mathrm{ch}}_{i} O^{\mathrm{ch}}_{i} O^{\mathrm{ch}\star}_{i}$, the eigenmode $O^{\mathrm{ch}}_{i}$ with a positive eigenvalue $\Xi^{\mathrm{ch}}_{i}$ corresponds to an attractive channel. The channel with the most divergent eigenvalue $\Xi$ is identified as the leading instability of the system.
	By  diagonalizing the effective interaction matrices in real space orbit basis, the eigenmode with largest eigenvalue is identified as the leading instability.
	
	\begin{figure}[t]
		\centerline{\includegraphics[width=0.75\textwidth]{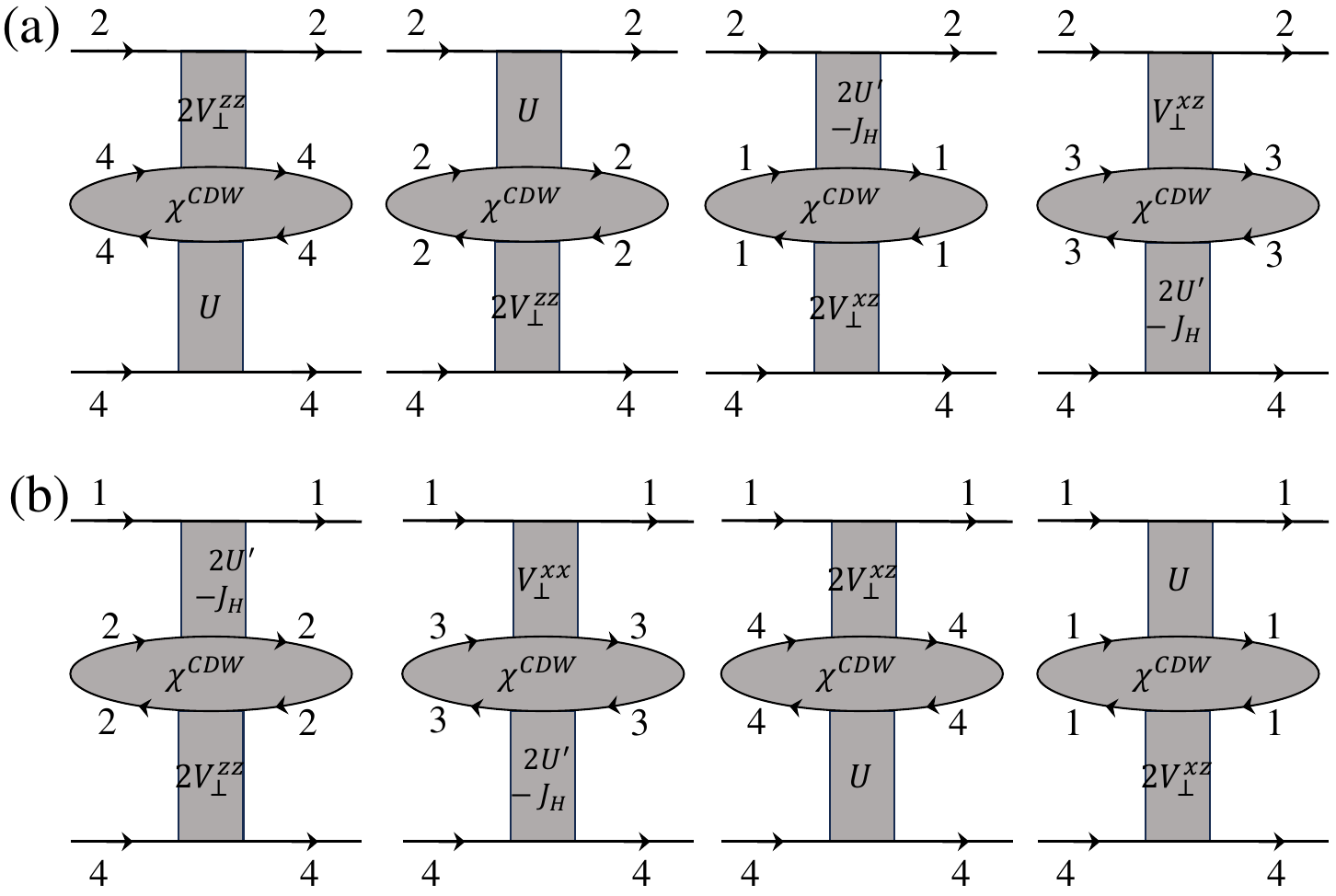}}
		\caption{Dominant contributions to effective pairing interaction $V^{\mathrm{P}}_{24,24}$ (a) and $V^{\mathrm{P}}_{14,14}$ (b) mediated by charge fluctuations. \label{fig994}}
	\end{figure}
	
	\section{Effective pairing interaction from spin and charge fluctuations} \label{apC}
	To obtain an effective pairing interaction mediated by spin and charge fluctuations, we can solve the TUFRG flow equations within RPA, where we neglect the crosschannel projections from P to C/D and interchannel projections between C and D in Eq. (\ref{Proj}),
	\begin{equation}\label{Proj}
		\begin{aligned}
			V^{\mathrm{P}}&=V^{\mathrm{P},0}+P+V^{\mathrm{P} \leftarrow \mathrm{C}}+V^{\mathrm{P} \leftarrow \mathrm{D}}, \\
			V^{\mathrm{C}}&=V^{\mathrm{C},0}+C, \,
			V^{\mathrm{D}}=V^{\mathrm{D},0}+D.
		\end{aligned}
	\end{equation}
	The charge channel vertex is given by $K=C-2D$ with flow equation
	\begin{equation} \label{TUfloweq1}
		\begin{aligned}
			\frac{\dd}{\dd \Lambda}K(\boldsymbol{q})=V^{\mathrm{K}}(\boldsymbol{q}) \left[ \frac{\dd}{\dd \Lambda}\chi^{\mathrm{ph}}(\boldsymbol{q}) \right] V^{\mathrm{K}}(\boldsymbol{q}).
		\end{aligned}
	\end{equation}
	where $V^{\mathrm{K}}=V^{\mathrm{C}}-2V^{\mathrm{D}}=V^{\mathrm{K},(0)}+K$, $V^{\mathrm{K},(0)}=V^{\mathrm{C},(0)}-2V^{\mathrm{D},(0)}$. Under RPA, the flow equations for $C$ and $K$ can be solved with following exact solutions
	\begin{equation} 
		\begin{aligned}
			&\left[  V^{\mathrm{C}} (\Lambda) \right]  ^{-1}-\left[  V^{\mathrm{C},0} \right]^{-1}=\chi^{\mathrm{ph}}(\Lambda_0)-\chi^{\mathrm{ph}}(\Lambda), \\
			&\left[  V^{\mathrm{K}} (\Lambda) \right]  ^{-1}-\left[  V^{\mathrm{K},0} \right]^{-1}=\chi^{\mathrm{ph}}(\Lambda_0)-\chi^{\mathrm{ph}}(\Lambda).
		\end{aligned}
	\end{equation}
	From initial condition $\chi^{\mathrm{ph}}(\Lambda_0)=0$, we obtain
	\begin{equation} 
		\begin{aligned}
			& V^{\mathrm{C}} (\Lambda) =\left( 1- V^{\mathrm{C},0} \chi^{\mathrm{ph}}(\Lambda)  \right) ^{-1}  V^{\mathrm{C},0}  , \\
			& V^{\mathrm{K}} (\Lambda) =\left( 1- V^{\mathrm{K},0} \chi^{\mathrm{ph}}(\Lambda)  \right) ^{-1}  V^{\mathrm{K},0}.
		\end{aligned}
	\end{equation}
	\begin{equation} 
		\begin{aligned}
			& C(\Lambda)=V^{\mathrm{C}} (\Lambda)-V^{\mathrm{C},0} = V^{\mathrm{C},0} \chi^{\mathrm{SDW}}(\Lambda) V^{\mathrm{C},0}, \\
			& K(\Lambda)=V^{\mathrm{C}} (\Lambda)-V^{\mathrm{K},0} = V^{\mathrm{K},0} \chi^{\mathrm{CDW}}(\Lambda) V^{\mathrm{K},0}. 
		\end{aligned}
	\end{equation}
	where SDW and CDW susceptibilities are 
	\begin{equation} 
		\begin{aligned}
			& \chi^{\mathrm{SDW}}(\Lambda) = \left( 1- V^{\mathrm{C},0} \chi^{\mathrm{ph}}(\Lambda)\right)^{-1} \chi^{\mathrm{ph}}(\Lambda) , \\
			& \chi^{\mathrm{CDW}}(\Lambda) = \left( 1- V^{\mathrm{K},0} \chi^{\mathrm{ph}}(\Lambda)\right)^{-1} \chi^{\mathrm{ph}}(\Lambda)  .
		\end{aligned}
	\end{equation}
	Then we use above solutions for  $C$, $K$ and $D$ to solve paring channel flow equation for $P$ 
	\begin{equation} 
		\begin{aligned}
			& \frac{\dd}{\dd \Lambda}P(\boldsymbol{q})=V^{\mathrm{P}}(\boldsymbol{q}) \left[ \frac{\dd}{\dd \Lambda}\chi^{\mathrm{pp}}(\boldsymbol{q}) \right] V^{\mathrm{P}}(\boldsymbol{q})
		\end{aligned}
	\end{equation}
	with  $V^{\mathrm{P}}=V^{\mathrm{P},0}+V^{\mathrm{P} \leftarrow \mathrm{C}}+V^{\mathrm{P} \leftarrow \mathrm{D}}+P$. Within RPA, 
	\begin{equation} 
		\begin{aligned}
			& V^{\mathrm{P} \leftarrow \mathrm{C}}= \hat{\mathrm{P}} \circ \hat{\mathrm{C}}^{-1} \left[ V^{\mathrm{C},0} \chi^{\mathrm{SDW}} V^{\mathrm{C},0} \right] , \\
			&V^{\mathrm{P} \leftarrow \mathrm{D}}= \hat{\mathrm{P}} \circ \hat{\mathrm{D}}^{-1} \left[ \frac{1}{2}V^{\mathrm{C},0} \chi^{\mathrm{SDW}} V^{\mathrm{C},0}  
			-\frac{1}{2} V^{\mathrm{K},0} \chi^{\mathrm{CDW}} V^{\mathrm{K},0}  \right] ,
		\end{aligned}
	\end{equation}
	which are effective pairing interaction from spin and charge fluctuations and diagrammatically shown in Fig.~\ref{fig992}. 
	
	The four orbits $(xt,zt,xb,zb)$ are labeled as $(1,2,3,4)$ in the following. For pure on-site multi Hubbard interactions, the dominated  contributions to interlayer pairing interaction $V^{\mathrm{P}}_{24,24}$ and $V^{\mathrm{P}}_{14,14}$ are shown in Fig.~\ref{fig993}. Due to nesting between bonding $\alpha$ pocket and antibonding $\beta$, $\chi^{\mathrm{SDW}}_{24,24}$ and $\chi^{\mathrm{SDW}}_{13,13}$ will be negative and first two diagrams in Fig.~\ref{fig993}(a) will induce a attractive interlayer pairing interaction on $d_{z^2}$ orbit, which is responsible for $s_{\pm}$ superconductivity. The interlayer interorbit pairing interaction is much smaller due to small interlayer interorbit hybridization.

	With the interlayer nonlocolocal Coulomb interaction, the charge fluctuations are enhanced. The dominant contributions to $V^{\mathrm{P}}_{24,24}$ and $V^{\mathrm{P}}_{14,14}$ from charge fluctuations are shown in Fig.~\ref{fig994}.
	The interlayer interorbit pairing  interaction $V^{\mathrm{P}}_{14,14}$ are greatly enhanced from the first two diagram in Fig.~\ref{fig994}(b). Note that $\chi^{\mathrm{SDW}}_{ii,ii}$ is positive and overall factor $-1/2$ is not shown. 
	Hence charge fluctuations can mediated a strong attractive pairing interaction for the interlayer interorbit $d_{x^2-y^2}$ superconductivity.

	\section{Effect of interlayer interorbital repulsion on pairing without $\gamma$ pocket} 
	\begin{figure}[t]
		\centerline{\includegraphics[width=0.4\textwidth]{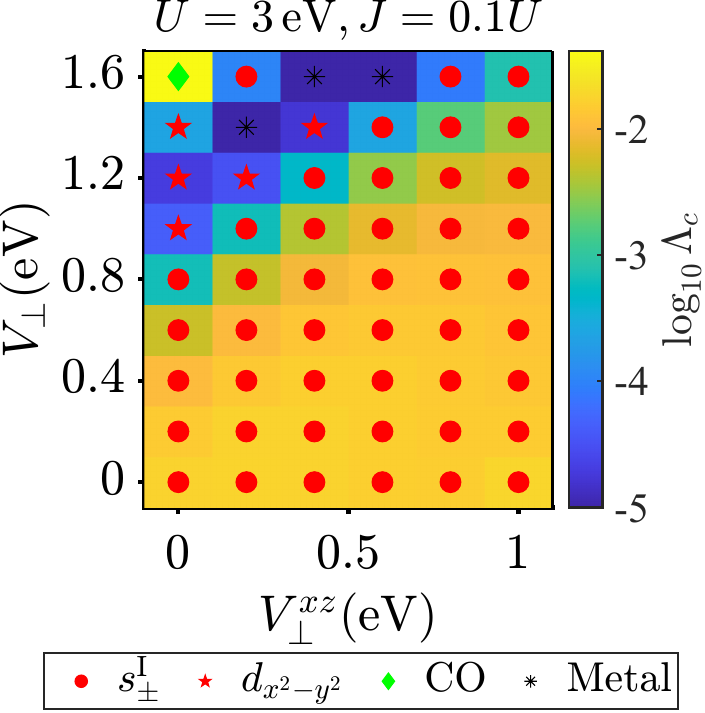}}
		\caption{ Interlayer-interorbit repulsion $V_{\perp}^{xz}$ versus interlayer-intraorbit repulsion $V_{\perp}$ phase diagram with $U=3$ eV, $J_{H}=0.1U$, $n=3.33$. In the region marked by black star, the RG flow exhibits no divergence above the scale of $10^{-5}$ so we label it as Metal. \label{fig000}}
	\end{figure}
	For the fermiology without the $\gamma$ pocket, the interlayer \( s^{\mathrm{I}}_{\pm} \)-wave pairing remains stable as long as the Hund's coupling is not too strong. The effects of interlayer intraorbital (\( V_{\perp} \)) and interorbital (\( V_{\perp}^{xz} \)) repulsions are shown in Fig.~\ref{fig000}. 
	It is evident that the presence of \( V_{\perp}^{xz} \) shifts the transition from \( s_{\pm} \) to \( d_{x^2 - y^2} \) pairing to larger values of \( V_{\perp} \), thereby enhancing the robustness of the \( s_{\pm} \)-wave pairing when both types of interlayer interactions are present.

\end{document}